\documentclass[11pt,a4paper]{article}
\usepackage[utf8]{inputenc} 			
\usepackage[T1]{fontenc} 			
\usepackage{csquotes}					
\usepackage{mathtools}
\usepackage{amsfonts}
\usepackage{amssymb}
\usepackage{amsmath}
\usepackage{amstext}
\usepackage{mathrsfs}
\numberwithin{equation}{section}
\usepackage{dsfont}
\usepackage{multicol} 
\usepackage{esdiff} 
\usepackage{multirow}
\usepackage{multicol}
\usepackage{booktabs}
\usepackage{geometry}
\usepackage{enumitem}
\usepackage[usenames,dvipsnames]{color}
\usepackage{tikz}
\usetikzlibrary{decorations.pathmorphing,decorations.markings,trees}
\usetikzlibrary{calc}
\usepackage{booktabs}
\usepackage{blindtext}
\usepackage{fancyhdr}
\usepackage{graphicx}
\usepackage{wrapfig}
\usepackage{float}
\usepackage{siunitx}
\usepackage{textcomp}
\usepackage[numbers,square,sort&compress]{natbib}
\usepackage{bm}
\usepackage{cancel}
\usepackage{cases}
 \usepackage{slashed}
\newcommand{\dd}{\mathrm{d}}

\renewcommand{\thanks}[1]{\footnote{#1}}

\usepackage{bm}
\usepackage{accents}
\usepackage{hyperref}
\newcommand{\be}{\begin{equation}}
\newcommand{\ee}{\end{equation}}
\newcommand{\im}{\mathrm{i}}
\newcommand{\e}{\operatorname{e}}
\usepackage{authblk}
\usepackage[onehalfspacing]{setspace}
\usepackage[font={stretch=1.1}, width=1.1\linewidth]{caption}
\begin{document}
\pagenumbering{roman} 
\begin{titlepage}
\begin{center}
\begin{figure}[H]
\centering
\includegraphics[width=5cm]{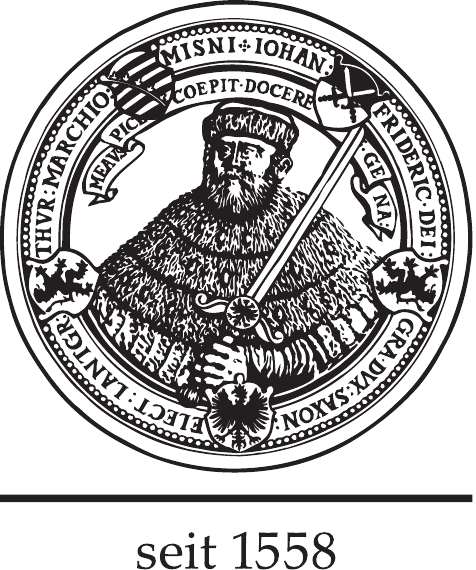}
\end{figure}

\vspace{1\baselineskip}

\textbf{\Huge{Holographic quenches and anomalous transport}}

\vspace{2\baselineskip}

\textsc{\Large{Masterarbeit}}

\vspace{2\baselineskip}

\Large{zur Erlangung des akademischen Grades\\ Master of Science (M. Sc.)\\ im Studiengang Physik}

\vspace{2\baselineskip}

\textsc{\Large{Friedrich-Schiller-Universität Jena}}
\textsc{\Large{Physikalisch-Astronomische Fakultät}}
\textsc{\Large{Theoretisch-Physikalisches Institut}}

\vspace{2\baselineskip}

\Large{Original eingereicht am 30.09.2016 von Sebastian Grieninger}%\\ geb. am 14.08.1991 in Hassfurt}
\end{center}
%\newpage
%\thispagestyle{empty}
%\begin{figure}
%\begin{minipage}{1cm}

%\end{minipage}
%\vspace{12.5cm}
%\begin{minipage}{1cm}

%\end{minipage}
%\end{figure}
%\noindent
%\textbf{\Large{Gutachter}}
%\vspace{1.5\baselineskip}\\
%Erstgutachter:
%\begin{figure}[H]
%\begin{minipage}{5cm}
%\end{minipage}
%\hspace{1cm}
%\begin{minipage}{8cm}
%{Jun. Prof. Dr. rer. nat. Martin Ammon}\\
%Theoretisch-Physikalisches Institut	\\
%Friedrich-Schiller-Universität Jena
%\end{minipage}

%\end{figure}

%\noindent
%Zweitgutachter:
%\begin{figure}[H]
%\begin{minipage}{5cm}
%\end{minipage}
%\hspace{1cm}
%\begin{minipage}{10cm}
%Dr. rer. nat. Amadeo Jiménez-Alba\\
%Theoretisch-Physikalisches Institut	\\
%Friedrich-Schiller-Universität Jena
%\end{minipage}

%\end{figure}
\end{titlepage} 
 \cleardoublepage
 \newpage%
 \section*{Abstract}
 In my master thesis, I investigated the chiral-magnetic effect in the context of holography; I focused in especially on the impact of the chiral anomaly at transport properties and non-equilbrium behaviour in response to an holographic quench. Concretely, I considered an  $U(1)_\text{A}\times U(1)_\text{V}$-Einstein-Maxwell bottom-up model consisting of two massless gauge fields, coupled by a Chern-Simons term in the fivedimensional AdS spacetime. The two gauge fields provide a time dependent electric field and a static magnetic field, parallel to it. As response of the system to quench, I investigated the electromagnetic current in direction of the magnetic field which is generated due to the CME. In the first part of the thesis, I characterised the initial response of the system, in a fixed Schwarzschild AdS background, subjected to a ``fast'' quench. The corresponding hyperbolic PDE is solved by means of a fully spectral code in spaces as well as in time direction. Note that this was the first application of a fully spectral code within holography. In the case of ``fast'' quenches, the system exhibits an universal scaling behaviour, independent of the external parameters as the strength of the anomaly and the magnetic field, respectively. The late time behaviour of the system shows, depending on the quench and external parameters, in some cases long lived oscillations in the current. In order to determine whether the oscillations are an anomaly driven phenomenon or an artefact of the magnetic field, I computed the quasi-normal modes of the systems, including the backreaction of the matter fields on the background metric. It turns out that the long lived oscillations appear only in presence of the anomaly and can be traced back to the presence of Landau levels in the system. The results of my master thesis were partly published in \cite{Ammon:2016fru}; however, the thesis contains a lot of interesting, and so far unpublished, results and can be viewed as extended version of the paper. 
\thispagestyle{empty}
\newpage
\newpage
\small{\tableofcontents}
\raggedbottom
\thispagestyle{empty}
\newpage
\pagenumbering{arabic}
\normalsize
\section{Introduction}
Dealing with strongly coupled field theories is a challenging topic even nowadays. Quantum field theories are best understood in their pertubative regime, since an expansion in the coupling constant $\lambda$ leads to significant simplifications in the calculation and makes them at least possible, respectively. However, in general strongly coupled systems are not accessible in the context of pertubation theory, since an expansion in terms of the coupling constant does not fulfil the requirement $\lambda\ll1$ and the corresponding power series expansion is not convergent at all. In order to tackle questions in strongly coupled field theories, we have to go beyond the pertubative regime. \newline \newline
The standard approaches to calculations in the non-pertubative regime are the (functional) renomalisation group, lattice field theories and the AdS/CFT correspondence. The acronym AdS/CFT is an abbreviation for \textbf{A}nti-\textbf{d}e \textbf{S}itter/\textbf{C}onformal \textbf{F}ield \textbf{T}heory, where the former refers to a spacetime and the latter to a certain type of field theory. Originally, the AdS/CFT correspondence was developed in context of string theory. Within string theory, the connection between gauge theories and gravitational theories is no surprise. Field theoretical degrees of freedom are captured by open strings ending on hypersurfaces embedded in a higher dimensional spacetime, whereas the gravitational degrees of freedom are described by closed strings. Maldacena \cite{Maldacena:1997re} pointed out that in certain situations the theories of closed and open strings have the same degrees of freedoms and are therefore dynamically equivalent. This shows the deep connection between gauge and gravitational theories, which is also present in the AdS/CFT correspondence. \newline\newline
In \textbf{Q}uantum \textbf{F}ield \textbf{T}heory (QFT) we know three types of symmetry breaking, namely explicit, spontaneous or anomalous. The former occurs by adding terms to the Lagrangian, which break the symmetry. We refer to the symmetry breaking as spontaneously, if the symmetry is respected at the level of the action, but not by the state itself. Thus, the theory is symmetric, but the vacuum state breaks the symmetry. Anomalous symmetry breaking ``happens'' during the quantisation procedure; namely, the full renormalised theory is not invariant under a certain symmetry of the classical theory. If we consider a process, which is based on external symmetries, anomalies will capture the core properties, which would be otherwise highly suppressed. For example, the scattering amplitude of the decay $\pi^0\to\gamma\gamma$ is completely determined by the so called triangle anomaly. 
Thus, anomalies are not only visible on microscopic scales but can also have a macroscopic manifestation, influencing transport phenomena.      
\newline\newline
In the recent years, macroscopic phenomena induced by anomalous effects gained a lot of interest. One of the macroscopic phenomena, which has its origin in the presence of anomalies in the underlying theory, is the \textbf{C}hiral \textbf{M}agnetic \textbf{E}ffect (CME) \cite{Fukushima:2008xe, Son:2009tf} (see \cite{Kharzeev:2013ffa} for a review). The CME requires the presence of an external magnetic field and may lead to three experimental results: the appearance of an electromagnetic current in direction of the magnetic field, a massless propagating mode, the so called chiral magnetic wave \cite{Kharzeev:2010gd}, and in presence of parallel electromagnetic field one observes an enhancement of the electric conductivity, known as negative magnetoresistence~\cite{Nielsen:1983rb}.
The requirement of massless fermions is a crucial restriction on possible experimental playgrounds. Nevertheless, there emerged two main systems, namely the \textbf{Q}uark-\textbf{G}luon \textbf{P}lasma (QGP) and Weyl/Dirac semimetals. The QGP is a macroscopic system, consisting of deconfined quarks and gluons. Although the QGP is an extreme state of matter, we have experimental access to such systems. Despite the high temperatures, the quark gluon plasma is strongly coupled \cite{Heinz:2008tv,Shuryak:2008eq}. Within \textbf{H}eavy-\textbf{I}on \textbf{C}ollisions (HIC) a QGP is produced for $10^{-24}$ s. Moreover, during the collision very strong magnetic fields emerge for a short time. Due to the high temperatures in the QGP, we can neglect the quark masses.
The second experimental playground, related to the CME, are Weyl/Dirac semimetals. Although it is not clear yet, whether Weyl/Dirac semimetals are strongly coupled or not, one of the abovementioned CME-ingredients, namely the negative magnetoresistence, was already measured in several kind of Weyl/Dirac semimetals \cite{Li:2014bha, Li2015, 2015PhRvX...5c1023H, 2016NatCo...710301L}. \newline\newline
A complete understanding of the out-of-equilibrium properties \cite{Yee:2009vw, Landsteiner:2013aba} requires real time studies of the system under consideration. However, it is very challenging to perform real-time calculations within lattice field theories. Nevertheless, the AdS/CFT correspondence provides a well suited tool, to study real-time dynamics in strongly coupled field theories. The CME in presence of time dependent electromagnetic fields has already been studied at weak coupling in \cite{Fukushima:2015tza, Iwazaki:2009bg}  and via holography during thermalization in \cite{Lin:2013sga}.
Throughout our studies, we especially have Weyl/Dirac semimetals in mind, since in these ``chiral'' condensed matter systems the CME vanishes in equilibrium \cite{Basar:2013iaa}. Therefore, we can ``push'' the system in an out-of-equilibrium state and monitor the out-of-equilibrium response. We can generate a current in these materials by introducing a chiral imbalance, realised by switching on a finite axial potential; this can be done by a time dependent axial charge in presence of parallel electric and magnetic fields
\begin{equation}
\dot{\rho}_5\sim\bm{E}\cdot\bm{B}.
\end{equation} 
Therefore, the local change rate in chirality depends on the product of the electric and magnetic field. Allowing for a time dependent axial charge density results in the non-conservation of axial charge. In order to obtain a non-trivial evolution of the axial charge density, either the electric field or the magnetic field (or both) has to be time dependent. Throughout this thesis, we choose a static magnetic field and a time dependent electric field. In particular, we are interested in fast quenches of the electric field for the following reason.\newline\newline
Recently, it was experimentally possible to gain insights into strongly coupled systems, subjected to a certain kind of quantum quenches. To model these systems theoretically, we have to consider a time dependent parameter in the Hamiltonian, for example the coupling constant, and perform real-time computations. This was already done for a certain class of holographic models, revealing an interesting feature; the authors of \cite{Buchel:2013lla} found an universal behaviour in the early time response for fast enough quenches. We are interested whether our holographic model shows this universal behaviour too and how the anomaly affects it. 
\newline\newline
The thesis is structured as follows. In chapter \ref{sec:an}, we present a short review of anomalies with focus on the triangle anomaly and its application in external fields. The third chapter reviews the essential ingredients of holography. Chapter \ref{sec:tdcih} is mainly based on the work \cite{Ammon:2016fru} of the author in collaboration with Martin Ammon, Amadeo Jiménez-Alba, Rodrigo P. Macedo, and Luis Melgar.
In section \ref{sec:timeholo} we discuss the holographic model and in particular our ansatz for the gauge fields. In order to characterise the non-equilibrium dynamics, we first have to understand the stationary solutions, outlined in section \ref{sec:statsol}. The stationary solutions are used as reference in order to characterise the non-equilibrium response. In section \ref{sec:nonstatsol}, we derive the time-dependent equation and build the numerical set-up. To get an intuition of the quench induced dynamics, we briefly discuss in section \ref{sec:gahtq} the qualitative differences for different magnetic fields, anomaly parameter, and abruptnesses of the quench. In section \ref{sec:uifq}, we investigate the early time response in regards to universal features. The late time response of the system is captured by the quasi-normal modes of the system. In section \ref{sec:qnmf}, we explain the necessary ingredients to compute QNMs in our set-up. The last part is dedicated to characterise the resonances, we observed in the current.
 In chapter \ref{sec:conclu}, we interpret our results and give an outlook for further investigations.
\newpage
\section{Anomalies}\label{sec:an}
In classical field theory Noether's theorem states that symmetries of the classical action lead to conserved currents. For example, Poincaré invariance leads to conservation of the stress-energy tensor, $T_{\mu\nu}$, whereas global internal symmetries lead to conserved currents, $J_\mu$. At the quantum level, the conservation of the currents results in relations between correlation functions containing the corresponding currents, the so called Ward identities. However, symmetries derived at the classical level are not necessarily present at the quantum level; anomalies are the failure of a classical symmetry to survive the quantisation procedure. They can appear because there is no regulator preserving all symmetries and removing the infinities  at the same time. In the path integral formalism anomalies appear, because the path integral measure is, due to the regularisation, not invariant under certain symmetry transformations. Anomalies are important and can in fact be seen experimentally, for example in the decay of a neutral pion to two photons. However, anomalies constrain the dynamics and provide consistency conditions for theories; for example, ``gauge symmetries'' are not allowed to be anomalous, since it would result in a break down of unitarity. %%%%%%%%%%%%%%%%%%%%%%%%%%%%%%%%%%%%%%%%%%%%%%
\newline\newline In order to discuss symmetries on the quantum level, we first review symmetries of the classical theory which are relevant for the discussion about anomalies. It is therefore customary, to understand the left-/right-handed basis and the corresponding connection to the axial/vector formulation \cite{Bertlmann:1996xk}. \newline\newline
We consider a prototype Lagrangian of left- and right-handed fermions,
\begin{equation}
\mathcal L_{\text{L,R}}=\bar\psi_\text{L}\,(\im \slashed\partial+\slashed A^\text{L})\,\psi_\text{L}+\bar\psi_\text{R}\,(\im\slashed\partial+\slashed A^\text{R})\,\psi_\text{R}.\label{eq:leftright}
\end{equation}
The Lagrangian is invariant under local $U_L(1)$ and $U_R(1)$ transformations and preservs therefore an $U_\text{L}(1)\times U_\text{R}(1)$ symmetry. The corresponding symmetry transformations are depicted in table \ref{tab:sym2}.
\begin{table}[H]\centering
\begin{tabular}{*{2}{c}}
	\toprule
	$U_\text{L}(1)$ & $U_\text{R}(1)$ \\
	\midrule
	$\psi_\text{L}\to\e^{\im\Omega_\text{L}(x)}\psi_\text{L}$ & $\psi_\text{R}\to\e^{\im\Omega_\text{R}(x)}\psi_\text{R}$ \\
	$A^\text{L}_\mu\to A^\text{L}_\mu+\partial_\mu\Omega_\text{L}(x)$ & $A^\text{R}_\mu\to A^\text{R}_\mu+\partial_\mu\Omega_\text{R}(x)$ \\
	\bottomrule\end{tabular}\caption{Symmetries of the Lagrangian eq. \eqref{eq:leftright}.\label{tab:sym2}}
	\end{table}
The corresponding left- and right-handed symmetry currents can be computed by
\begin{align}
	J_\mu^\text{L}&=\bar\psi_\text{L}\gamma_\mu\psi_\text{L}
\ \ \ \text{and}\ \ \	J_\mu^\text{R}=\bar\psi_\text{R}\gamma_\mu\psi_\text{R}. 
\end{align}
We can define a notion of chirality by considering the eigenvalue equation $\gamma_5\psi_\pm=\pm\psi_\pm$ for $\psi_\text{L,R}\equiv\psi_\pm$. The eigenvalue $\pm 1$ of $\gamma_5$\footnote{We define $\gamma_5$ by $\gamma_5=\im\gamma_0\gamma_1\gamma_2\gamma_3$.} with respect to $\psi_\pm$ denotes the chirality.\newline\newline
The left-handed $\psi_\text{L}$ and right-handed $\psi_\text{R}$ components can be united to a single spinor $\psi$
\begin{equation}
\psi_\text{L,R}\equiv	\psi_\pm= P_\pm \psi,
\end{equation}
where we defined the projection operator
\begin{equation}
	P_\pm=\frac 12(1\pm\gamma_5).
\end{equation}
Similar to the spinors, the left- and right-handed gauge field can be combined, resulting in an axial and vector gauge field
\begin{equation}
	A_\mu^\text{L}=V_\mu+A_\mu, \ A_\mu^\text{R}=V_\mu-A_\mu.
\end{equation}
\begin{table}[H]\centering
\begin{tabular}{*{2}{c}}
	\toprule
	$U_\text{V}(1)$ & $U_\text{A}(1)$ \\
	\midrule
	$\psi\to\e^{\im\alpha(x)}\psi$ & $\psi\to\e^{\im\beta(x)\,\gamma_5}\psi$ \\
	$V_\mu\to V_\mu+\partial_\mu\alpha(x)$ & $A_\mu\to A_\mu+\partial_\mu\beta(x)$ \\
	\bottomrule\end{tabular}
\caption{Symmetries of the Lagrangian eq. \eqref{eq:lagrmod}.\label{tab:sym}}
\end{table}
We can rewrite the Lagrangian \eqref{eq:leftright} in terms of two external gauge fields, namely the axial gauge field, $A_\mu$, and the vector gauge field, $V_\mu$,
\begin{equation}
	\mathcal L(V_\mu,A_\mu)=\bar\psi\,(\im \slashed \partial+\slashed V+\slashed A\,\gamma_5)\,\psi.\label{eq:lagrmod}
\end{equation}
In the vector/axial basis the $U_\text{L}(1)\times U_\text{R}(1)$ symmetry translates into an $U_\text{V}(1)\times U_\text{A}(1)$ symmetry, depicted in table \ref{tab:sym}.
The corresponding symmetry currents can be computed by
\begin{align}
	J_\mu&=\bar\psi\gamma_\mu\psi =J_\mu^\text{L}+J_\mu^\text{R},\label{eq:lh}\\
 J^5_\mu&=\bar\psi\gamma_\mu\gamma_5\psi= -J_\mu^\text{L}+J_\mu^\text{R}\label{eq:rh}.
\end{align}
\subsection{The Abelian anomaly}\label{sec:aban}
The existence of anomalies was noticed when Sutherland \cite{SUTHERLAND1967433} and Veltman \cite{10.2307/2415932} worked out a theorem, which forbade the decay of a neutral pion into two photons. However, this theorem was in contradiction to the experimental results; \textbf{A}dler, \textbf{B}ell, and \textbf{J}ackiw (ABJ) \cite{PhysRev.177.2426,Bell:1969ts} resolved the contradiction by taking anomaly effects into account. In this way they were able to calculate the decay rate in agreement with the one found experimentally. Concretely, they considered the triangle Feynman diagram of the decay, consisting of two vector and one axial current and revealed how to cure the UV divergence correctly. They found that not both symmetry currents, the axial and the vector current, can be conserved at the same time and therefore one of the Ward identities has to be violated.\newline\newline
In the physical meaningful case, where the vector symmetry is gauged and the axial not, the Ward identity, corresponding to the vector current holds, whereas the axial current is anomalous; namely (in the massless case),
\begin{equation}
\partial_\mu J_\textbf{V}^\mu=0, \ \partial_\mu J_5^\mu=\mathcal A, \label{eq:ano}
\end{equation}
where the anomalous contributions is denoted by (with $\tilde\varepsilon$ being the Levi-Civita symbol),
\begin{equation}
\mathcal A=\frac{e^2}{16\pi^2}\tilde\varepsilon^{\mu\nu\alpha\beta}F_{\mu\nu}F_{\alpha\beta}.\label{eq:anomop}
\end{equation}
The anomalous contribution to eq. \ref{eq:ano}, $\mathcal A$, is referred to as the famous ABJ anomaly. As we will see throughout this chapter, eq. \eqref{eq:ano} is not the only possibility to express the divergences of the current.\newline\newline
In this section, we derive the result motivated in eq. \eqref{eq:ano}. Therefore, we consider the QED-Lagrangian \cite{Bertlmann:1996xk}
\begin{equation}
\mathcal L_\text{QED}=\bar\psi\,(\im\slashed\partial-m+\text{e} \slashed A)\,\psi-\frac 14 F_{\mu\nu}F^{\mu\nu}.
\end{equation}
We can construct a vector current $J_\mu=\bar\psi\gamma_\mu\psi$, an axial current $J^5_\mu=\bar\psi \gamma_\mu\gamma_5\psi$, and a pseudo-scalar $P=\bar\psi\gamma_5\psi$, obeying the corresponding conservation laws
\begin{equation}
\partial^\mu J_\mu=0\ \ \text{and} \ \ \partial^\mu J_\mu^5=2\im mP.\label{eq:klass}
\end{equation}
The classical conservation laws imply, at the quantum level, relations among Green's functions, the so called Ward identities. They are given by the derivative of the vacuum expectation value of the time ordered product of a current $J^\mu$ and generic operators $\mathcal O^i$, namely
\begin{align}
\partial^x_\mu\langle 0|\mathcal T J^\mu(x)\,&\mathcal O^1(y_1)\,\ldots\,\mathcal O^n(y_n)|0\rangle=\langle 0|\mathcal T\partial^x_\mu J^\mu(x)\,\mathcal O(y_1)\,\ldots\,\,\mathcal O(y_n)|0\rangle\nonumber\\
&+\sum\limits_{i=1}^{n}\langle 0 |\mathcal T\,[J^0(x),\mathcal O^i(y_i)]\,\delta(x-y_i)\,\mathcal O^1\ldots\mathcal O^{i-1}\mathcal O^{i+1}\,\ldots\mathcal O^n|0\rangle.
\end{align}
If the Ward identities are violated, the renormalisability of the quantum field theory will be lost. In our case, we are interested in the three point functions, consisting of the above mentioned currents and pseudo-scalar, respectively,
\begin{align}
\tilde M_{\mu\nu\lambda}&\equiv\langle 0|\mathcal T J_\mu(x)J_\nu(y)J_\lambda^5(z)|0\rangle,\\
\tilde M_{\mu\nu}&=\langle 0|\mathcal T J_\mu(x)J_\nu(y)P(z)|0\rangle,
\end{align}
which correspond to the triangle graphs. For convenience regarding the further discussion, we transform the amplitudes $\tilde M_{\mu\nu\lambda}$ and $\tilde M_{\mu\nu}$ to momentum space, for example 
\begin{equation}M_{\mu\nu}=\int\dd^4x\,\dd^4y\,\dd^4z\,\e^{\im (k_1x+k_2y-qz)}\,\tilde M_{\mu\nu}.\end{equation} Afterwards, we can re-express the quantities using the classical conservation laws eq. \eqref{eq:klass} and integrating by parts, which leads us to
\begin{align}
q^\lambda M_{\mu\nu\lambda}&=2m\,M_{\mu\nu},&\text{axial Ward identity (AWI)}\\
k^\mu M_{\mu\nu\lambda}&=0=k^\nu M_{\mu\nu\lambda}&\text{vector Ward identity (VWI).}
\end{align}
\begin{figure}[H] 
	\centering
	\includegraphics[width=9cm]{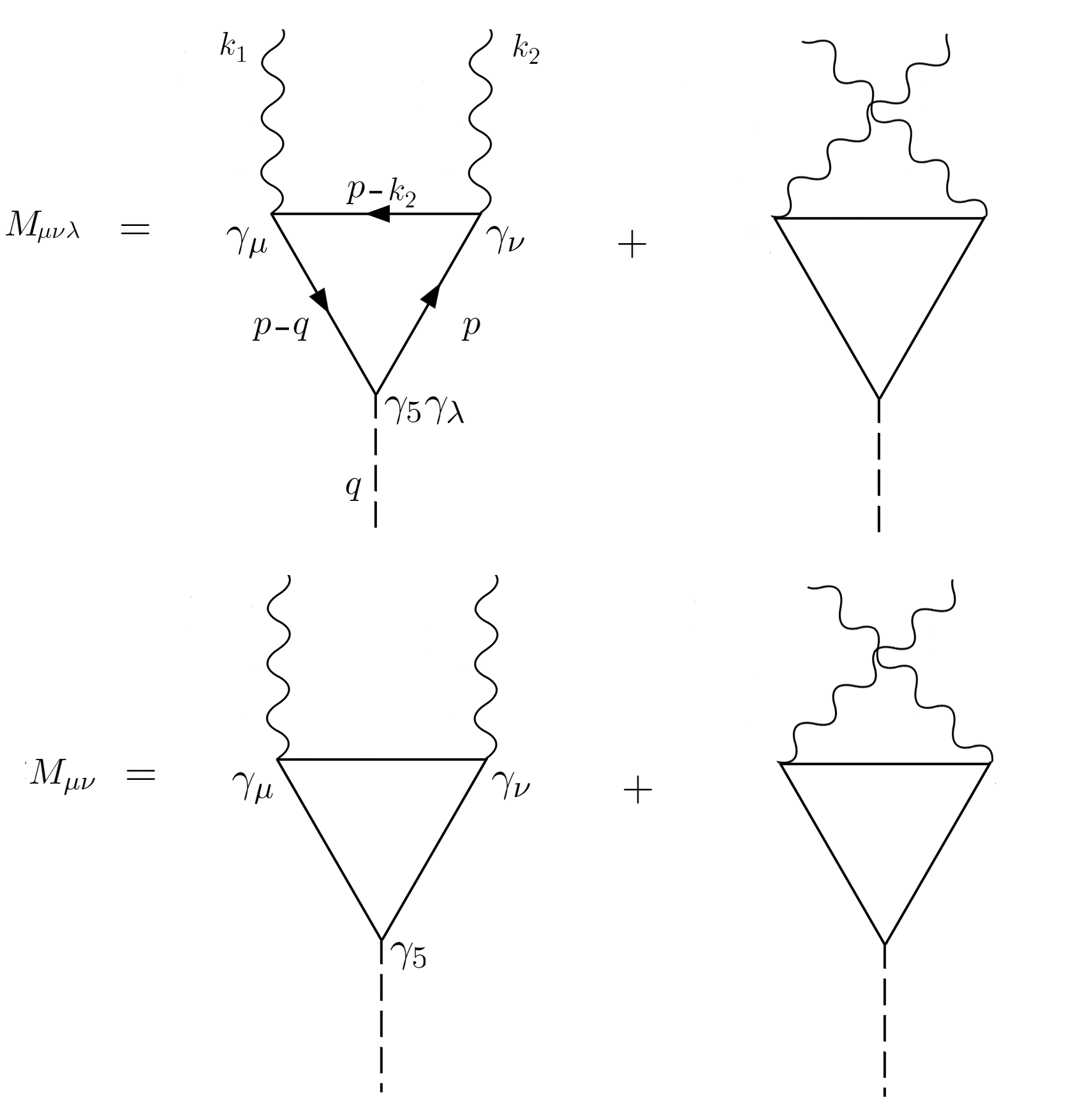}
	\caption{\label{fig:CME}Triangle diagrams consisting of vector-vector-axial currents and vector-vector-pseudoscalar, respectively, and a charged fermion loop.}
\end{figure}
However, the Feynman graphs shown in figure \ref{fig:CME} imply that the WIs do not hold. If the VWI is fulfilled, the AWI will be anomalous and vice versa. But what is going wrong in our derivation? The answer is, we were not allowed to integrate by parts, because the Green's functions are divergent and have to be regulated. The regularisation yields via a ``surface`` term, arising in the integration by parts, to the anomalous contribution.
In order to derive this contribution, we regulate the triangle graphs of the vector-vector-axial currents and vector-vector-pseudoscalar currents, depicted in figure \ref{fig:CME}. 
Using Feynman rules, we can read of the amplitudes using the abbreviation $G(p)=\im/(\slashed p-m)$
\begin{align}
M_{\mu\nu\lambda}&=\im\int\frac{\dd^4p}{(2\pi)^4}(-)\,\text{tr}\left[G(p)\gamma_\lambda\gamma_5 G(p-q)\gamma_\nu G(p-k_1)\gamma_\mu+\begin{pmatrix} k_1\leftrightarrow k_2\\ \mu\leftrightarrow \nu\end{pmatrix}\right],\\
M_{\mu\nu}&=\im\int\frac{\dd^4p}{(2\pi)^4}(-)\,\text{tr}\left[G(p)\gamma_5 G(p-q)\gamma_\nu G(p-k_1)\gamma_\mu+\begin{pmatrix} k_1\leftrightarrow k_2\\ \mu\leftrightarrow \nu\end{pmatrix}\right],
\end{align}
where we set $q=k_1+k_2$ due to momentum conservation. Contracting the first identity with $q^\lambda$ and using $\slashed q\gamma_5=\gamma_5\,(\slashed p-\slashed q)-m
+(\slashed p-m)\,\gamma_5+2m\,\gamma_5$ results in
\begin{equation}
q^\lambda M_{\mu\nu\lambda}=2m\,M_{\mu\nu}+R^1_{\mu\nu}+R^2_{\mu\nu},
\end{equation}
where \begin{align}
R^1_{\mu\nu}(k_1,k_2)&=-\int\frac{\dd^4p}{(2\pi)^4}\,\text{tr}\left[G(p-k_2)\,\gamma_5\gamma_\nu\,G(p-q)\gamma_\mu-G(p)\,\gamma_5\gamma_\mu\,G(p-k_1)\,\gamma_\nu\right]\\&\equiv R^2_{\nu\mu}(k_2,k_1). 
\end{align}
Note that the AWI holds when the rest terms $R^i_{\mu\nu}$ vanish. This happens when we shift the momenta in the first term of the rest terms by $p\to p+k_2$ and by $p\to p+k_1$, respectively. However, the integrals are linear divergent\footnote{The quadratic divergence vanishes since $\tilde\varepsilon_{\mu\nu\alpha\beta}p^\alpha p^\beta=0$.} and hence we are not allowed to do a shift. Nevertheless, we have to evaluate the integrals; to do this we switch to spherical momentum coordinates and introduce an radial momentum cut-off $r$. After evaluating the integral, we take the limit $r\to\infty$. For linear divergent integrals we can use the following useful formula
\begin{equation}
\Delta(a)=\int\dd^4x\,[f(x+a)-f(x)]=\im\,2\pi^2a^\mu\lim\limits_{r\to\infty}r_\mu\,r^2f(r),
\end{equation}
where $r$ is the radius of a $S^3$ sphere. Making use of this formula and \newline tr$\,\gamma_5\gamma_\beta\gamma_\nu\gamma_\alpha\gamma_\mu=4\im\,\tilde\varepsilon_{\beta\nu\alpha\mu}$, we obtain, in the symmetric momentum limit,
\begin{equation}
R^1_{\mu\nu}=-\frac{1}{8\pi^2}\tilde\varepsilon_{\mu\nu\alpha\beta}k_1^\alpha k_2^\beta\equiv R^2_{\mu\nu}.\label{eq:rest}
\end{equation}
Regularising, as outlined above, leads to an ambiguity of the amplitude $M_{\mu\nu\lambda}$, since we chose the momentum route arbitrarily. A shift of the internal momentum integration \mbox{$p\to p+a,$} with $a=a_1 k_1+(a_1-a_2)k_2$, alters the value of the integral. Since we are interested in a cut-off independent result we have to calculate the difference of the amplitudes with and without the shift, namely
\begin{align}
\Delta_{\mu\nu\lambda}(a)&=M_{\mu\nu\lambda}(a)-M_{\mu\nu\lambda}(0)=-\frac{a_2}{8\pi^2}\,\tilde\varepsilon_{\mu\nu\lambda\alpha}(k_1^\alpha-k_2^\alpha),
\end{align}
which we obtain by using the same tools as for eq. \eqref{eq:rest}. The final amplitude is given by $M_{\mu\nu\lambda}(a)=M_{\mu\nu\alpha}(0)+\Delta_{\mu\nu\lambda}(a)$ resulting, after contraction with $q^\lambda$, in
\begin{equation}
q^\lambda M_{\mu\nu\lambda}(a_2)=2m\,M_{\mu\nu}-\frac{1-a_2}{4\pi^2}\tilde\varepsilon_{\mu\nu\alpha\beta}k_1^\alpha k_2^\beta.\label{eq:AWI}
\end{equation}
Analogously, we can derive the result for the VWI given by
\begin{equation}
k_1^\mu M_{\mu\nu\lambda}(a_2)=\frac{1+a_2}{8\pi^2}\tilde\varepsilon_{\nu\lambda\alpha\beta}k_1^\alpha k_2^\beta.\label{eq:VWI}
\end{equation}
The important insight is, we cannot find an $a_2$ so that both, the AWI \eqref{eq:AWI} and the VWI \eqref{eq:VWI}, hold at the same time! We can choose $a_2$ in a way, where the VWI holds and the AWI is anomalous, or the AWI holds and the VWI is anomalous, or both are anomalous. The VWI is satisfied in the case $a_2=-1$, which we have to impose in order to gauge the vector symmetry; then the AWI is anomalous, reproducing the result of the previous section. The choice of the coefficient $a_2$ is, on the level of actions, done by finite counter-terms, the so called Bardeen counter-terms. 
After the calculation one question remains: we obtained the result via an one-loop pertubation calculation; do radiative corrections of higher order pertubation theory alter the result? The answer is no; the Adler-Bardeen theorem \cite{PhysRev.182.1517} states that the anomaly expression is exact on the operator level and therefore the numerical coefficient of the anomalous term is not touched by higher order corrections.\newline\newline
An important concept, when dealing with anomalies, is the framework of consistent and covariant currents. The former are obtained by variation of the local vacuum functional with respect to the corresponding gauge fields. We refer to these currents, for example the axial current eq. \eqref{eq:ano}, as consistent currents since they fulfil the Wess-Zumino consistency condition. However, the consistent currents transform not covariant under the anomalous symmetry. When we are interested in a covariant formulation, we have to add Bardeen-Zumino terms \cite{BARDEEN1984421} to the currents and obtain the covariant form, which are in the case of the model with two gauge fields, as presented in eq. \eqref{eq:lagrmod}, connected by
\begin{equation}J^\mu_{\text{cov,A}}=J^\mu_{\text{cons,A}}+\frac{1}{24\pi^2}\tilde\varepsilon^{\mu\nu\rho\lambda}V_\nu H_{\rho\lambda} \text{ and } J^\mu_{\text{cov,V}}=J^\mu_{\text{cons,V}}+\frac{1}{12\pi^2}\tilde\varepsilon^{\mu\nu\rho\lambda}A_\nu H_{\rho\lambda},\label{eq:covcons}
\end{equation}
where $F=\dd A$ and $H=\dd V$. In the covariant formulation, all terms, depending directly on gauge fields, are absent and the currents contain only gauge invariant quantities as the field strength. 
\subsection{Anomalous transport: The Chiral Magnetic Effect}\label{CMEchapter}
\begin{figure}[H] 
	\centering
	\includegraphics[width=7.2cm]{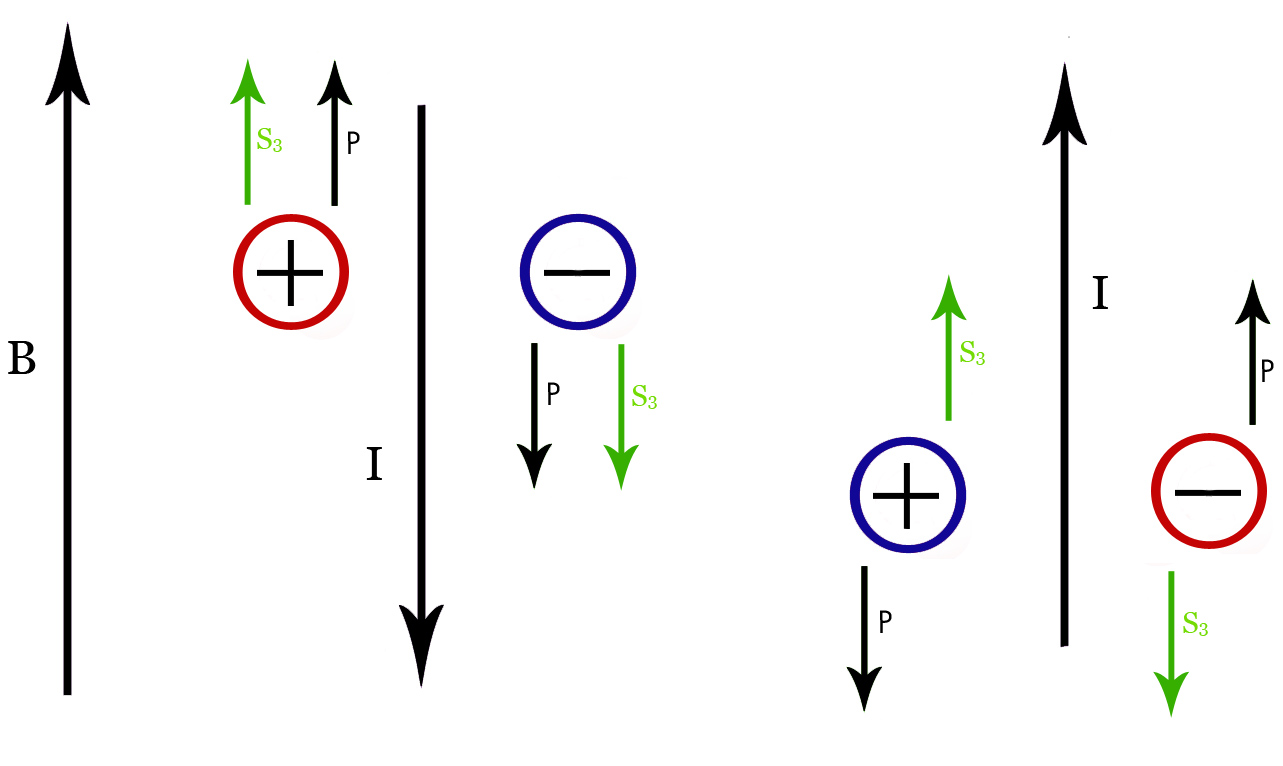}
	\caption{Pictoral demonstration of the CME in four spacetime dimensions. The magnetic field $B$ aligns the spins of the particles (+) and antiparticles (-). The chirality is denoted by red and blue, respectively. Furthermore, the helicity is characterised by the orientation of spin and momentum. An excess of right- or left chirality (more particles of the left two and right two species, respectively), resulting in a chiral imbalance, implies a non-zero current. For a similar graphic see \cite{amadeo}.\label{fig:CME4}}
\end{figure}
Ever since their discovery, anomalies play a crucial role in the development of QFT. Recently, the interest focused on the macroscopic manifestation of quantum anomalies in systems with chiral fermions. One of the effects relating anomalies to transport properties is the \textbf{C}hiral \textbf{M}agnetic \textbf{E}ffect (CME).
The CME describes the generation of an electric current parallel to a magnetic field induced by a chirality imbalance. \newline\newline
In this section, we consider a system possessing chiral fermions, following \cite{Kharzeev:2013ffa,Fukushima:2008xe,Fukushima:2012vr}. To get an intuitive understanding, we first discuss the effect in the weak coupling limit where the quasi-particle picture is appropriated.
 \newline\newline
 We consider a Lagrangian, as presented in eq. \eqref{eq:lagrmod}, with two external background fields providing an electric and a magnetic field.
The triangle graphs, with two external gauge fields, namely a composite current and a charged fermion loop, are depicted in figure \ref{fig:CME}.
The contributions of the fermions to the triangle diagram depend on their chirality. Changing the chirality leads to a contribution with opposite sign. Therefore, the vector current is given by $J_\mu^\text{V}=\bar\psi\gamma_\mu\psi=J_\mu^\text{L}+J_\mu^\text{R}$ (see eq. \eqref{eq:lh} and \eqref{eq:rh}) and the electric charge is conserved; on the other side, the axial current is given by $J_\mu^{A}=\bar\psi\gamma_\mu\gamma_5\psi=-J_\mu^\text{L}+J_\mu^\text{R}$. The additional $\gamma_5$ matrix cancels the change of the sign due to $\gamma_5\psi_\pm=\pm\psi_\pm$ and the divergence of the current gets an anomalous contribution
\begin{equation}
\partial_\mu J^\mu_\text{A}=\frac{e^2}{2\pi^2}\,\bm{E}\cdot \bm{B},\label{eq:axan}
\end{equation}
where $e$ is the fermion charge. Nevertheless, the contributions to the divergence of the vector current do not cancel when the number of left and right handed particles is not equal, namely in the case of a chirality imbalance (which can be captured by an axial chemical potential $2\mu_5\equiv\mu_\text{R}-\mu_\text{L}$). Due to eq. \eqref{eq:axan} the axial anomaly may generate an electric current in presence of a magnetic field.\newline\newline
In order to illustrate the CME, which is depicted in figure \ref{fig:CME4}, let us consider a Dirac sea of massless fermions.
In absence of external fields, chirality is conserved and we have two disconnected Fermi surfaces, consisting of left- and right-handed fermions. The chirality of the fermions can be changed by adiabatically switching on external fields, in particular an electric field parallel to a magnetic field. Then, the equilibrium state between the Fermi surfaces is violated and left-handed anti-particles will transform into right-handed particles and vice versa, depending on the sign of $\bm{E}\cdot\bm{B}$, as we will see in the following section.\newline\newline
Thus, the first important ingredient for the CME is a magnetic field\footnote{In general, magnetic fields are crucial for the understanding of topological effects in condensed matter systems, for example the Quantum Hall Effect. The CME can be interpreted as a 3+1 dimensional version of the Quantum Hall Effect.}. Due to the magnetic field there is a preferred direction for the spins of the fermions.
The magnetic field adjusts the spins of positive (negative) fermions in direction (opposite direction) of the magnetic field. In presence of an electric field, positive fermions feel the force $e\bm{E}$ and move in direction of $\bm{E}$. 
After a time $t$, positive fermions have Fermi-momentum $p_\text{R}^\text{F}=e\,E\,t$, whereas negative fermions decrease their Fermi momentum to $p^\text{F}_\text{L}=-p_\text{R}^\text{F}$. The right-handed fermions moving in the direction of the electric field create an electric current, which is usually compensated by the left-handed fermions moving in the other direction.
The density of states of positive fermions in $z$-direction is given by $\dd N_\text{R}/\dd z=p_\text{R}^{F}$, whereas the fermions in the transverse directions populate Landau levels  $\dd^2 N_\text{R}/(\dd x\,\dd y)= eB/2\pi$, leading to a total change rate of
\begin{equation}
\frac{\dd^4 N_\text{R}}{\dd t\,\dd V}=\frac{e^2}{(2\pi)^2}\,\bm{E}\cdot\bm{B}=-\frac{\dd^4 N_\text{L}}{\dd t\,\dd V}.
\end{equation}
The local change rate in chirality, $N_5=N_R-N_L$, is then given by
\begin{equation}
\frac{\dd^4 N_5}{\dd t\,\dd V}=\frac{e^2}{2\pi^2}\,\bm{E}\cdot\bm{B},\label{eq:axan2}
\end{equation}
in agreement with eq. \eqref{eq:axan}.
Note that this local change rate is purely an effect of the electromagnetic axial anomaly.\newline\newline
The transformation of a left-handed fermion into a right-handed fermion requires, in the Dirac sea picture, the transfer of a particle from the left-handed Fermi surface to the right-handed Fermi surface, involving an energy cost of $2\mu_5$ and $\mu_5\,\dd N_5$, respectively. Hence, the energy, required per unit time, is given by the energy cost times the chirality change. This energy stems from the power of the current or in other words
\begin{equation}
\int\dd^3x\,\bm{j}_\text{V}\cdot\bm{E}=\mu_5\frac{\dd N_5}{\dd t}=\frac{e^2\mu_5}{2\pi^2}\int\dd^3x\,\bm{E}\cdot\bm{B},
\end{equation}
and in the limit $\bm{E}\to0$, respectively,
\begin{equation}
\bm{j}_\text{V}=\frac{e^2\mu_5}{2\pi^2}\bm{B},
\end{equation}
which is known as the density of the CME current\footnote{There is a dual version, the Chiral Seperation Effect, which gives rise to an axial current $\bm{j}_\text{A}=\frac{e^2\mu}{2\pi^2}\bm{B}$.}.
\newpage
\section{The AdS/CFT correspondence}\label{sec:ads}
 \begin{figure}[h] 
 	\centering
 	\includegraphics[width=6cm]{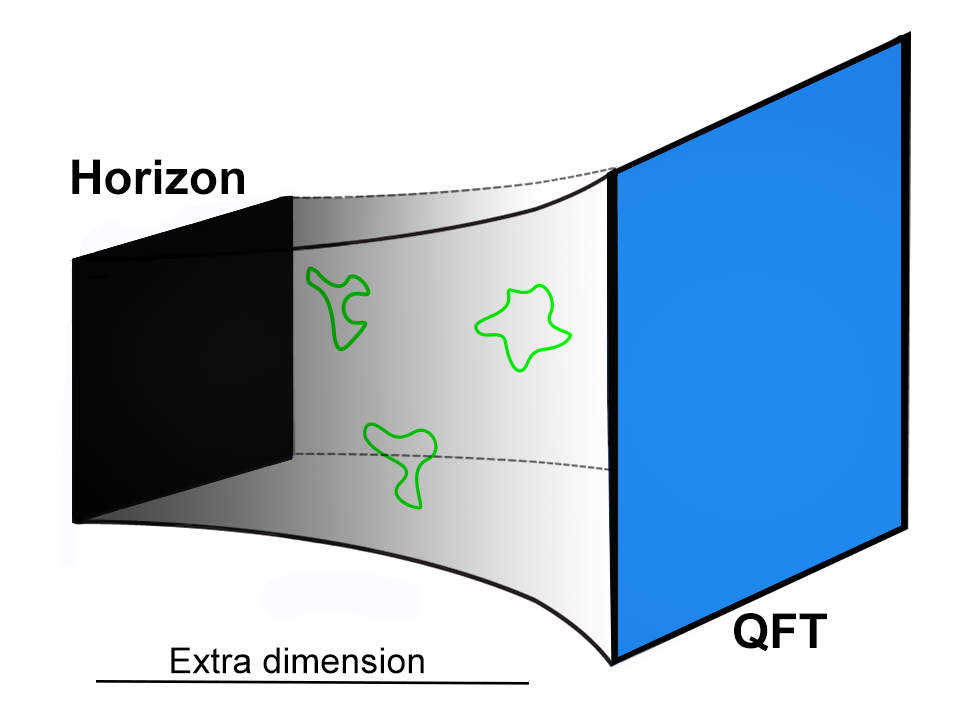}
 	\caption{\label{fig:Ads} A cartoon of the holographic duality. The strongly coupled quantum field theory lives on the boundary of spacetime, i.e. the blue hyperplane, whereas the gravity degrees of freedom (depicted in terms of green closed strings) can propagate into the extra dimension, i.e. into the bulk.}
 \end{figure}
\noindent In 1994, Leonard Susskind gave a string theory interpretation of 't Hooft's work, stating, we can store the information of a three dimensional world on an two dimensional surface, as realised in a holographic image, implying quantum gravitational theories could be described by a lower dimensional quantum field theory \cite{Susskind:1994vu,Stephens:1993an}. \newline This idea is based on the observations of Hawking \cite{1975CMaPh..43..199H} and Bekenstein \cite{PhysRevD.7.2333} according to which the entropy of black holes scales with the area of the black hole horizon and not with its volume. Let us assume we take some region of space and put some information in it; then the maximum number of bits we could throw into this region should intuitively be proportional to the volume of the region. However, let us shrink the region down to a size where it collapses to a black hole (while keeping the information in it fixed); then its information content can also be described by the area of the black hole horizon, which is a surface rather than a volume. We conclude that the maximum amount of information that can be stored in a volume is proportional to its area. \newline The AdS/CFT correspondence is a realisation of the holographic principle; in 1997, Juan Maldacena \cite{Maldacena:1997re} conjectured the dynamical equivalence between four-dimensional $\mathcal N=4$ Super-Yang-Mills (SYM) theory and type IIB superstring theory compactified on AdS$_5\times \text{S}^5$ (see also \cite{Gubser:1998bc,Witten:1998qj} and \cite{Ammon:2015wua,Nastase,Schalm} for textbooks). As depicted in figure \ref{fig:Ads}, the SYM is located on the conformal boundary surrounding the higher-dimensional AdS space. The acronym AdS refers to \textbf{A}nti-\textbf{d}e \textbf{S}itter spacetime which is the maximal symmetric solution of the Einstein equations with negative cosmological constant. \newline\newline
Furthermore, in a particular regime the AdS/CFT correspondence is a so called\newline \textit{strong/weak} duality relating a strongly coupled theory to a weakly coupled theory, which is much easier manageable. Throughout this thesis, we relate a strongly coupled CFT to a weakly coupled gravitational theory. At the conformal boundary there is a one to one map, the holographic dictionary, which contains the information of how to translate the gravity quantities into field theory quantities. Although still a conjecture, overwhelming evidence in favor of holographic dualities was collected over the last two decades. 
\subsection{The large $N$ limit and different versions of AdS/CFT}
In the previous chapter, we outlined the dynamical equivalence between certain CFTs and gravitational theories. On the field theory side, we have two free parameters, the Yang-Mills coupling constant $g_\text{YM}$ and the parameter $N$ of the gauge group $SU(N)$ \cite{Ammon:2015wua}; on the gravitational theory side the free parameters are the string coupling constant $g_\text{s}$ and the ratio AdS curvature $L$ divided by the string length $\sqrt{\alpha'}.$ In the context of AdS/CFT, the parameters are mapped onto each other in the following way
\begin{equation}
g_\text{YM}^2=2\pi\,g_\text{s}, \ \ 2\lambda\equiv 2g_\text{YM}^2 N=L^4/\alpha'^2,\label{eq:AdSCFTstrong}
\end{equation}
where we introduced the 't Hooft coupling $\lambda$. Since it is very difficult to do explicit calculations in the strongest form of the correspondence, it is necessary to simplify the theories by approximations. The central approximation done in the context of AdS/CFT is the large $N$ limit. String theory is currently best understood in its pertubative regime; in order to perform a pertubative expansion in the string coupling, we have to restrict the coupling constant to small values, i.e. $g_s\ll 1$. However, the string coupling is also connected to the Yang-Mills coupling via the AdS/CFT correspondence, requiring $g_\text{YM}\ll 1$. Keeping the ration of AdS curvature $L$ over the string length $\sqrt{\alpha'}$, and therefore the 't Hooft coupling $\lambda$, fixed requires $N\to\infty$; hence we have to consider a particular limit of AdS/CFT which is referred to as the strong form of the AdS/CFT correspondence. \newline
Within this thesis, we are interested in strongly coupled field theories; therefore we chose to work in the limit $\lambda\to\infty$, which is referred to as the weak form of AdS/CFT. On the gravity side, this limit translates (keeping the AdS curvature $L$ fixed) to $\alpha'\to0$, reducing classical string theory to classical supergravity. \newline\newline
Throughout this thesis, we will use an \textit{bottom-up} approach. In this case, we do not have to deal with the ten-dimensional supergravity fields and compactify them with a Kaluza-Klein reduction on the $S^5$; we already start with an effective toy model, where the gravitational fields are defined in AdS$_5$. In the so called \textit{bottom-up} approach, we write down a toy model Lagrangian for the gravitational theory with minimal ingredients, which captures the core properties of the dual field theory model under consideration. After solving the corresponding equations of motion in the context of relativity, we extract the observables, we are interested in, from the asymptotic behaviour of the gravity fields.
\subsection{Scalar fields in AdS}\label{sec:scalar}
For pedagogical reasons we choose the following procedure: we first consider the example of a scalar field to outline the main purpose and give then a brief generalisation to vector fields. The action of a massive free scalar field $\phi$ in a $(d+1)$-dimensional curved spacetime, with coordinates $X^\mu$ (where $\mu=0,1,\ldots,d$), reads \cite{Ammon:2015wua,Ramallo:2013bua,Hartnoll:2009sz}
\begin{equation}
S=-\frac C2\int\dd^{d+1}X\,\sqrt{-g}\,(g^{\mu\nu}\partial_\mu\phi\,\partial_\nu\phi+m^2 \phi^2).\label{eq:scalarwirk}
\end{equation}
The corresponding equations of motion are obtained by a variational principle and given by
\begin{equation}
\frac{1}{\sqrt{-g}}\partial_\mu\,(\sqrt{-g}g^{\mu\nu}\partial_\nu\phi)-m^2\phi=0,\label{eq:scalar}
\end{equation}
where the coordinates $X$ are denoted by $X^\mu=(u,x^m)$ with $m=1,\ldots,d$. After a Wick rotation of the time coordinate, the metric of Euclidean AdS in the Poincaré patch is given by
\begin{equation}
\dd s^2=g_{\mu\nu} \dd X^\mu\dd X^\nu=\frac{L^2}{u^2}(\dd u^2+\delta_{mn}\dd x^m\dd x^n). \label{eq:lineeucl}
\end{equation}
It is well known that eq. \eqref{eq:scalar} evaluated for the metric in eq. \eqref{eq:lineeucl} has a regular singular point at $u=0$. In this case Frobenius' theorem
states that the differential equation has two independent, well defined (power series) solutions in the neighbourhood of this point with leading power $\Delta_-$ and $\Delta_+$ in $u$
  \begin{equation}
  \phi(u,x) \sim \,u^{\Delta_-}\,(\phi_\text{s}(x)+\phi_\text{s}^{(1)}(x)\,u+\ldots) +u^{\Delta_+}\,(\phi_\text{v}(x)+\phi_\text{v}^{(1)}(x)\,u+\ldots),\label{eq:unksol}
  \end{equation}
  where the subscripts will be explained later\footnote{In the case of $\Delta_+-\Delta_-\in\mathbb N_0$ there are logarithmic contributions to the power series. For simplicity, we chose $\Delta_+$ and $\Delta_-$ such that $\Delta_+-\Delta_-\not\in\mathbb N_0$ for now.}. By solving eq. \eqref{eq:scalar} close to $u=0$, the coefficients $\phi_\text{s}(x)$ and $\phi_\text{v}(x)$ remain undetermined. However, all higher order coefficients $\phi^{(i)}_\text{s}$ and $\phi^{(i)}_\text{v}$ are completely determined in terms of $\phi_\text{s}(x)$, $\phi_\text{v}(x)$ and their derivatives. To determine the exponents of the solution in eq. \eqref{eq:unksol}, we make the ansatz $\phi(u,x)\sim u^\Delta$ and solve eq. \eqref{eq:scalar} to leading order in $u$ using $\sqrt{g}=\frac{L^{d+1}}{u^{d+1}}$
  \begin{align}
  0&=\frac{1}{\sqrt{g}}\,\partial_\mu\,\left (\sqrt{g}\,g^{\mu\nu}\,\partial_\nu\phi\right )-m^2\phi
  =\frac{u^{d+1}}{L^{d+1}}\,\partial_u\left(\frac{L^{d+1}}{u^{d+1}} g^{uu}\,\partial_u u^\Delta\right )-m^2\,u^{\Delta}+\mathcal O\left(u^{\Delta+2}\right)\nonumber\\&=\frac{u^{d+1}}{L^{d+1}}\,\partial_u\left(\frac{L^{d-1}\,\Delta}{u^{d-1}} \,u^{\Delta-1}\right )-m^2\,u^{\Delta}+\mathcal O\left(u^{\Delta+2}\right)\nonumber\\
  &=\left(\frac{\Delta\,(\Delta-d)}{L^2}-m^2\right)\,u^{\Delta}+\mathcal O\left(u^{\Delta+2}\right),\label{eq:mrel}
  \end{align}
  where derivatives with respect to $x$ are contained in $\mathcal O\left(u^{\Delta+2}\right)$. To fulfil eq. \eqref{eq:mrel} to leading order in $u$, the coefficient of $u^\Delta$ has to vanish, resulting in a quadratic equation in $\Delta$ with solutions
  \begin{equation}
  \Delta_\pm=\frac d2 \pm \sqrt{\frac{d^2}{4}+m^2L^2}.\label{eq:confmass}
  \end{equation}
  Evaluating the action \eqref{eq:scalarwirk} for the solution $\phi_+\sim u^{\Delta_+}(\phi_\text{v}(x)+\ldots)$, gives a finite result; hence, we refer to $\Delta_+$ as normalisable solution.
  The undetermined coefficients $\phi_\text{s}(x)$ and $\phi_\text{v}(x)$ of eq. \eqref{eq:unksol} are related via the AdS/CFT correspondence to quantities on the field theory side. In the so called standard quantisation, $\phi_\text{v}$ corresponds to the \textbf{v}acuum expectation value of the dual gauge invariant operator $\mathcal O$, with dimension $\Delta\equiv\Delta_+$\footnote{An interesting feature reveals for masses in the range
  	$-\frac{d^2}{4}<m^2L^2\le -\frac{d^2}{4}+1$.
  	In this case both solutions are normalisable and can be used to quantise the theory resulting in two different CFTs corresponding to the same classical AdS action.}. The second solution $\phi_\text{s}$ functions as a \textbf{s}ource term on the dual field theory side. Adding source terms to the action $S$ results in the generating functional $W[\phi_s]$ for connected Green's functions
  \begin{equation}
  S'=S-\int \dd^dx\,\phi_\text{s}(x)\mathcal O(x).\label{eq:green}
  \end{equation}
To understand the importance of the previous statement better, we look at the behaviour of the scalar field under diffeomorphisms. Since the scalar field is a Lorentz scalar, it has to be invariant under diffeomorphisms
\begin{equation}
\phi(u,x)=\tilde{\phi}(u',x'),\label{eq:diffinv}
\end{equation}
where $u'=\lambda u$ and $x'=\lambda x$. Using the asymptotic expansion \eqref{eq:unksol} and the invariance under diffeomorphisms \eqref{eq:diffinv} 
  \begin{equation}
  \tilde\phi(\lambda x,\lambda u)=(\lambda u)^{d-\Delta_+}\,(\tilde\phi_\text{s}(\lambda x)+\ldots)+\ldots\overset{!}{=}
  \phi(u,x) =u^{d-\Delta_+}\,(\phi_\text{s}(x)+\ldots)+\ldots
  \end{equation}
results in
  \begin{equation}
  \tilde\phi_\text{s}(\lambda x)=\lambda^{\Delta_+-d}\,\phi_\text{s}(x).\label{eq:scalartr}
  \end{equation}
  Furthermore, the action iteself has to be invariant under diffeomorphisms; plugging the scaling behaviour of the scalar field \eqref{eq:scalartr} in eq. \eqref{eq:green} yields to
  \begin{align}
  \int \dd^dx\,\phi_\text{s}(x)\,\mathcal O(x)&=
  \int\dd^d x'\,\tilde\phi_\text{s}(x')\,\tilde{\mathcal O}(x')\overset{x'=\lambda x}{=}\lambda^d\,\int \dd^dx\,\tilde\phi_\text{s}(\lambda x)\,\tilde{\mathcal O}(\lambda x)\\&\overset{\eqref{eq:scalartr}}{=}\int \dd^dx\,\tilde\phi_\text{s}(x)\,\lambda^{\Delta_+}\,\tilde{\mathcal O}(\lambda x).
  \end{align}
  This requires for the scaling behaviour of the dual field theoretical operator $\mathcal O(x)$, since $\phi_\text{s}(x)$ was chosen arbitrarily, \begin{equation}
  \tilde{\mathcal O}(\lambda x)=\lambda^{-\Delta_+}\mathcal O(x).
  \end{equation}
 Hence, we conclude that $\mathcal O(x)$ has the conformal dimension $\Delta_+$. Comparing the representations of the field theory operators with the representations of the dual gravitational operators results, in the case of a massive scalar field with $\Delta=\Delta_+$, in the important statement
  \begin{equation}
  m^2L^2=\Delta(\Delta-d).\label{eq:defconf}
  \end{equation}
  On the right hand side we have only quantities of the field theory side, namely the scaling dimension under dilatations $\Delta$, whereas on the left hand side only quantities of the gravitational theory  are present, namely the mass $m$ of the scalar field and the curvature radius of AdS $L$. Obviously, the conformal dimension of the field theory operator is linked to the mass of the scalar field in the gravitational theory. Since the conformal dimension has to be a real quantity, the mass in eq. \eqref{eq:confmass} is subjected to the so called Breitenlohner-Freedman bound
  \begin{equation}
  m^2\ge-\left(\frac{d}{2L}\right)^2.
  \end{equation}
  For negative $m^2$ the corresponding dual operator is relevant (with conformal dimension $\Delta\equiv\Delta_+\in[d/2,d)$), for vanishing masses marginal ($\Delta_+=d $) and for positive $m^2$ irrelevant ($\Delta=\Delta_+>d$).\newpage
\subsection{Massless vector fields in AdS}
Since the model under consideration throughout this thesis contains massless gauge fields coupled by a Chern-Simons term in the bulk, we briefly adapt the procedure, explained for the scalar field, to massless gauge fields.
The action of a massless U(1) gauge field $A_\mu$ with a Maxwell-Chern-Simons term  is given by
\begin{equation}
S=\int\dd^{d+1}X\,\sqrt{-g}\,\left(-\frac 14 F_{\mu\nu}F^{\mu\nu}+\frac{\kappa}{2}\,\varepsilon^{\mu\alpha\beta\gamma\delta}A_\mu F_{\alpha\beta}F_{\gamma\delta}\right),
\end{equation}
where $\varepsilon$ is the Levi-Civita tensor.
The equations of motion follow directly from a variational principle and read (see appendix \ref{app::gaugeeq} for further details)
\begin{equation}
\nabla_\mu F^{\mu\nu}+\frac{3\kappa}{2} \,\varepsilon^{\nu\alpha\beta\gamma\rho}F_{\alpha\beta}F_{\gamma\rho}=0.
\end{equation}
With the metric, defined in eq. \eqref{eq:lineeucl}, we find to leading order in $u$, with $A_m\sim u^{\Delta_A}$,
\begin{equation}
\Delta_A(\Delta_A+d-2)=0\,\Leftrightarrow \Delta_{A,\pm}=-\frac{d-2}{2}\pm\frac{d-2}{2}.
\end{equation}
Since the gauge field is an one-form, we have to consider the behaviour of $A_\mu(u,x)\,\dd x^\mu$ under diffeomorphisms\footnote{We work in the axial gauge where $A_u=0$.}
\begin{equation}
\tilde A_{m}(u,x)\, \dd \tilde x^m=\lambda^{0}u^{0}\,\tilde A_{m,\text{s}}(x) \,\dd(\lambda x^m)\overset{!}{=} A_m(u,x)\, \dd x^m=u^{0}\,A_{m,\text{s}}(x) \,\dd x^m,
\end{equation}
and therefore
\begin{equation}
\tilde A_m(x)=\lambda^{-1} \,A_m(x).
\end{equation}
 Again, we can interpret the non-normalisable mode $A_{m,\text{s}}\sim u^0$ as source term for the dual field theory operator $J^m$
  \begin{equation}
  S'=S-\int \dd^dx\,A_{m,\text{s}}(x) J^m(x).\label{eq:greenvec}
  \end{equation}
  Since the action has to be invariant under diffeomorphisms 
   \begin{align}
   \int \dd^dx\,A_{m,\text{s}}(x) J^m(x)=
   \int\dd^d x'\,\tilde A_{m,\text{s}}(x') \tilde J^m(x')\overset{x'=\lambda x}{=}\lambda^{d-1}\,\int \dd^dx\,A_{m,\text{s}}(\lambda x)\,\tilde{J^m}(\lambda x),
   \end{align}  
   we conclude for the scaling dimension of $J^m(x)$
   \begin{equation}
   J^m(x)=\lambda^{d-1}\tilde J^m(x),
   \end{equation}
   implying that $J^m$ has conformal dimension $\Delta=d-1$.
  
\subsection{Correlation functions in AdS/CFT}
In chapter \ref{sec:scalar}, we established an one to one map between the mass of the scalar field and the conformal dimension of the dual operator. Furthermore, we noticed that the non-normalisable mode $\phi_\text{s}$ can be interpreted as source term for the dual operator $\mathcal O$. This indicates a duality between the generating functionals of both theories \cite{Ammon:2015wua,D'Hoker:2002aw,Freedman:1998tz,Ramallo:2013bua}.\newline\newline
In Euclidean signature, we can formulate the partition function $Z_\text{CFT}[\phi_\text{s}]$ of the action $S'$,  given by eq. \eqref{eq:green}, as
\begin{equation}
Z_\text{CFT}[\phi_\text{s}]=\e^{-W_\text{CFT}[\phi_\text{s}]}=\left\langle\operatorname{exp}\left(\int\dd^dx\,\phi_\text{s}(x)\,\mathcal O(x)\right)\right\rangle_\text{CFT}.
\end{equation}
In its strongest form, the AdS/CFT correspondence states the equality of the partition function of the CFT and the partition function of the gravitational theory
\begin{equation}
Z_\text{CFT}[\phi_\text{s}]=Z_\text{AdS}[\phi_\text{s}]. \label{eq:ads}
\end{equation}
The partition function of string theory $Z_\text{AdS}$ is difficult to calculate; however, in the large $N$ limit, we can perform a saddle point approximation of $Z_\text{AdS}$
\begin{equation}
Z_\text{AdS}\sim e^{-S_\text{sugra}[\phi]}\label{eq:saddle}.
\end{equation}
The action of the supergravitational fields $\phi$ in AdS$_5$ is given by $S_\text{sugra}[\phi]$ containing in general massive scalar fields as described in eq. \eqref{eq:scalarwirk}. Using the equality of the partition functions in \eqref{eq:ads} and the saddle point approximation established in eq. \eqref{eq:scalarwirk}, we get the connection between the supergravity action $S_\text{sugra}[\phi]$ and the generating functional on the CFT side $W[\phi_\text{s}]$
\begin{equation}
W_\text{CFT}[\phi_\text{s}]=S_\text{sugra}[\phi]\big|_{\lim\limits_{u\to 0}\,(\phi(u,x)z^{\Delta-d})=\phi_\text{s}(x)},\label{eq:adscft}
\end{equation}
where we already assumed that $\Delta\equiv\Delta_+$ is the conformal dimension of the operator $\mathcal O$ and $\phi_\text{s}$ the corresponding source. Eq. \eqref{eq:adscft} is a non trivial statement since it claims the equality of a CFT generating functional (in the large $N$ limit) and a generating functional of a gravitational theory; this statement is precisely the weak form of the AdS/CFT correspondence. As usual, the correlation functions can be computed by differentiating the generating functional with respect to the sources $\phi_\text{s}^i$
\begin{equation}
\langle \mathcal O_1(x_1) 
\ldots\mathcal O_n(x_n)\rangle_\text{CFT}=-\left.\frac{\delta^nW_\text{CFT}}{\delta\phi^1_\text{s}(x_1) 
	\ldots\delta\phi^n_\text{s}(x_n)}\right|_{\phi^i_\text{s}=0},\label{eq:vev}
\end{equation}
where $\phi^i_\text{s}(u,x)$ are the corresponding sources of the operators $\mathcal O_i$.
\subsection{Holographic renormalization}\label{sec:holo}
With eq. \eqref{eq:vev}, we have a formula at hand which tells us how to compute correlation functions. However, the supergravity action is divergent, for example because of the infinite volume of AdS leading to infrared divergences. From the quantum field theory point of view, this is a well known phenomenon; in QFT, the computation of correlation functions gives rise to contact terms resulting from ultraviolet divergences. These divergences are renormalisation scheme dependent and for this reason unphysical. \newline\newline The AdS/CFT correspondence relates the infrared divergences of the supergravity action to the ultraviolet divergences of the dual field theory. In order to compute physical reasonable results, we have to remove the divergences. This is done by the so called holographic renormalization.\newline\newline 
In this thesis, we will give a sketchy introduction to the Hamiltonian approach to holographic renormalization \cite{Papadimitriou:2004ap,papadimitriou2005aspects}\footnote{For the standard approach to holographic renormalization see \cite{Henningson:1998ey,deHaro:2000vlm,Bianchi:2001kw,Skenderis:2002wp}.}; a more detailed calculation of the counter-terms can be found in appendix \ref{app::ren1}. Within this approach, the radial coordinate of AdS plays the role of the time coordinate in the ADM formalism (R. \textbf{A}rnowitt, S. \textbf{D}eser and C.W. \textbf{M}isner, \cite{PhysRev.116.1322}) for gravity. We assume that the reader is familiar with the standard approach to the holographic renormalization and the ADM formalism for gravity.\newline\newline
Consider a toy model with a massive free scalar field $\phi$ and a gauge field $A_\mu$ in a curved spacetime, whose action reads (in Euclidean signature)
\begin{equation}
 S_\text{matter}=\int\dd^{d+1}X\,\sqrt{g}\left(-\frac 14 F_{\mu\nu}F^{\mu\nu}+\frac 12 (\partial\phi)^2\right),
\end{equation}
coupled to gravity 
\begin{equation}
S_\text{gravity}=-\frac{1}{2\kappa_{d+1}^2}\,\left(\int\dd^{d+1}X\,\sqrt{g}R+\int \dd^dx\,\sqrt{\gamma}\,2K\right),\end{equation}
where $\kappa_{d+1}^2=8\pi G_{d+1}$. \newline\newline
We are interested in hypersurfaces of constant values of the radial coordinate $r=1/u$. On this hypersurfaces $\Sigma_r$, we can formulate the action in terms of the induced metric and the extrinsic curvature. Such a hypersurface can be viewed as a radial cut-off at $r_0$; therefore, the regulated manifold is bounded by $\Sigma_{r_0}$. On this hypersurface, the induced quantities are the induced metric $\gamma$ and $K$, the trace of the extrinsic curvature of the hypersurface.\newline
 The hypersurface $\Sigma_{r_0}$ works as boundary condition for the action; hence, the momenta on the regulating surface can be computed by variation of the on-shell action with respect to boundary values of the induced fields given by
\begin{equation}
\Pi^{\mu\nu}(r_0,\sigma)=\frac{\delta S_\text{on-shell}}{\delta \gamma_{\mu\nu}(r_0,x)}, \ \ \Pi^{\mu}(r_0,x)=\frac{\delta S_\text{on-shell}}{\delta A_{\mu}(r_0,x)}, \ \ \Pi(r_0,x)=\frac{\delta S_\text{on-shell}}{\delta \phi(r_0,x)}.
\end{equation} 
 Furthermore, we fix the lapse and shift freedom so that the bulk metric has the form
\begin{equation}
\dd s^2=\dd r^2+\gamma_{mn}(r,x)\,\dd x^m\,\dd x^n 
\end{equation}
and choose the axial gauge, where $A_r=0$. From now on, we consider the hypersurface which is a radial distance $r_0=e^{r}$ apart from the boundary and suppress the index of $r_0$. In this section, the radial derivative will be denoted by a dot.\newline\newline
We can rewrite the regulated on-shell action, using Einstein equations and the Gauss-Codazzi identities, and introducing a new variable $\lambda$, 
\begin{equation}
I_\text{reg}=-\frac{1}{\kappa_{d+1}^2}\int_{\Sigma_r}\dd^{d}x\,\sqrt{\gamma}(K-\lambda),
\end{equation}
provided $\lambda$ satisfies the equation
\begin{equation}
\dot{\lambda}+K\lambda-\kappa_{d+1}^2\,\left(\mathcal L_\text{matter}+\frac{1}{d-1} T^\sigma_\sigma\right)=0,
\end{equation}
where $T^\sigma_\sigma$ is the trace of the energy-momentum tensor.\newline\newline
The leading terms of the asymptotic expansion read
\begin{equation}
\gamma_{mn}\sim\e^{2r},\ \ \ \ \ \ A_m\sim\e^{-\Delta_A r}=A_{(0)m}(x),\ \ \ \ \ \ \phi\sim\e^{-\Delta_-r},
\end{equation}
where $\Delta_-$ and $\Delta_A$ denote the non-normalisable modes, derived in the previous section. Hence, the radial derivatives of this expansions are given by
\begin{equation}
\dot{\gamma}_{mn}\sim 2\gamma_{mn},\ \ \ \ \ \ \dot A_m=\mathcal O(\e^{-r}),\ \ \ \ \ \ \dot{\phi}=-\Delta_- \phi.
\end{equation}
The canonical momenta are functionals of the induced fields on the hypersurface $\Sigma_r$ and read in the case of our toy model
\begin{equation}
\Pi^{mn}=-\frac{\sqrt{\gamma}}{2\kappa_{d+1}^2} \,(K\gamma^{mn}-K^{mn}),\ \ \ \ \ \ \Pi^m=-\frac {\sqrt{\gamma}}{4}\,\dot A^m, \ \ \ \ \ \ \Pi=\frac{\sqrt{\gamma}}{2}\,\dot{\phi}.
\end{equation}
Since the momenta are functionals of the induced fields, we can re-express the radial derivative by the functional differential operator
\begin{equation}
\partial_r=\int\dd^dx\,\left(\dot{\gamma}_{mn}[\gamma,A,\phi]\ \frac{\delta}{\delta\gamma_{mn}}+\dot{A}_m[\gamma,A,\phi]\ \frac{\delta}{\delta A_m}+\dot{\phi}[\gamma,A,\phi]\ \frac{\delta}{\delta\phi}\right).\label{eq:rad}
\end{equation}
We identify the radial derivative asymptotically with the dilatation operator $\delta_\text{D}$\footnote{On the field theory side it is worth mentioning the following: usually one parametrises a spacetime by Cartesian coordinates $(t,x)$; however, in Euclidean signature time and space are treated at the same footing. Note, that we can map a complex plane by a conformal transformation to a cylinder (r,$\varphi$). On the cylinder the states live on spatial slices of constant $\varphi$ and evolve by the Hamiltonian $H=\partial_r$. After the mapping to the plane the Hamiltonian becomes the dilatation operator and states live on circles with constant radius evolving by the dilatation operator.} by
\begin{equation}
\partial_r=\delta_\text{D}+\mathcal O(\e^{-r}).
\end{equation}
We can now re-express the asymptotic expansions in eigenfunctions of the dilatation operator
\begin{align}
\Pi^{i}_{j}&=\sqrt{\gamma}\,(\Pi_{(0)j}^{\textcolor{white}{(0)}{i}}+\Pi_{(2)j}^{\textcolor{white}{(0)}i}+\ldots+\Pi_{(d)j}^{\textcolor{white}{(d)}{i}}+\tilde\Pi_{(d)j}^{\textcolor{white}{(d)}{i}} \log\e^{-2r}),\\
\Pi^{i}&=\sqrt{\gamma}\,(\Pi_{(3)}^{\textcolor{white}{(0)}{i}}+\Pi_{(4)}^{\textcolor{white}{(0)}i}+\ldots+\Pi_{(d)}^{\textcolor{white}{(d)}{i}}+\tilde\Pi_{(d)}^{\textcolor{white}{(d)}{i}} \log\e^{-2r}),\\
\Pi&=\sqrt{\gamma}\,(\Pi_{(\Delta_-)}+\Pi_{(\Delta_-+1)}+\ldots+\Pi_{(\Delta_-+d)}+\tilde\Pi_{(\Delta_-+d)} \log\e^{-2r}),\\
\lambda&=\lambda_{(0)}+\lambda_{(2)}+\ldots+\lambda_{(d)}+\tilde\lambda_{(d)} \log\e^{-2r}.
\end{align}
The terms with subscript $(i)$, where $i<d$, transform according to their scaling dimension under the dilatation operator, as well as the terms proportional to the logarithms, for example
\begin{equation} \delta_\text{D}\Pi_{(n)}^{\textcolor{white}{(n)}i}=-n\Pi_{(n)}^{\textcolor{white}{(n)}i}, 3\le n <d, \ \ \delta_\text{D}\tilde\Pi_{(d)}^{\textcolor{white}{(n)}i}=-d\tilde\Pi_{(d)}^{\textcolor{white}{(n)}i}.
\end{equation}
Identifying the dilatation operator with the radial derivative determines the transformation law of the missing terms, namely
\begin{equation}
\delta_\text{D}X_{(k)}=-k X_{(k)}-2 \tilde X_{(k)},
\end{equation}
where $k=d$ for $X=\{\Pi^i_j,\Pi^i,\lambda\}$ and $k=\Delta_-$ for $X=\Pi$.\newline
In a similar fashion, we can write down the covariant expansion of the radial derivative
\begin{equation}
\partial_r=\delta_\text{D}+\delta_{(1)}+\ldots+\tilde{\delta}_{(d)}\log\e^{-2r}+\ldots,
\end{equation}
where $\delta_{(n)}$ are covariant functional operators of successively higher dilatation weight. Note that the on-shell action is of first order whereas the e.o.m. of the fields are of second order in the radial derivative. The order lowers by one when we write the equations in terms of the canonical momenta. \newline\newline In order to proceed, we have to determine the explicit expansions of the momenta and the radial derivative in terms of eigenfunctions of the dilatation operator. The crucial part is that both expansions have to be done with respect to the dilatation weight order by order at the same time. After determining the expansions, we can solve the equation for the on-shell action order by order. Hence, the counter-term is given by
\begin{equation}
I_\text{c.t.}=\frac{1}{\kappa_{d+1}^2}\int_{\Sigma_{r_0}}\dd^dx\sqrt{\gamma}\,\left(\sum\limits_{n=0}^{d-1}(K_{(n)}-\lambda_{(n)})+(\tilde K_{(d)}-\tilde{\lambda}_{(d)})\,\log\e^{-2r_0}\right).
\end{equation}
With this at hand, the renormalised action is defined as
\begin{equation}
I_\text{ren}=\lim\limits_{r_0\to\infty}(I_{r_0}+I_\text{ct})=-\frac{1}{\kappa_{d+1}^2}\int_{\partial M}\dd^dx\sqrt{\gamma}\,(K_{(d)}-\lambda_{(d)})
\end{equation}
and the renormalised correlation functions by
\begin{equation}
\langle T_{mn}\rangle=-\frac{1}{\kappa_{d+1}^2}(K_{(d)mn}-K_{(d)}\gamma_{mn}), \ \ \langle J^n\rangle_\text{ren}=\Pi_{(d)}^{\textcolor{white}{(d)}n}, \ \ \langle\mathcal O\rangle_\text{ren}=\Pi_{\Delta_-}.
\end{equation}
\subsection{Black holes in holography}\label{sec:QNM}
One important ingredient in the discussion of AdS/CFT is still missing: black holes; black holes are thermodynamical objects to whom we can associate a certain temperature, the Hawking temperature. Via holography, we can identify the temperature of the black hole with the temperature of the dual field theory. Since black holes have a certain temperature they provide a scale; hence, they break the conformal invariance in the dual field theory.
From the classical point of view, we can view black holes, loosely speaking, as a one way door. Everything that crosses the black hole horizon, will be lost; hence, we do not have to take those things into account which are hidden behind the black-hole horizon. In this way, the horizon functions as IR cut-off since we only have to integrate over the spacetime region outside the horizon. On the QFT side, we note the following: in a QFT at finite temperature, we impose a certain periodicity on the imaginary time $0<\tau<\beta, \beta=1/(k_\text{B}T)$, where the temperature $T$ functions as cut-off for long times, which correspond, via holography, to low energies.\newline\newline
In non-horizon penetrating coordinates, the analytic extension (to Euclidean AdS spacetime) of a black hole is conformally equivalent to a cone. If we want to avoid conical singularities, we have to impose a certain periodicity. This periodicity is identified with the periodicity resulting from the Euclidean time formalism in QFT, leading to the above mentioned result that the temperature of the black hole is equal the temperature of the dual field theory.
We see that in context of AdS/CFT the ``thermal circle`` is implemented in a natural way in the generating functionals, namely via the presence of a black hole in the gravitational theory.\newline\newline
We conjectured that a black hole induces a particular temperature in the dual CFT. But how effects a small pertubation on top of the black hole horizon the dual field theory? The answer to this question can be found in the discussion of \textbf{Q}uasi-\textbf{N}ormal-\textbf{M}odes (QNMs) \cite{lrr-1999-2,Kovtun:2005ev,Sachs:2003zj,Horowitz:1999jd,Berti:2009kk}. If we perturb a black hole, the surrounding geometry will ``ring`` and settle down back in equilibrium. The relaxation time, as well as the frequencies of the ringing, are independent of the pertubation and totally determined by the properties of the black hole. This phenomenon is well known from closed systems where the frequencies are referred to as normal modes. However, in presence of a classical black hole, the system is not closed since the black hole may absorb the pertubations 
and the modes will decay. \newline\newline QNMs are the solutions to the linearised equations of gauge field pertubations and pertubations of the gravitational background, subjected to certain boundary conditions. On the boundary the solutions have to fulfil a Dirichlet condition; at the horizon they are subjected to an ingoing wave boundary condition.  By virtue of the restriction to ingoing waves, the condition at the future horizon is determined since classical black holes do not emit radiation. Furthermore, omitting the solution with outgoing waves yields to a non hermitian boundary value problem; hence, the corresponding eigenfrequencies are complex. In QFT, small deviations from thermal equilibrium are captured by dispersion relations related to damped frequencies. Dispersion relations are formally the singularities of retarded Green's functions in the complex frequency plane. Via the AdS/CFT correspondence, the quasi normal frequencies of electromagnetic and gravitational pertubations in an asymptotic locally AdS (AlAdS) spacetime are related to the poles of real time Green's functions within current or energy-momentum of the dual CFT at finite temperature. This implies that the QNMs, corresponding to electromagnetic and gravitational pertubations, are related to conserved symmetry currents (or anomalous currents) on the CFT side. We conclude that the AdS/CFT correspondence implies the following identifications:
 	\begin{center}
 	AlAdS spacetime with event horizon $\Leftrightarrow$ thermal states in the dual CFT \\ \textcolor{white}{..}\\
 	Small pertubation of the black hole or black brane background $\Leftrightarrow$ small deviations from the equilibrium in the CFT with definite relaxation time.
 	\end{center}
This implies that we can obtain information about time-scales in the thermalization process of strongly coupled systems at finite temperature via the QNM spectrum of the dual gravitational theory. 
\newpage
\section{Time dependent CME in holography and anomaly induced Landau resonances}\label{sec:tdcih}
Throughout the following chapters, we want to investigate the CME in holography. Especially, we  focus on time dependencies and the effect of dynamical charge generation. Concretely, we consider an $U(1)\times U(1)$ model which was previously studied in the literature within the context of anomalous transport \cite{Gynther:2010ed, Jimenez-Alba:2014iia, Jimenez-Alba:2014pea}. This is in contrast to the usual approach where the net charge is introduced by an axial chemical potential \cite{ Yee:2009vw, Gynther:2010ed}. The $U(1)\times U(1)$ model allows us to model the CME in massless QED.

\subsection{Holographic Setup}\label{sec:timeholo}
We consider the 5-dimensional matter Lagrangian
\begin{equation}\label{eq::lagr}
\mathcal{L}= -\frac{1}{4}F^{\mu\nu}F_{\mu\nu}-\frac{1}{4}H^{\mu\nu}H_{\mu\nu}+ \frac{\kappa}{2}\varepsilon^{\mu\alpha\beta\rho\lambda}A_\mu\left(F_{\alpha\beta}F_{\rho\lambda}+
3H_{\alpha\beta}H_{\rho\lambda}\right)\,,
\end{equation}
consisting of two gauge fields $A_\mu,V_\mu$ ($F=\dd A,H=\dd V)$ coupled by a \mbox{Chern-Simons (CS)} term in the bulk. In the first part of this thesis, we restrict ourselves to the probe approximation and consider a Schwarzschild-AdS$_5$ blackhole in infalling Eddington-Finkelstein coordinates as fixed background metric
\begin{equation}
\dd s^2=\frac{1}{u^2}\,\left(-f(u)\dd v^2-2\,\dd v \dd u +\dd x^2+\dd y^2+\dd z^2\right),\hspace{2cm}f(u)=1-u^4,
\end{equation}
where we fixed the diffeomorphisms by setting the black-hole horizon to $u$=1; furthermore, we set the AdS radius to $L=1$. 
In infalling Eddington-Finkelstein coordinates, we find that the equations of motion are first order in time. In this coordinates, the surfaces of constant time penetrate, from a geometrical point of view, the black-hole horizon. Therefore, the ingoing condition is trivially realised by geometry, i. e. we do not have to impose regularity of gauge fields at $u=1$.\newline\newline
The equations of motion for the gauge fields are obtained by a variational principle (see appendix \ref{app::gaugeeq} for further details)
\begin{align}
\nabla_\mu F^{\mu \nu}+\frac{3\kappa}{2}\varepsilon^{\nu\alpha\beta\rho\lambda}
\left(F_{\alpha\beta}F_{\rho\lambda}+H_{\alpha\beta}H_{\rho\lambda}\right)=0\,,\label{eq:eom11}\\
\nabla_\mu H^{\mu \nu}+3\kappa\,\varepsilon^{\nu\alpha\beta\rho\lambda}
F_{\alpha\beta}H_{\rho\lambda}=0\,.\label{eq:eom22}
\end{align}
In order to motivate the specific ansatz for the Lagrangian in \eqref{eq::lagr}, we compute the divergence of the dual current operators (for details see appendix \ref{app::ren}). The holographic Ward identities read
\begin{equation}
\label{eq::ward}
\langle\partial_i J^i_V \rangle=0\,,\hspace{2cm}\langle\partial_i J^i_A \rangle=-\frac{\kappa}{2}\tilde\varepsilon^{\,ijkl}\left( F_{ij}F_{kl}+3H_{ij}H_{kl} \right),
\end{equation}
where $\tilde\varepsilon^{\,ijkl}$ denotes the epsilon symbol in the boundary theory\footnote{\label{kaapa} A comparison to the QCD expression, given in eq. \eqref{eq:QCDcurrent}, yields to $\kappa=-1/(8\pi^2)$ for the anomaly parameter.}.
Recall that the abelian part of the QCD singlet anomaly for the toy model Lagrangian considered in eq. \eqref{eq::lagr} is given by \cite{Bertlmann:1996xk}
\begin{equation}
\partial_i J_\text{A}^i=\frac{3}{16\pi^2}\tilde\varepsilon^{\,ijkl}\,\left(H_{ij}H_{kl}+\frac 13 F_{ij}F_{kl}\right).\label{eq:QCDcurrent}
\end{equation}
To mimic this feature, we chose the prefactor of the term $\sim H_{\alpha\beta}\,H_{\rho\lambda}$ in eq. \eqref{eq::lagr} equal to 3. This leads to a relative factor of 3 between the terms $ H_{ij}\,H_{kl}$ and $ F_{ij}\,F_{kl}$ in equation \eqref{eq::ward}. The finite Bardeen counter-terms\footnote{The finite counter-term is necessary since otherwise we would implicitly assume that the regularisation scheme treats left- and right handed particles on the same footing \cite{kharzeev2014strongly}.} have been chosen so that the VWI is fulfilled and the AWI (in eq. \eqref{eq::ward}) is anomalous; with this choice, we can identify the vector current with the electromagnetic current. Therefore, we refer to $A_\mu$ as the axial gauge field and $V_\mu$ as the vector gauge field, providing the external fields required for the CME.
\newline\newline
The dynamics of the axial charge is captured by $\dot\rho_5\sim\bm{E}\cdot\bm{B}$, where the dot denotes the time derivative $\partial_v$. As outlined in chapter \ref{CMEchapter}, at least one of the electricmagnetic fields has to betime dependent in order to dynamically generate axial charge. Concretely, we consider a time dependent electric field parallel to a constant magnetic field (both homogenous and pointing in $z$-direction). This is covered by the ansatz
\begin{equation}
A_v(v,u),\ \ \ \ V_y=Bx,\ \ \  V_z(v,u)\label{eq:ansatz}
\end{equation}
with boundary condition
\begin{equation}
\dot V_z(v,u\to 0)=E(v)
\end{equation}
and regularity of $A_v(v,u)$ at the horizon. The equations of motion resulting from this ansatz for the gauge fields and the metric read (with $'\equiv\partial_u$)
\begin{align}
A_v''-\frac{1}{u}A_v'- \lambda\,u\, V_z' =0\,,\label{eq:eom1}\\
V_z''+ \left(\frac{f'}{f}-\frac{1}{u} \right)V_z' -\frac{2}{f}\dot V_z '+\frac{1}{u f}\dot V_z-\lambda\, \frac{u}{f} A_v'=0\,,\label{eq:eom2}\\\
\dot A_v'-\lambda\, u\,  \dot V_z=0\,,\label{eq:eom3}
\end{align}
where we introduced $\lambda=12\,\kappa B$ for simplicity.
The differential eqs. \eqref{eq:eom1},\,\eqref{eq:eom2} have a regular singular point at $u=0$; thus, the solution to the equations has to behave at this point like
\begin{align}\label{eq:expansion}
V_z&\sim V_{0}+u\, \dot V_{0} + u^2\tilde V +\frac{1}{2} u^2\,\log(u) \ddot V_{0}+ \mathcal{O}(u^3),\\
A_v&\sim A_{0}+u^2 A_2 + \mathcal{O}(u^3).
\end{align} 
We will see that the normalisable mode of $V_z$ is proportional to the covariant vector current and the non-normalisable mode of $A_v$ to the axial charge density $\rho_5$.
Integrating eq. \eqref{eq:eom3} in time
\begin{equation}\label{eq:av}
A_v'=\lambda\, u\,   V_z+C_1(u)\,
\end{equation}
and substituting the resulting expression into eq. \eqref{eq:eom1} yields
\begin{equation}
 u\,C_1'(u)+C_1(u)=0.
\end{equation}
This restricts the constant $C_1$ to be at most linear in $u$, namely $C_1(u)=Cu$. With this at hand, we can replace $A'_v$ in eq. \eqref{eq:eom2} and end up with a single hyperbolic PDE for $V_z$
\begin{equation}\label{eq:vzeq}
V_z''+ \left(\frac{f'}{f}-\frac{1}{u} \right)V_z' -\frac{2}{f}\dot V_z '+\frac{1}{u f}\dot V_z-\frac{\left(\lambda u\right)^2}{f} V_z-\lambda u^2\frac{C}{f} =0.
\end{equation}
Since we are interested in one point functions, we have to specify the counter-terms in order to compute the dual correlators (see appendix \ref{app::ren})
\begin{equation}
S_\text{ct}=-\int_{\partial \mathcal{M}} \dd^4x\, \sqrt{-\gamma}\, \frac{1}{4} F_{ij}F^{ij}  \log{u}-\int_{\partial \mathcal{M}}\dd^4x\,\sqrt{-\gamma}\, \frac{1}{4} H_{ij}H^{ij} \left( \log{u}-\frac{1}{2} \right).\label{eq:counter}
\end{equation}
The computation of the one point function (see eq. \eqref{eq:logcontr}) reveals that the logarithmic term in eq. \eqref{eq:expansion} contributes to the correlator. In order to avoid this contribution, we have to add a finite (gauge invariant) counter-term to the action. This is manifest in the last term of eq. \eqref{eq:counter}. In this regularisation scheme, the one-point function, corresponding to the consistent current, is given by (see appendix \ref{app::ren})
\begin{equation}\label{eq:1pcons}
\langle J^z_\text{V}\rangle_\text{cons}=2 \tilde V - \lambda A_{0}.
\end{equation}
Imposing a vanishing $A_v$ at the horizon results in a vanishing consistent current in the time independent limit \cite{Jimenez-Alba:2014iia,Gynther:2010ed}; hence, it is customary to use the covariant current. The covariant current is connected to the consistent current via eq. \eqref{eq:covcons} and $\kappa=-1/(8\pi^2)$ as mentioned in footnote \ref{kaapa}. In our case we obtain, using the ansatz eq. \eqref{eq:ansatz}, \begin{equation} \langle J^z_V\rangle_\text{cov}=\langle J_V^z\rangle_\text{cons}+ \lambda A_{0}.\end{equation}Apart from that, regularity of the axial gauge field in Eddington-Finkelstein coordinates does not require a vanishing $A_v$ at the horizon. We can fix the gauge freedom by setting the non-normalisable mode of $A_v$ to zero. Therefore, the consistent and covariant current are indistinguishable and we can omit the subscript. Furthermore, we drop the subscript $V$ since we integrate out the axial gauge field and the computations contain only the $z-$component of the vector gauge field. The current can be obtained by extracting the coefficient of the non-normalisable mode of the vector gauge field $V_z$ at the conformal boundary
\begin{equation}
\label{eq:Current}
\langle J_z \rangle\equiv\langle J^z \rangle=2 \tilde V. 
\end{equation}
\subsection{Stationary solutions}\label{sec:statsol}
In order to study the time dependent response of the system, we have to construct initial data;
we start with an equilibrium solution, obtained from the time independent version of eq. \eqref{eq:vzeq}. Since we want to excite the system out of equilibrium, we switch on a time dependent electric field; this is done, as usual in holography, via a boundary condition. As outlined in chapter \ref{sec:ads}, the boundary value of the gauge fields are source terms in the dual field theory; hence, the boundary value of the vector gauge field induces an external source in the dual field theory, namely the electric field. In order to study the non-equilibrium response of the system, we have to integrate in time direction; 
the time independent form of eq. \eqref{eq:vzeq} reads
\begin{equation}
f\,u\, V_z''+ \left(f'u-f \right)V_z' -\left(\lambda u\right)^2 u\,V_z-\lambda u^3\,C =0.\label{eq:timeind}
\end{equation}
 We can transform eq. \eqref{eq:timeind} to the Legendre differential equation by introducing a new radial variable $\rho\equiv u^2$ and shifting the field $\tilde V_z\equiv V_z+\frac{C}{\lambda}$,
\begin{align}\label{eq:leg}
(1-\rho^2) \tilde V_z''- 2 \rho \tilde V_z' -\frac 14\, \lambda^2 \tilde V_z=0.
\end{align}
The solutions of eq. \eqref{eq:leg}, which are regular at $\rho=1$, are given in terms of the Legendre functions $\mathcal P_l$  
\begin{equation}\label{eq:regularsolultions}
V_z=\tilde V_z-\frac{C}{\lambda}=C_2\mathcal{P}_l(u^2)-\frac{C}{\lambda},\hspace{2cm} l=\frac{1}{2}\left(-1\pm\sqrt{1-\lambda^2} \right)\,.
\end{equation}%
As mentioned beforehand, the constant $C$ is just a shift of the field $V_z$ and we gauge fix it to zero. To interpret the second constant $C_2$ we integrate eq. \eqref{eq:av} using eq. \eqref{eq:regularsolultions}
\begin{equation}
\mu_A\equiv A_v(u=0)-A_v(u=1)=\lambda C_2 \int^0_1 \dd u \,\mathcal{P}_l(u^2)\,u\,;
\end{equation}%
this implies that the constant determines the absolute value of the chemical potential. The asymptotic expansion of the stationary solution \eqref{eq:regularsolultions} reads
\begin{equation}
V_z\sim V_0-\frac{\lambda^2}{8}\frac{\Gamma\left(\frac 14 (3-\sqrt{1-\lambda^2})\right)\,\Gamma\left(\frac 14 (3+\sqrt{1-\lambda^2})\right)}{\Gamma\left(\frac 14 (5-\sqrt{1-\lambda^2})\right)\,\Gamma\left(\frac 14 (5+\sqrt{1-\lambda^2})\right)}\,V_0 u^2+\ldots, 
\end{equation}
where we identified \begin{equation}
V_0\equiv - \frac{C_2 \cos\left(\frac \pi 2 \sqrt{1-\lambda^2} \right) \,\Gamma\left(\frac 14 (1-\sqrt{1-\lambda^2})\right)\,\Gamma\left(\frac 14 (1+\sqrt{1-\lambda^2})\right)}{2\pi^{\frac 32}},
\end{equation}with $\Gamma$ being the Euler gamma functions. Note that in the stationary case the time derivatives of the source in eq. \eqref{eq:expansion} vanish. Therefore, the coefficient of the term quadratic in $u$ is the required expression for the current in the stationary case, given by
\begin{equation}\label{eq:adcurrent}
\langle J^z \rangle =- \frac{\lambda^2}{4}\frac{\Gamma\left(\frac 14 (3-\sqrt{1-\lambda^2})\right)\,\Gamma\left(\frac 14 (3+\sqrt{1-\lambda^2})\right)}{\Gamma\left(\frac 14 (5-\sqrt{1-\lambda^2})\right)\,\Gamma\left(\frac 14 (5+\sqrt{1-\lambda^2})\right)}\,V_0 \overset{\lambda\gg1}{\sim}-\lambda\,V_0.
\end{equation} %valid ???
The last term is valid since $\lambda$ times the factor containing the gamma functions asymptotes, for large $\lambda$, 4. The value of the current depends on the integration constant $C_2$ (via $V_0$). A vanishing $V_0$ at the beginning requires therefore a vanishing $C_2$ and yields to a trivial solution for $V_z$.
\subsection{Non-equilibrium solutions}\label{sec:nonstatsol}
In the following, we want to study the non-equilibrium response of the system which requires that we integrate eq. \eqref{eq:vzeq} in time.
For numerical convenience we rescale the function $V_z$, using the asymptotic expansions. This leads to explicit  $V_0$ dependent terms in the differential equation. In section \ref{sec:statsol} we explained that the stationary solutions, we are interested in, are independent of $V_0$. To get rid of $V_0$ at the initial time we choose $V_0(v)$ in such a way that the function and all its derivatives vanish in the beginning of the simulation.
Concretely, we introduce an auxiliary field $U(v,u)$ defined by
\begin{equation}
\label{eq:AuxField_U}
V_z(v,u) \equiv V_{0}(v)+u\, \dot V_{0}(v) + u^2\,U(v,u) +\frac{1}{2}u^2\,\log(u) \left(\ddot V_{0}(v)+u\, V^{(3)}_{0}(v) \right).
\end{equation}%
Rewriting eq. \eqref{eq:vzeq} in terms of the auxiliary field $U(v,u)$ leads to
\begin{equation}
\label{eq:DynEq_TimeDom}
\left[- u\,(1\!-\!u^4)\frac{\partial^2 }{\partial u^2 }\! -\! \left( 3\! -\! 7u^4\right) \frac{\partial}{\partial u}\! +\! (8\! +\! \lambda^2)\,u^3 \right]U(v,u)  + \left[  2u\, \frac{\partial }{\partial u }\! +\!  3\ \right] \dot{U}(v,u) + S(v,u) = 0,
\end{equation}
where we defined
\begin{equation}
S(v,u) = a_4(u)\, V^{(4)}_0(v) + a_3(u)\,  V^{(3)}_0(v) + a_2(u)\, \ddot V_0(v) + a_1(u)\, \dot V_0(v) + a_0(u)\, V_0(v)+a_{-1}(u),
\end{equation}
with
\begin{align}
\label{eq:Source_ak}
a_4(u) =& \frac{1}{2}\left(2u +  5u\log{u} \right), \quad a_3(u) = \frac{15}{2}u^4\log{u}+ \frac{\lambda^2}{2}u^4\log(u)+4u^4-1,  \\
a_2(u) =& u^3\left(\!4\log{u}\!+\!3\!+\!\frac{\lambda^2}{2}\log{u}\! \right), \quad
a_1(u)\! =\! \left(3+\lambda^2\right)u^2, \quad  a_0(u)\! =\! \lambda^2u, \quad 
a_{-1}(u)\! =\! \lambda C u.\nonumber
\end{align}
On a time slice of constant $v_0$ the differential equation \eqref{eq:DynEq_TimeDom} is a second order ODE with respect to the radial coordinate requiring two integration constants. The first integration constant is fixed using the leading term of the asymptotic behaviour $V_z\sim V_0$. The second integration constant is incorporated by
setting the shift $C$ in eq. \eqref{eq:Source_ak} to zero. %Check this???
Therefore, the boundary conditions enter the initial configuration in the following way (inverting the boundary expansion and solving for the auxiliary field)
\begin{equation}
\label{eq:ID_U}
U(0,u) \equiv U_{\rm in}(u) = \frac{V_0}{u^2}\left[ \frac{\mathcal P_l(u^2)}{\mathcal P_l(0)} - 1\right].
\end{equation}
A brief review of the numerical methods can be found in appendix \ref{num} and the convergency tests in appendix \ref{conv}.
\subsection{Gaussian and hyperbolic tangent quenches}\label{sec:gahtq}
%%%%%%%%%%%%%%%%%%%%%%%%%%%%%%%%%%%%%%%%%%%%%%
\begin{figure}[t!] 
	\centering
	\includegraphics[width=6.7cm]{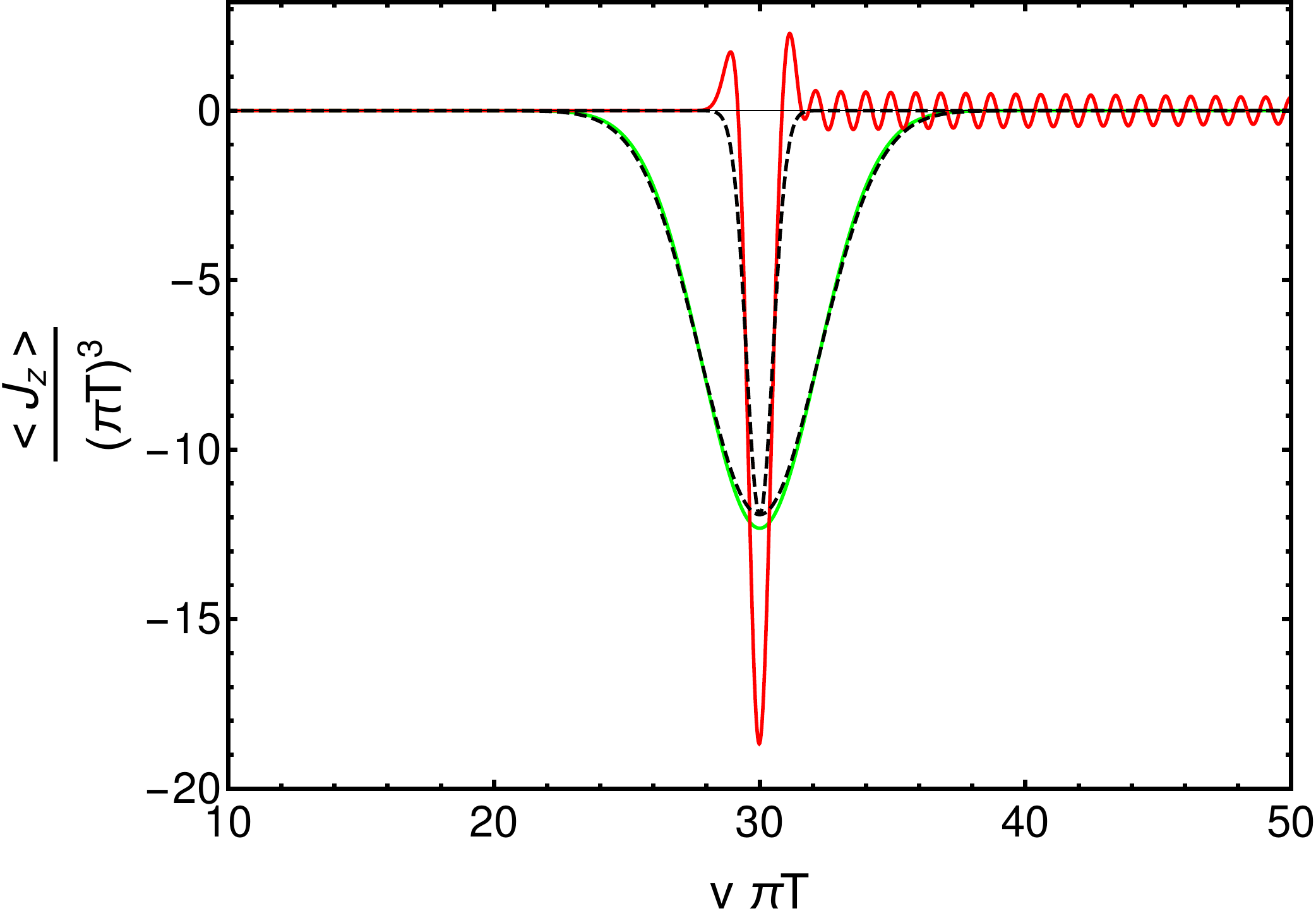}
	\hspace{1cm}
	\includegraphics[width=6.7cm]{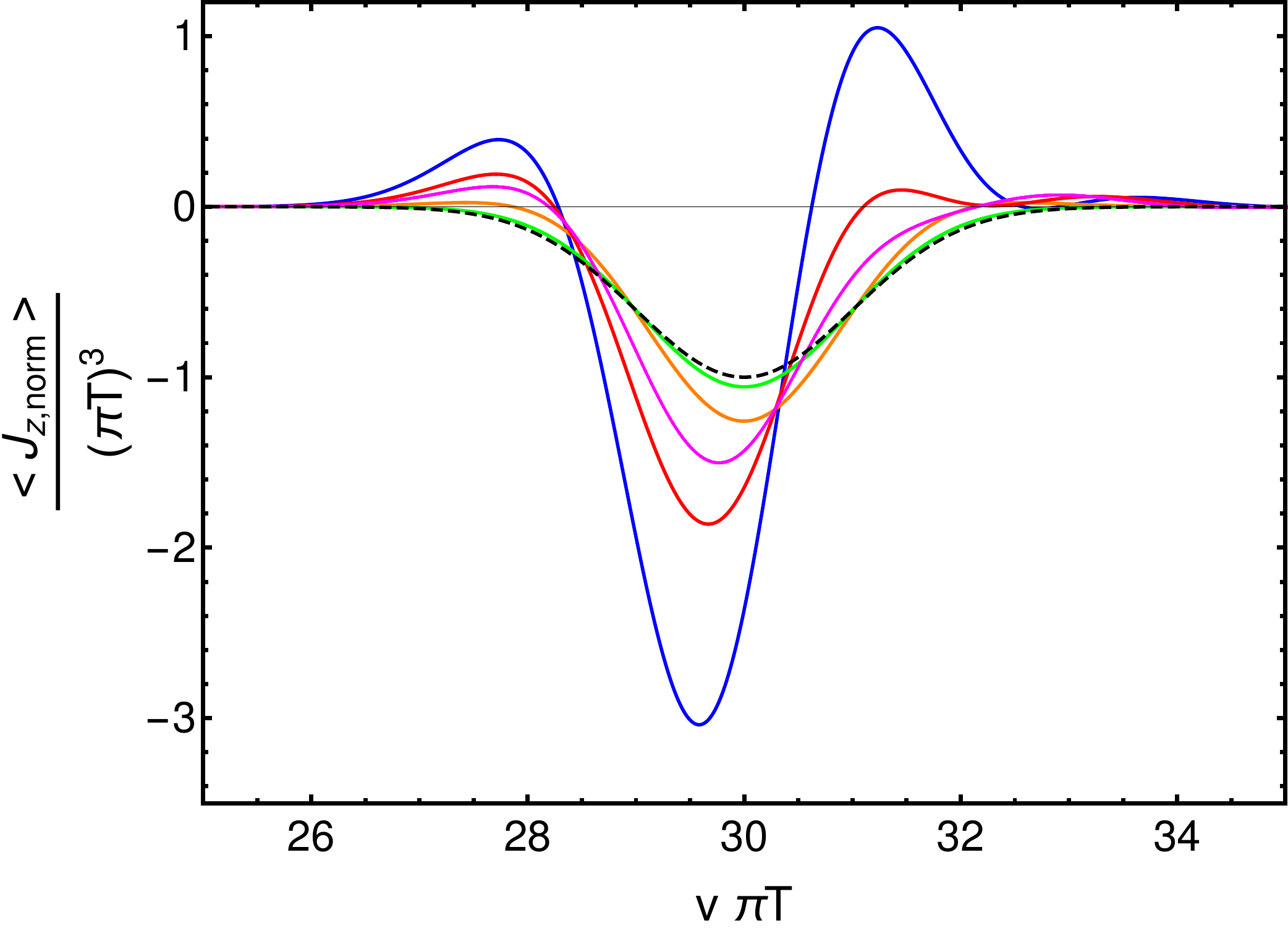}
	\caption{\label{fig:adiabvsnoadiab}Left: Current against time for Gaussian quenches with same height and different widths $\Lambda=0.1 (\pi T)^2$ ``slow'' (green) and $\Lambda=2 (\pi T)^2$ ``fast'' (red) at fixed magnetic field $\kappa \tilde{B}=1$, where $\tilde{B}\equiv B/(\pi T)^2$. Dashed black lines show the adiabatic response given by eq. \eqref{eq:adcurrent}. Right: Current against time for fixed Gaussian source with $\Lambda =0.5(\pi T)^2$ and several values of the magnetic field $\kappa \tilde{B}\in\{0.1,0.15,0.2,0.5,4\}$ (blue-green). For an easier comparison, the currents have been rescaled such that the corresponding adiabatic response (black, dashed) is the same.}
\end{figure}
%%%%%%%%%%%%%%%%%%%%%%%%%%%%%%%%%%%%%%%%%%%%%%

\begin{figure}[t!] 
	\centering
	\includegraphics[width=6.75cm]{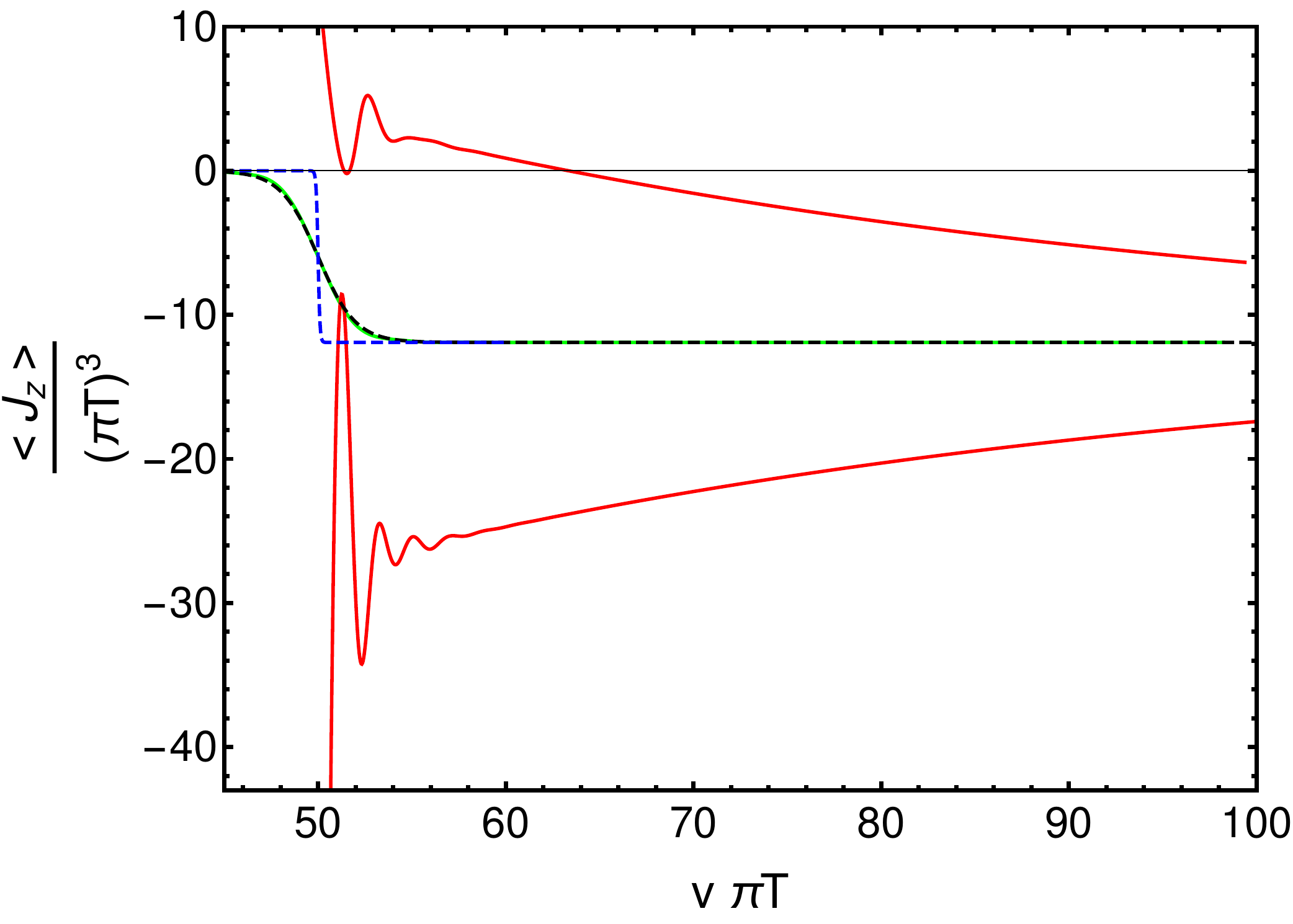}
	\hspace{1cm}
	\includegraphics[width=6.6cm]{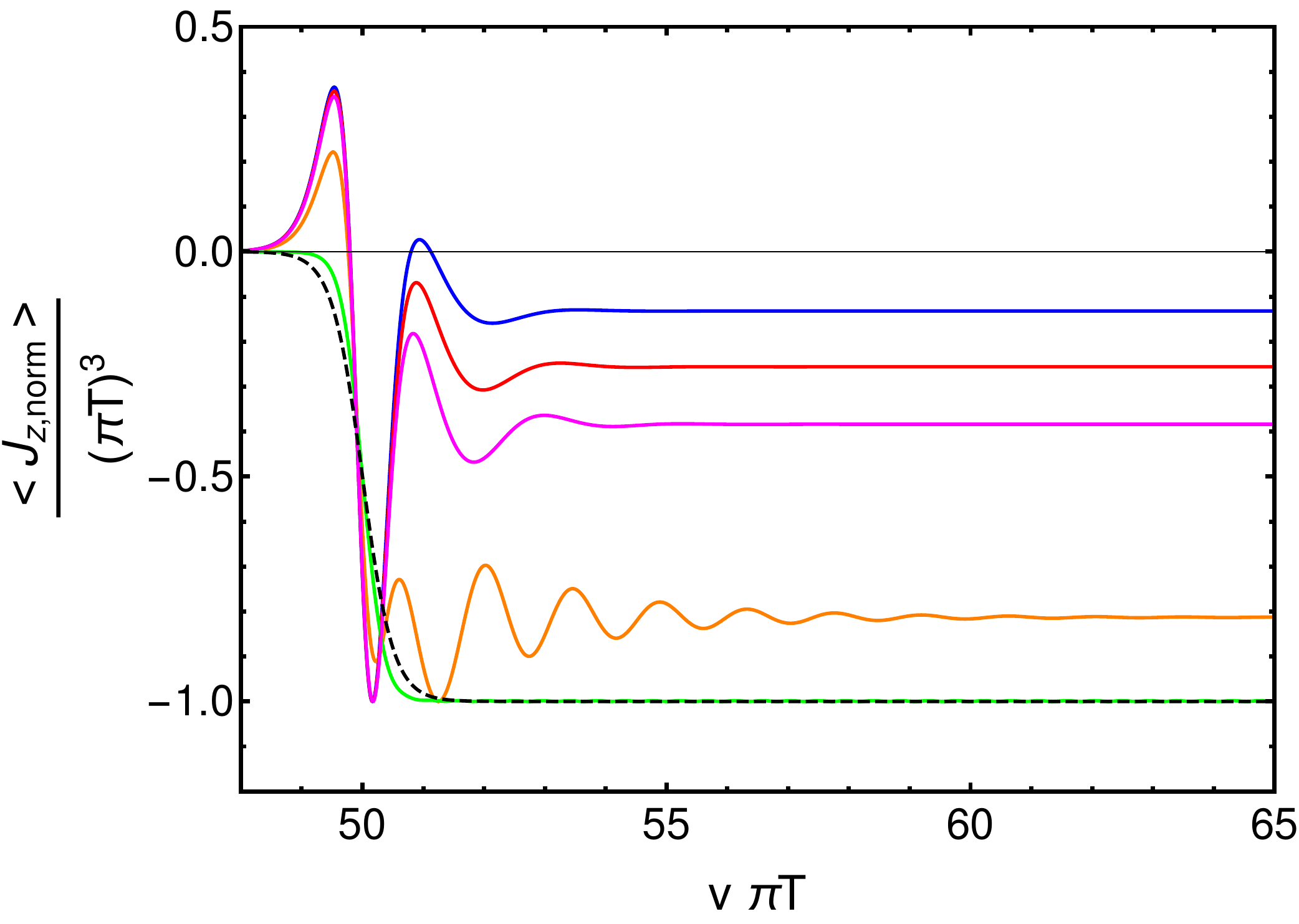}
	\caption{\label{fig:adiabvsnoadiab2}Left: Current against time for hyperbolic tangent quenches at fixed magnetic field $\kappa \tilde{B}=1$ with different widths $\Lambda=0.1 (\pi T)^2$ ``slow'' (green) and $\Lambda=10 (\pi T)^2$ ``fast'' (red). Dashed back lines show the adiabatic response given by eq. \eqref{eq:adcurrent}. Right: Current against time for hyperbolic tangent source several values of the magnetic field $\kappa \tilde{B}\in\{0.1,0.15,0.2,0.5,4\}$ (blue-green) with fixed $\Lambda =2 (\pi T)^2$. For a better comparison, we normalised the minimum of the current to -1.}
\end{figure}

In the previous section, we explained the framework within which we can study the non-equilibrium behaviour of the model under consideration. We excite the system in a non-equilibrium state via an abruptly switched on electric field; with the electric field, we are able to ``pump'' energy in the system. The more the electric field tends to a delta function like shape, the more the system gets out of equilibrium. In the first part of this thesis, we are interested in the dependence of the response on the abruptness of the quench.  
Therefore, we will focus on two types of quenches, namely the Gaussian and the hyperbolic tangent quench 
\begin{equation}
\label{eq:Quenches}
V_0(v) = \left\{
\begin{array}{ccc}
\text{exp} \left \lbrace-(v-v_i)^2\Lambda\right\rbrace & & {\rm (Gaussian \, quench)} \\
\left[ 1+ \tanh\left \lbrace (v-v_i)\Lambda\right\rbrace\right]/2\,  & & {\rm (tanh \,  quench)},
\end{array}\right.
\end{equation}%
%
%%%%%%%%%%%%%%%%%%%%%%%%%%%%%%%%%%%%%%%%%%%%%%
where $\Lambda$ is the inverse width of the source and $v_i$ fixes the center. The parameter $\Lambda$ can be viewed as abruptness of the quench, implying a ``fast quench'' refers to $\Lambda\gg1$. There is one big difference between the two kind of quenches; recall that the electric field is given by the time derivative of the source. A Gaussian type source results in an electric field with a positive and negative pulse of same area; this means that we do not induce net axial charge in the system after the quench. In contrast, the derivative of the hyperbolic tangent source is a single pulse resulting in a final net axial charge. This is obvious because the total axial charge of the system behaves as
\begin{equation}
\rho_5\sim\int\dd v\,E\cdot B=B\,V_0(v)\Big|^{v_\text{final}}_{v_\text{inital}}.
\end{equation}
As a first step, we want to focus on the initial ``response'' of the system. In order to get an intuition how the system reacts in response to the quenches, we considered different values of the product $\kappa B$ and different abruptnesses $\Lambda$. The response of the system subjected to the Gaussian quench is depicted in figure \ref{fig:adiabvsnoadiab}. In reaction to a ``slow'' quench (green) the current follows the corresponding adiabatic response, given by eq. \eqref{eq:adcurrent}, whereas the response to a ``fast'' quench behaves different compared to the adiabatic response. In order to discuss dimensionless quantities, we introduce the rescaled magnetic field $\tilde B\equiv B/(\pi T)^2$. In the r.h.s. of figure \ref{fig:adiabvsnoadiab}, we present the response to the same Gaussian source for different values of $\kappa \tilde B$. Note that we normalised the adiabatic response associated to a particular value of $\kappa\tilde B$ so that the adiabatic response is equal for all different values of $\kappa\tilde B$. Obviously, the lower the value of $\kappa\tilde B$, the more differs the response of the system compared to the adiabatic response. \newline\newline In addition, we did a similar study for the hyperbolic tangent quench. The corresponding results are depicted in figure \ref{fig:adiabvsnoadiab2}. Whereas the response to the ``slow'' quench follows the adiabatic response,  the ``fast'' quench induces a damped oscillation of the current.
We notice again that for the smallest value of $\kappa\tilde B$ the response of system to the quench deviates the most from the equilibrium response. In all cases, the current is non zero after the quench, since the quench introduces net axial charge to the system.
%%%%%%%%%%%%%%%%%%%%%%%%%%%%%%%%%%%%%%%%%%%%%%%%%%%%%%%%%%
\subsection{Universality in fast quenches}\label{sec:uifq}
\begin{figure}[h] 
 	\centering
 	\includegraphics[width=6.7cm]{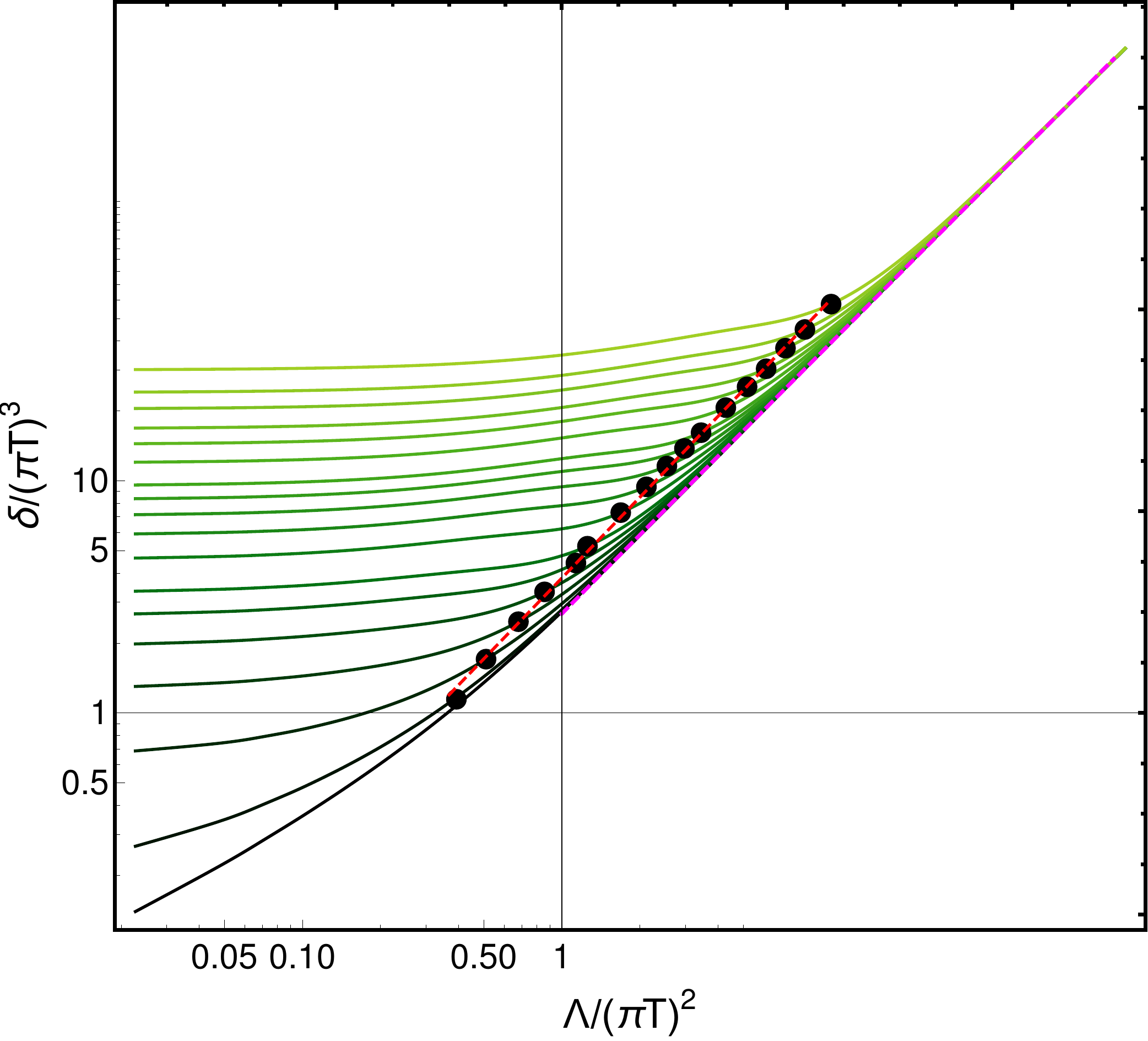}
 	\hspace{1cm}
 	\includegraphics[width=6.7cm]{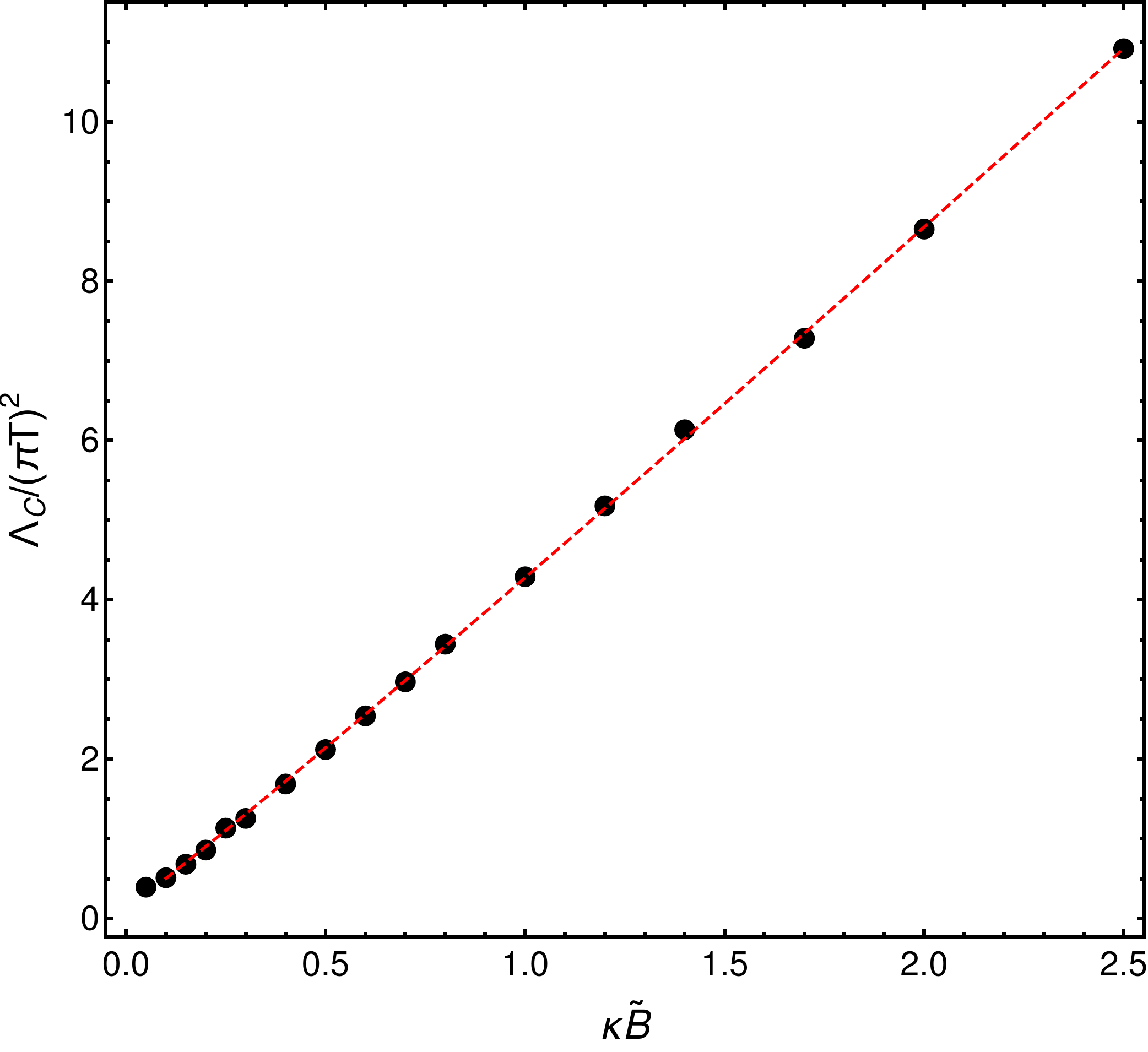}
 	\caption{\label{fig:max-min}Left: Double logarithmic plot of $\delta$ (see equation \eqref{eq:delta}) against the abruptness of the Gaussian quench $\Lambda$ for several values of $\kappa \tilde{B} \in [0,2.5]$ (black-light green). For fast enough quenches the system shows a response, independent of $\kappa\tilde B$. In this regime the logarithmic data fits to a straight line with slope 1.128 (magenta, dashed). Black points highlight the critical $\Lambda_C$ as defined in 
 		the text and fit in the logarithmic plot to a straight line (red, dashed). Right: Critical abruptness $\Lambda_C$ against $\kappa \tilde{B}$ (black) which can be fitted by a straight line (dashed, red).}
 \end{figure}

\begin{figure}[h] 
     	\centering
     	\includegraphics[width=6.7cm]{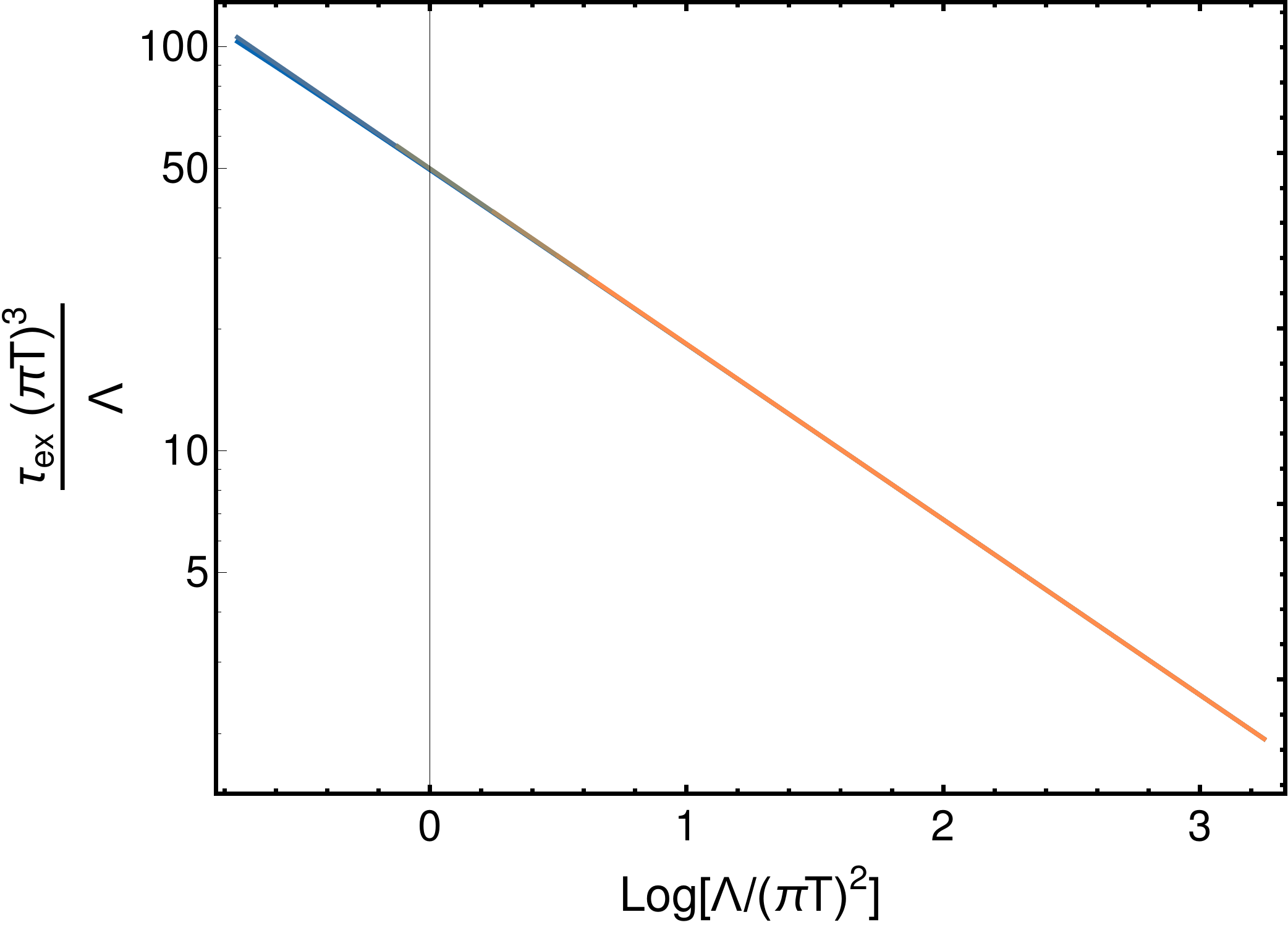}
     	\hspace{1cm}
     	\includegraphics[width=6.7cm]{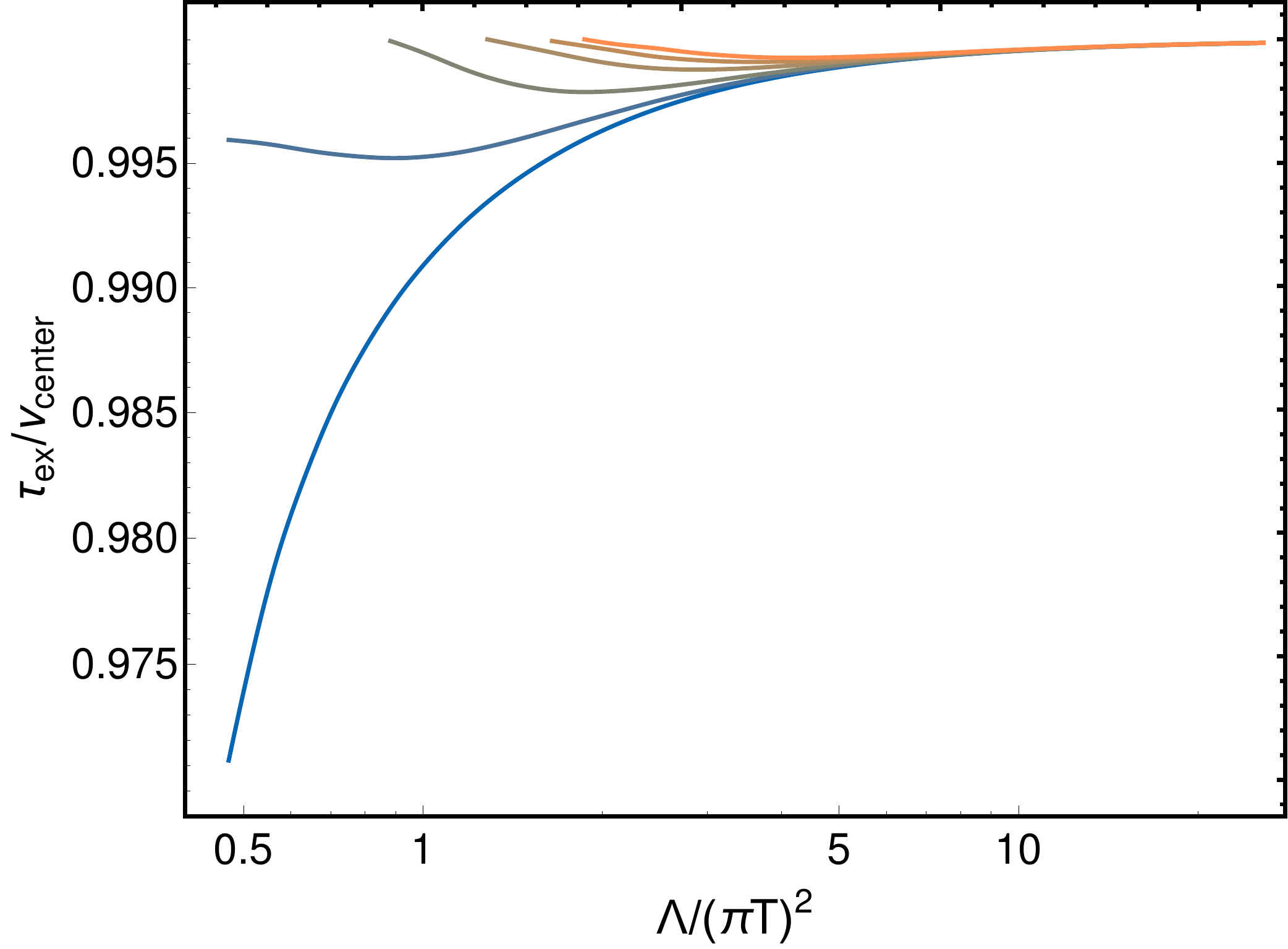}
     	\caption{\label{fig:excitime_tanh}Left: Excitation time divided by the abruptness $\tau_\text{ex}/\Lambda$ against the logarithm of the abruptness $\Lambda$ of a hyperbolic tangent quench for several values of $\kappa \tilde{B} \in \{0.125,0.25,0.5,1,1.5,2\}$ (blue-orange). Right: Logarithmic plot of the relative excitation time $\tau_\text{ex}/v_\text{center}$ against the abruptness of the quench $\Lambda$ for several values of $\kappa \tilde{B} \in \{0.125,0.25,0.5,1,1.5,2\}$ (blue-orange). For high enough $\Lambda$ all cases converge to a straight line.}
     \end{figure}

 Inspired by recent work about quantum quenches in strongly coupled field theories within holography \cite{Buchel:2012gw,Buchel:2013lla,Das:2014hqa}, we are especially interested in whether our system shows an universal regime in response to fast quenches. With regard to our model, our main focus is the dependence on the anomaly, namely whether the response to fast quenches is independent of the Chern-Simons coupling $\kappa$. In order to characterise the initial response of the system, we found it useful to define the quantity
 \begin{equation}
 \delta\equiv|\langle J_z(v_1)\rangle|-|\langle J_z(v_2)\rangle|,\label{eq:delta}
 \end{equation}
 where $|\langle J_z(v_1)\rangle|$ and $|\langle J_z(v_2)\rangle|$ are the first minimum and the first maximum, respectively, of the current. We consider now the dependence of the quantity $\delta$  on different values of the anomaly parameter $\kappa \tilde B$ and quench abruptness $\Lambda$.\newline
 On the left hand side of figure \ref{fig:max-min} the universal behaviour is apparent; in the case of fast enough quenches the difference is independent of the value of $\kappa \tilde B$ and fits perfectly to a straight line (magenta) in the double logarithmic presentation \begin{equation}\log(\delta/(\pi T)^3)=1.128\, [\log(\Lambda/(\pi T)^2)]^{0.997}+0.982,\end{equation} with a fitting error of $10^{-3}$. 
 %%%%%%%%%%%%%%%%%%%%%%%%%%%%%%%%%%%%%%%%%%%%%%
 %%%%%%%%%%%%%%%%%%%%%%%%%%%%%%%%%%%%%%%%%%%%%%
 %%%%%%%%%%%%%%%%%%%%%%%%%%%%%%%%%%%%%%%%%%%%%%
 %%%%%%%%%%%%%%%%%%%%%%%%%%%%%%%%%%%%%%%%%%%%%%
 \begin{figure}[h] 
 	\centering
 	\includegraphics[width=6.7cm]{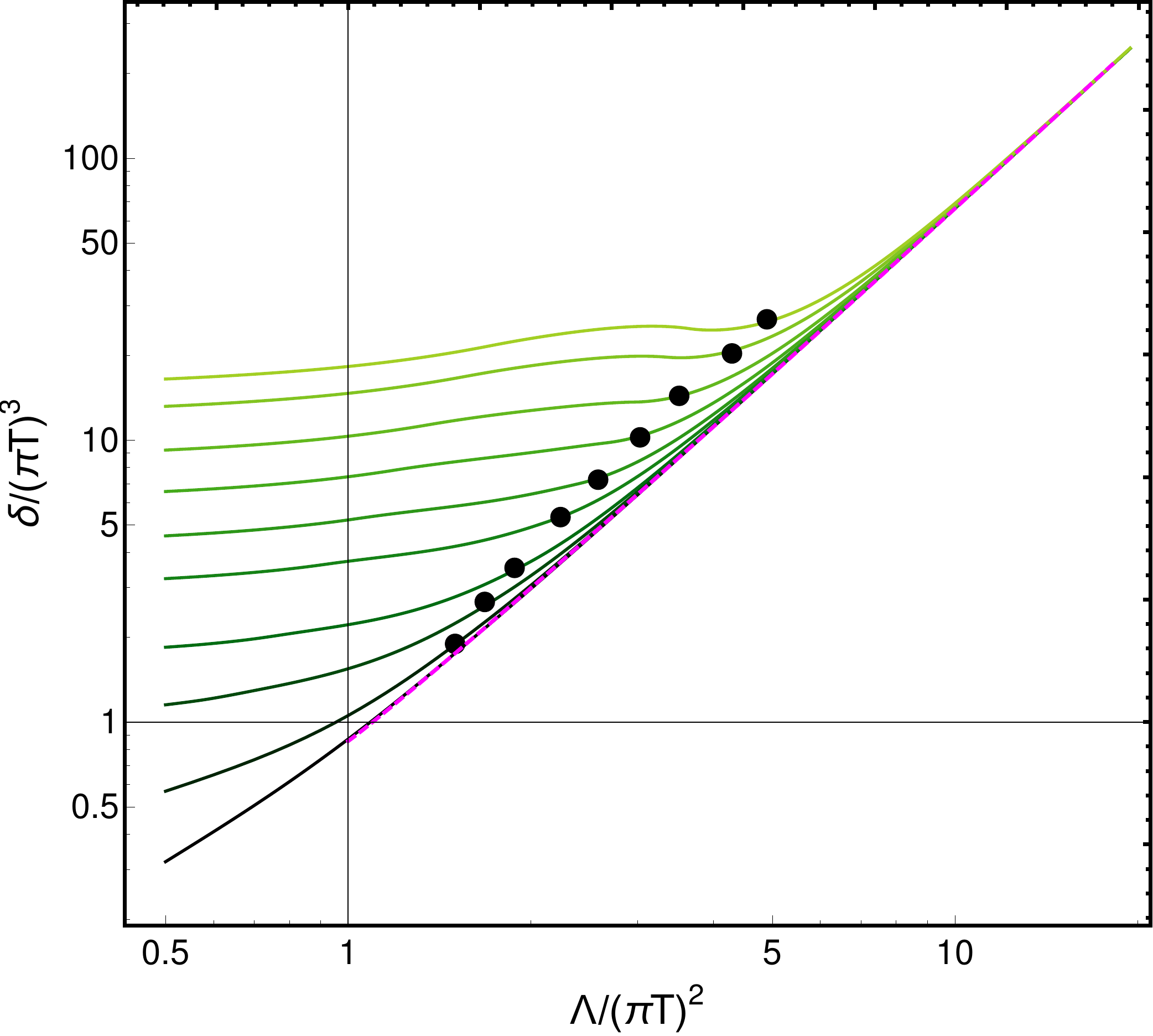}
 	\hspace{1cm}
 	\includegraphics[width=6.7cm]{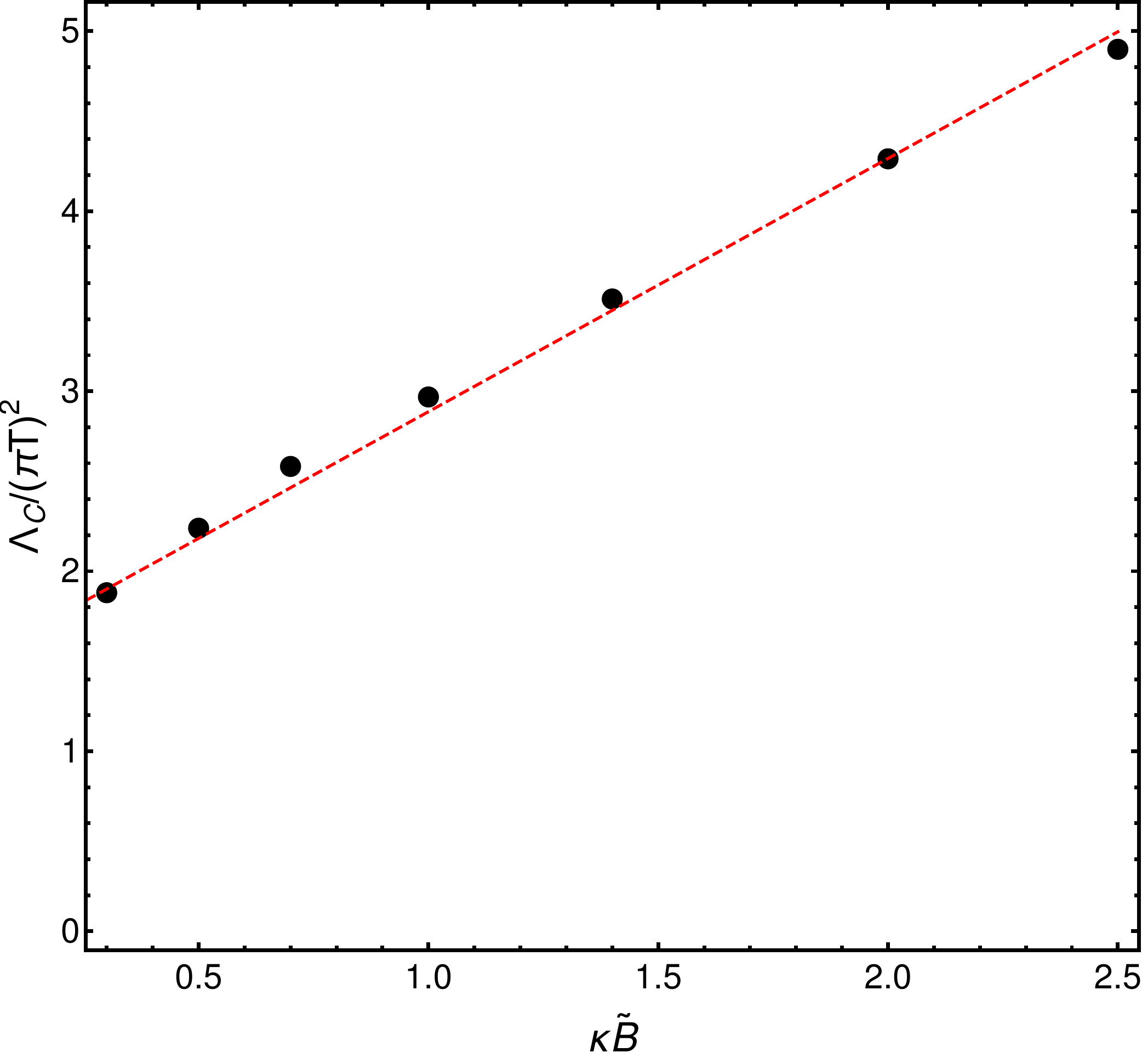}
 	\caption{\label{fig:tanhmax-min}Left: Double logarithmic plot of $\delta$ (see equation \eqref{eq:delta}) against the abruptness of the hyperbolic tangent quench $\Lambda$ for several values of $\kappa \tilde{B}\in[0-2.5]$ (black-light green). For high enough $\Lambda$ all cases converge to a straight line with slope 1.828. Black points highlight the critical $\Lambda_C$ as defined in 
 		the text. Right: Critical inverse width $\Lambda_C$ against $\kappa \tilde{B}$ (black) which can be fitted by a straight line (red).}
 \end{figure}
 %%%%%%%%%%%%%%%%%%%%%%%%%%%%%%%%%%%%%%%%%%%%%%
 %%%%%%%%%%%%%%%%%%%%%%%%%%%%%%%%%%%%%%%%%%%%%%
 The bigger the value of $\kappa \tilde B$, the abrupter the quench has to be in order to get to the universal regime. Furthermore, we can define a critical abruptness $\Lambda_\text{cr}$ of the quench, as the abruptness where $\delta''(\Lambda)$ has its global maximum, i.e. $\delta'''(\Lambda_\text{cr})=0$. To determine the critical abruptness we use splines for the data. The critical $\Lambda$ are depicted by black dots; we see that they fit in the logarithmic representation to a straight line (red). The dependence of $\Lambda_\text{cr}$ on the value of $\kappa \tilde B$ is shown in the r.h.s. of figure \ref{fig:max-min}.
 We find a linear dependence of the critical abruptness on $\kappa \tilde B$; concretely, we find
 \begin{equation}
 \Lambda_\text{C}/(\pi T)^2=4.15\, (\kappa \tilde B)^{1.04}+0.13,
 \end{equation}
 with a fitting error of $10^{-2}$. After considering the Gaussian quench, we did an analogous study with the hyperbolic tangent, depicted in figure \ref{fig:tanhmax-min}. The discussion of the results is more qualitatively than for the Gaussian quench, since the first minimum of the hyperbolic tangent quench can be altered by the oscillation of the current; this causes inaccuracies but the behaviour stays qualitatively the same. In the case of fast enough quenches, we found again that $\delta$ is independent of the value of $\kappa \tilde B$, concretely
 \begin{equation}
\log(\delta/(\pi T)^3)=1.828\, [\log(\Lambda/(\pi T)^2)]^{1.04}-0.155,\end{equation} with a fitting error of $10^{-3}$. \newline 
The dependence of the critical abruptness is again linear dependent on $\kappa \tilde B$ but we observe an obvious variance of the points.\newline\newline
  \begin{figure}[H] 
    	\centering
    	\includegraphics[width=6.7cm]{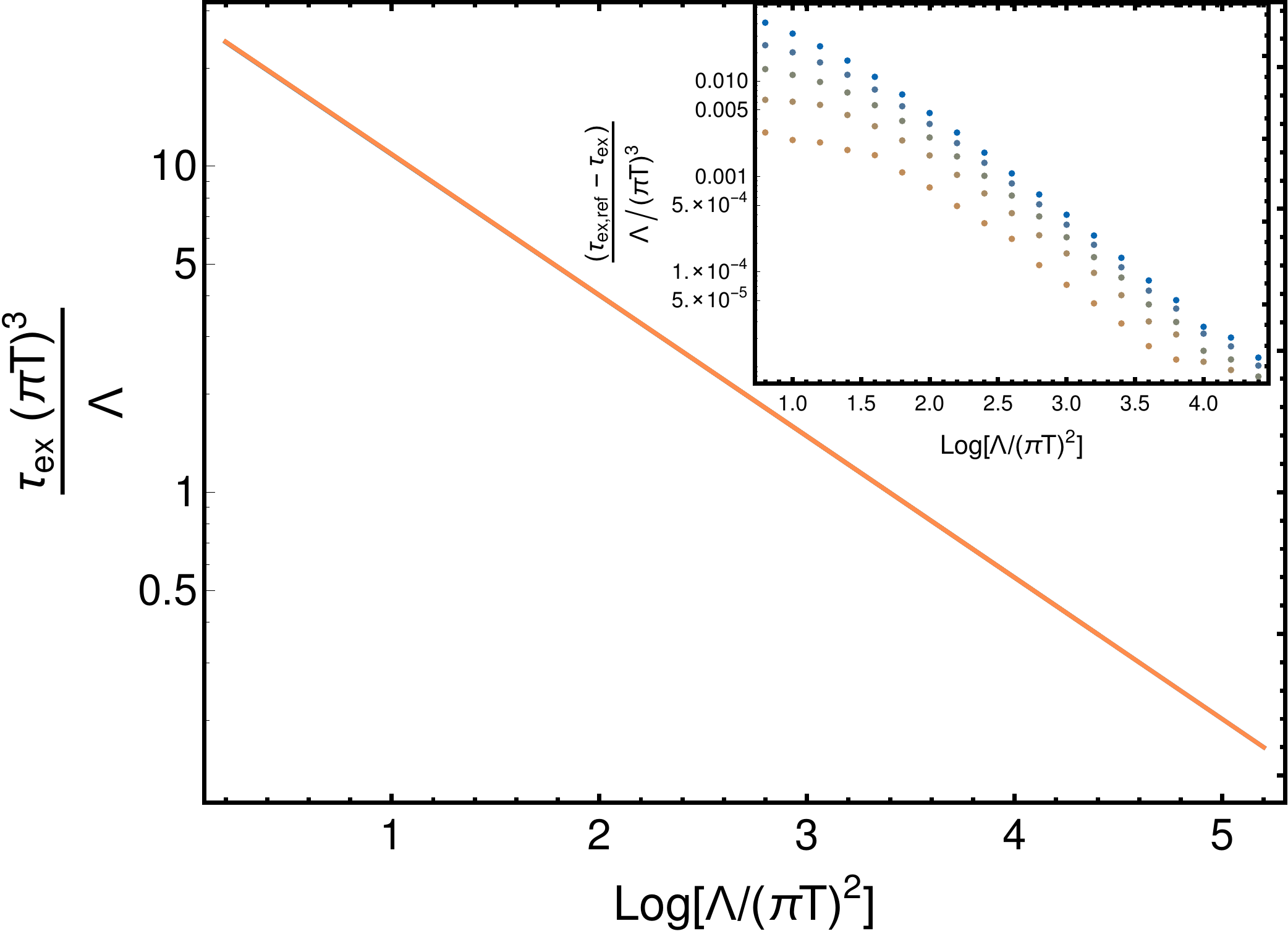}
    	\hspace{1cm}
    	\includegraphics[width=6.7cm]{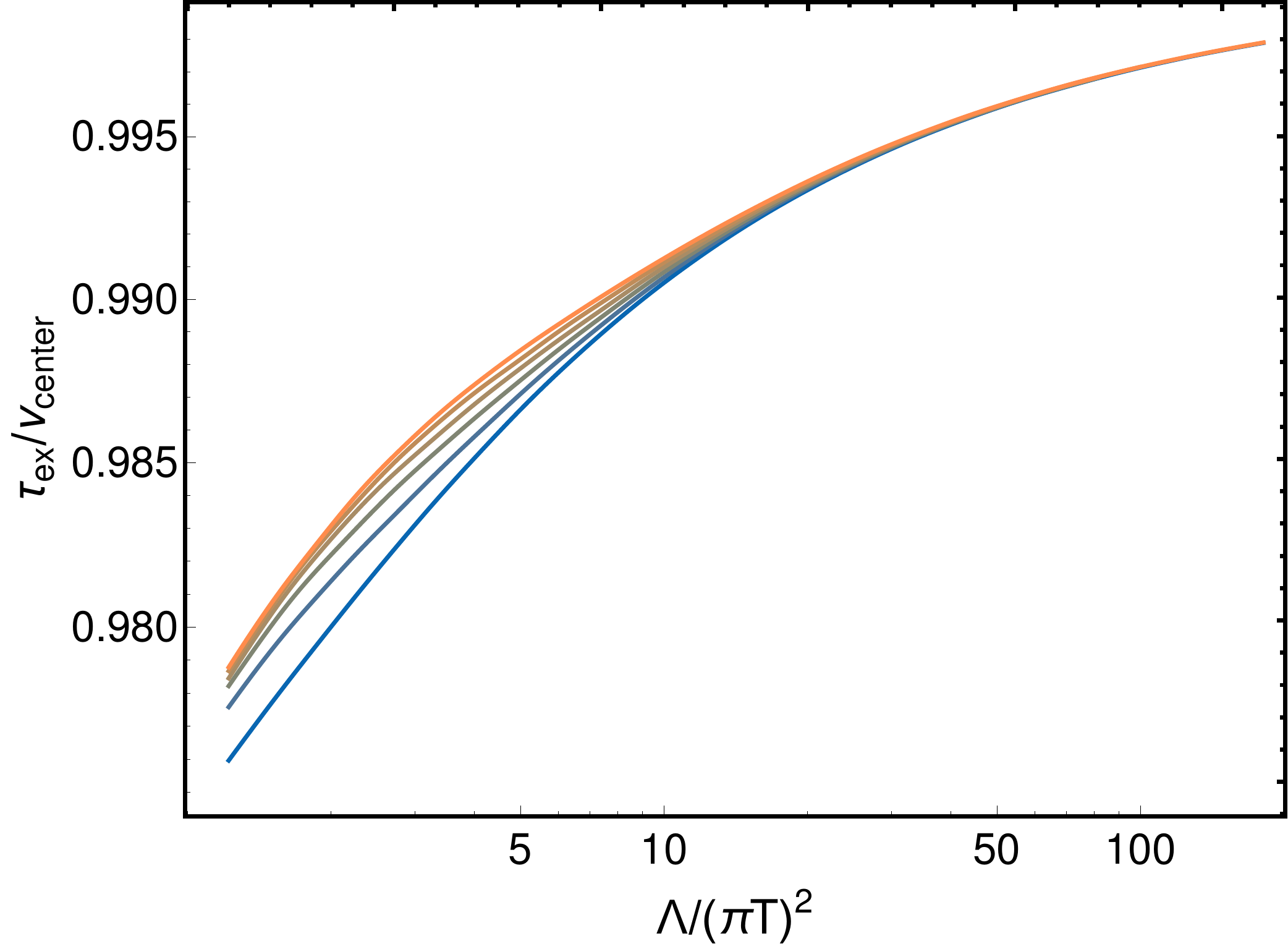}
    	\caption{\label{fig:excitime}Left: Excitation time divided by the abruptness $\tau_\text{ex}/\Lambda$ against the logarithm of the abruptness $\Lambda$ of a Gaussian quench for several values of $\kappa \tilde{B} \in \{0.5,0.75,1,1.25,1.5,1.75\}$ (blue-orange). Inset: The biggest $\kappa \tilde{B}=1.75$ functions as reference to make the difference between the lines visible. Right: Logarithmic plot of the relative excitation time $\tau_\text{ex}/v_\text{center}$ against the abruptness of the quench $\Lambda$ for several values of $\kappa \tilde{B} \in \{0.5,0.75,1,1.25,1.5,1.75\}$ (blue-orange). For high enough $\Lambda$ all cases converge to a straight line.}
    \end{figure}\newpage\noindent
 In order to compare the results to \cite{Buchel:2013lla}, we did a similar analysis defining an ``excitation time'' $\tau_\text{ex}$ by
 \begin{equation}
 \left.\frac{\langle J_z\rangle- F(v)}{F(v)}\right|_{v=v_k}=0.01,\label{eq:extime}
 \end{equation}
 where $\tau_\text{ex}$ is the first $v_k$, fulfilling eq. \eqref{eq:extime}, and $F(v)=\langle J_{z,\text{ad}}\rangle$ is given by the adiabatic response eq. \eqref{eq:adcurrent}. In other words: the excitation time is the first time where the current deviates 1\% from the adiabatic response. The results are shown in figure \ref{fig:excitime}. As the authors of \cite{Buchel:2013lla} we found an universal behaviour for fast enough quenches. The straight lines in the l.h.s. of figure \ref{fig:excitime} were hard to distinguish by eye. As shown in the inset, the differences between the lines get's smaller for bigger values of $\Lambda$, which is already clear from the picture on the r.h.s. of figure \ref{fig:excitime}. Hence, we notice that the smaller the value of $\kappa \tilde B$, the faster the system gets out of equilibrium. \newline\newline
 Further on, we did the same analysis for the hyperbolic tangent quench which we present in figure \ref{fig:excitime_tanh}. Again, in the case of fast enough quenches we can observe an universal behaviour, independent of $\kappa \tilde B$. As figure \ref{fig:adiabvsnoadiab} already indicates, we notice that the value of $\kappa \tilde B$ plays an important role in the slow quench regime. The smaller $\kappa \tilde B$, the more different the system behaves compared to the adiabatic response. This is mirrored in the excitation times which are in the slow quench regime much smaller for smaller values of $\kappa \tilde B$. The current reaches in this regime, dependent on the value of $\kappa \tilde B$, not for all quenches the 1\% deviation. The bigger the value of $\kappa \tilde B$, the bigger values of $\Lambda$ are necessary in order to push the system out of equilibrium.
    %%%%%%%%%%%%%%%%%%%%%%%%%%%%%%%%%%%%%%%%%%%%%%
 \subsection{Quasi-normal mode formalism}\label{sec:qnmf}
In section \ref{sec:QNM}, we presented a qualitative intuition about QNMs. In order to compute the QNM frequencies in our system, we briefly review an appropriate framework \cite{Kokkotas:1999bd,PhysRevD.45.2617}.
 We rewrite eq. \eqref{eq:DynEq_TimeDom} in the form
 \begin{equation}
\bm{\alpha}[U]+\bm{\beta}[\dot U]+S=0,
 \end{equation}
 for given initial data $U_\text{in}(u)=U(u,0)$. Applying a Laplace transformation\footnote{The parameter $s$ of the Laplace transformation is related to the Fourier parameter by $s=-\im \omega$.}
 \begin{equation}
 \bar{U}(u;s) = {\cal L}[U(v,u)](s) = \int\limits_0^{\infty} \\\dd v \, U(v,u) \e^{-sv}
 \end{equation}
 and taking into account that for a generic function $f(v,u)$
 \begin{equation}
 {\cal L}\left[\frac{\partial^{n}}{\partial v^n}f(v,u)\right](s) = s^n\,\bar{f}(u;s) - \sum_{k=0}^{n-1} s^{n-k-1} \frac{\partial^{k}}{\partial v^k}f(0,u),
 \end{equation}
 we obtain
 \begin{align}
 {\boldsymbol \alpha} [ \bar{U}] + s\, {\boldsymbol \beta} [ \bar{U}]  = {\boldsymbol \beta} [ U_{\rm in}] - \bar{S}, \label{eq:LapTransEq} \end{align}
 where \begin{align}
 \bar{S}(u;s) =  \bar{V}_0(s) \sum_{i=0}^{N_S} a_i(u)  s^i\, -  \sum_{i=0}^4 \sum_{k=0}^{i-1} s^{i-k-1} \frac{\dd^{k}}{\dd v^k}V_0(0). \label{eq:LapTransSource}
\end{align}
Note that in our case the last term has to be approximately zero, as mentioned in the previous section.
In order to obtain the QNM, we have to solve the homogeneous version of eq. \eqref{eq:LapTransEq}. 
We can formulate the problem as Generalised Eigenvalue problem (GEVP), where the complex eigenvalues $s_n$ correspond to the QNMs  and the regular solutions $\phi_n(u)$ are the corresponding eigenvectors
\begin{equation}
\bm{\alpha}[\phi_n]+s_n\,\bm{\beta}[\phi_n]=0.
\end{equation}
In our model, the problem is linear in $s_n$ due to our choice of ingoing Eddington-Finkelstein coordinates. In Poincaré coordinates the problem would be in general quadratic in $s_n$. In our set-up the operators read
\begin{align}
\bm{\alpha}&=-u(1-u^4)\,\frac{\dd^2}{\dd u^2}-(3-7u^4)\frac{\dd}{\dd u}+(8+\lambda^2)u^3,\\
\bm{\beta}&=-\left(2u\frac{\dd}{\dd u}+3\right).
\end{align}
Within this framework, we are able to compute the QNM frequencies required in the next section.
\subsection{Resonances}\label{sec:res}
\begin{figure}[H] 
	\centering
	\includegraphics[width=8cm]{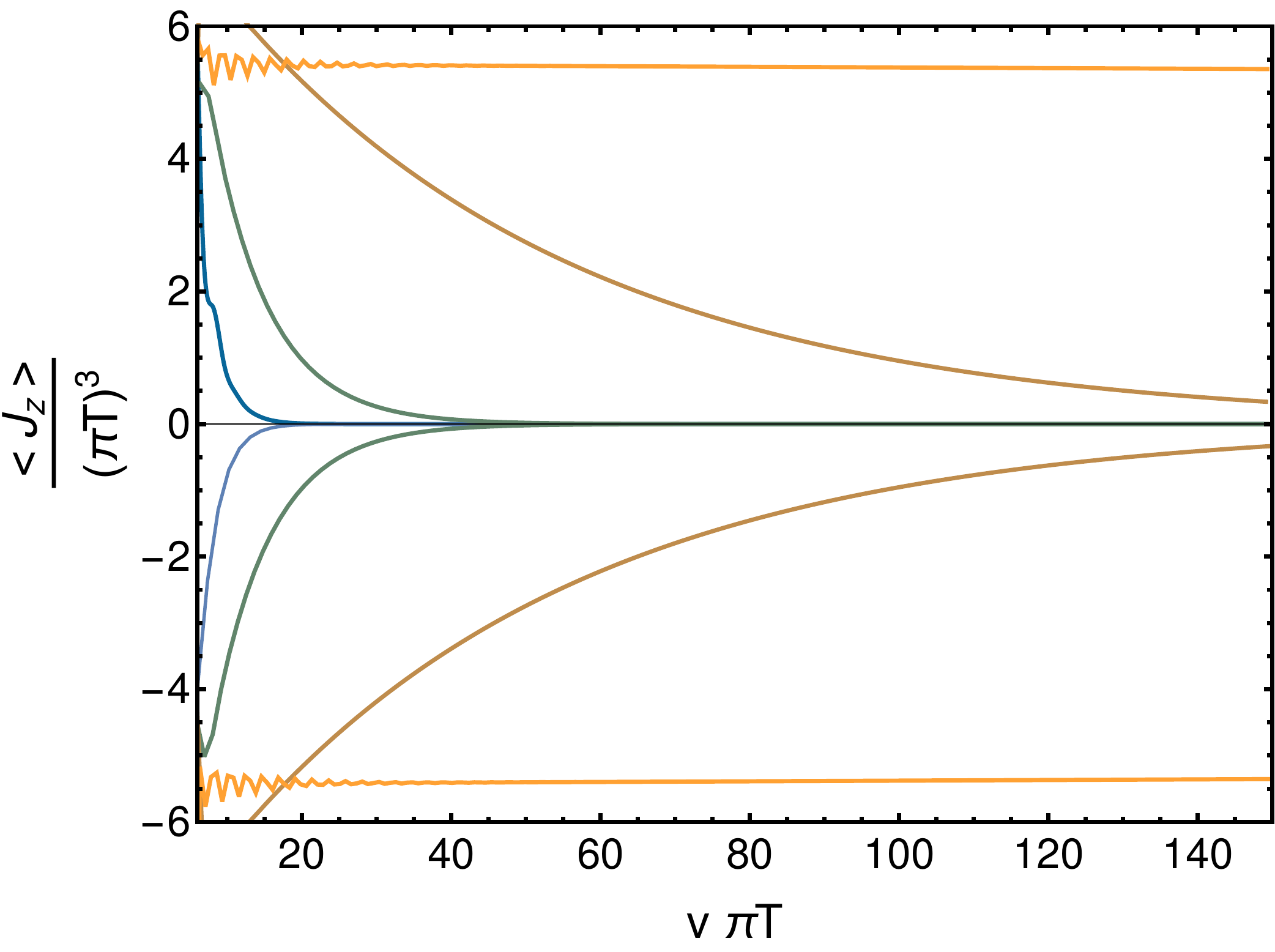}
	\caption{\label{fig:latetime} Envelope of the current against time for fixed Gaussian width $\Lambda=6(\pi T)^2$ and fixed center $v_0=5/(\pi T)$ for several values of $\kappa \tilde{B}\in\{0.5,0.75,1,3/2\}$ (blue - orange). Late time oscillations decay faster for smaller $\kappa \tilde{B}$.}
\end{figure}

We now focus on the late time behaviour of the current; concretely, we are looking at the current of a fixed Gaussian source for several values of the magnetic field. Increasing the value of $\lambda$ results in a lower decay rate of the late time oscillations, as shown in figure \ref{fig:latetime}. The relaxation time gets very small compared to all other scales of the system for $\lambda/12=\kappa \tilde B\gtrsim 1$. The late time behaviour is, as we will show in this section, dominated by the QNMs of the system. A small decay rate corresponds, in the QNM picture, to a small imaginary part. Therefore, the decreasing damping rate indicates the existence of a QNM approaching the real axis for increasing $\kappa \tilde B$.   
%%%%%%%%%%%%%%%%%%%%%%%%%%%%%%%%%%%%%%%%%%%%%%%
%%%%%%%%%%%%%%%%%%%%%%%%%%%%%%%%%%%%%%%%%%%%%%
%%%%%%%%%%%%%%%%%%%%%%%%%%%%%%%%%%%%%%%%%%%%%%
\begin{figure}[H] 
	\centering
	\includegraphics[width=6.7cm]{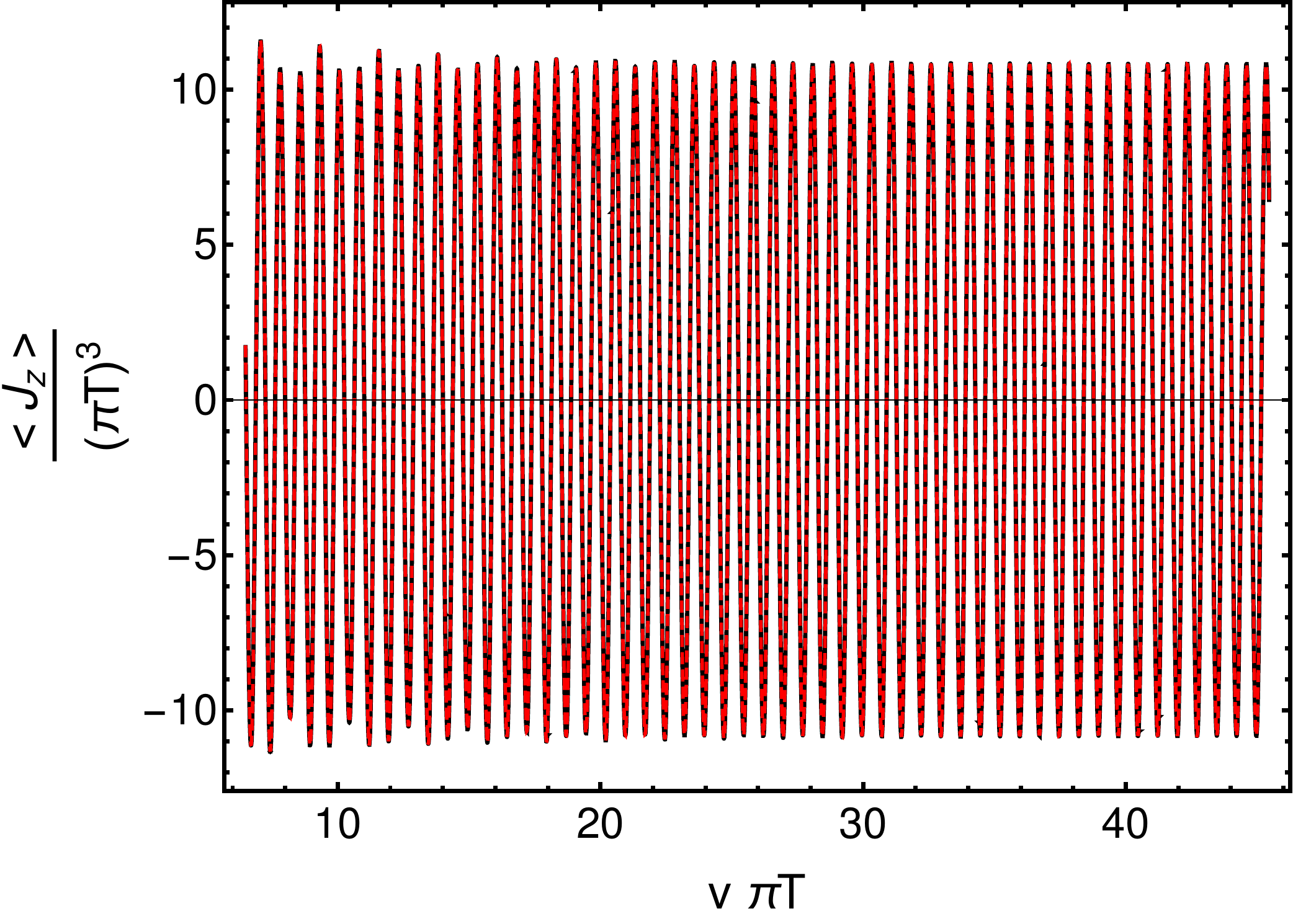}
	\hspace{1cm}	\includegraphics[width=6.7cm]{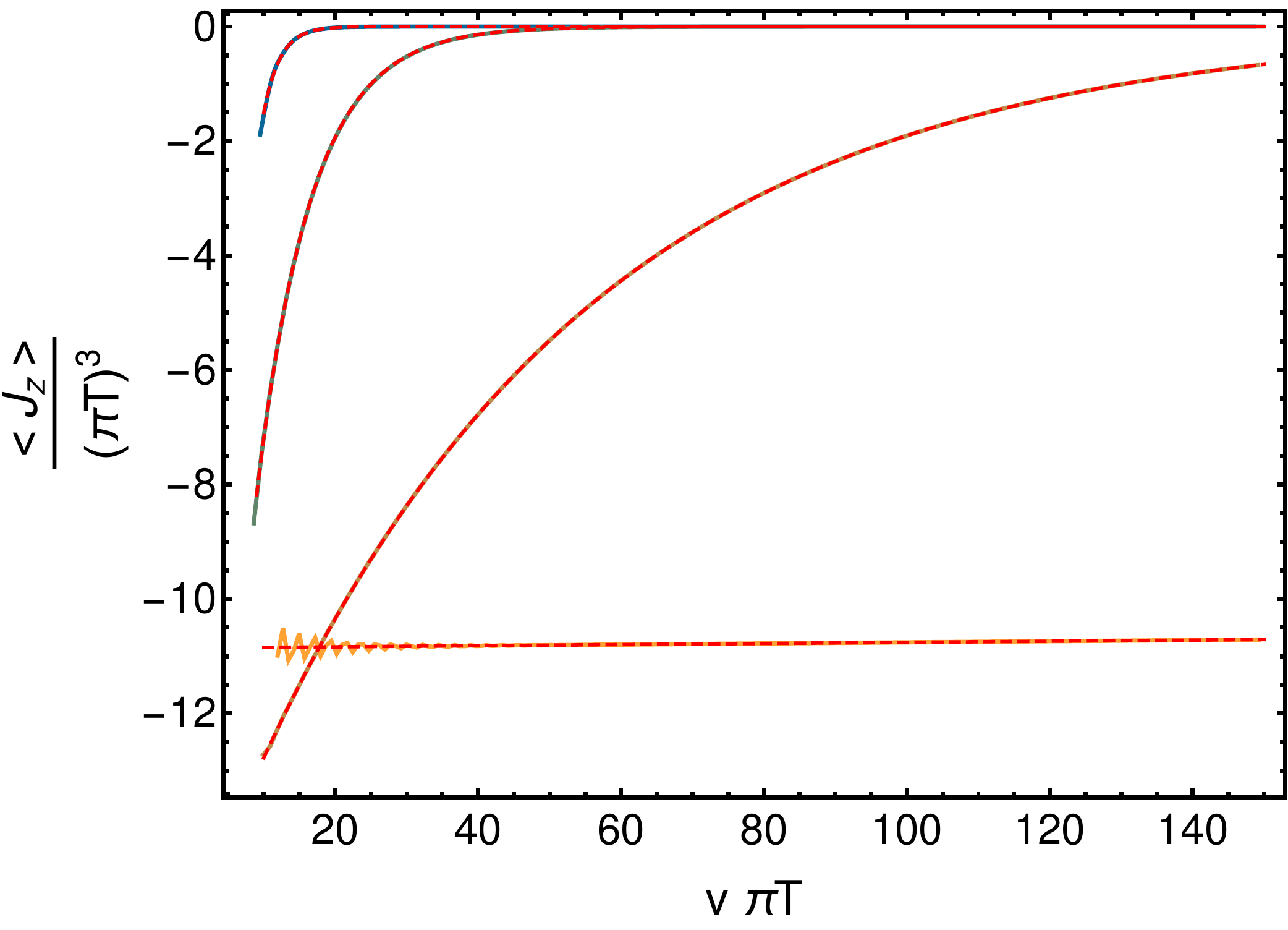}	
	\caption{\label{fig:fitqnm}Left: Zoom in the current shown in figure \ref{fig:latetime} (red). The fit \ref{eq:fit2} is depicted as black dashed line. Late time oscillations decay faster for smaller $\kappa \tilde{B}$. Right: The fit eq. \eqref{eq:fit3} to the data, presented in figure \ref{fig:latetime}.}
\end{figure}
To determine, whether the oscillations in the current correspond to the QNMs, we consider a damped oscillation, consisting of the two lowest QNMs $\omega_1+\im\,\delta_1$ and $\omega_2+\im\,\delta_2$
\begin{equation}
f(v)=A_1\,\e^{\delta_1 v}\,\sin(\omega_1v+B_1)+A_2\,\e^{\delta_2 v}\,\sin(\omega_2v+B_2).\label{eq:fit2}
\end{equation}
Thereby, the damping factor $\delta_i$ is given by the imaginary part of the QNM and the oscillation frequency $\omega_i$ by the real part. Since higher QNMs are damped very fast, we found it sufficient to fit to the first two QNMs. The result is depicted in the r.h.s. of figure \ref{fig:fitqnm}. The parameters of the fit, which matches perfectly the obtained current, are shown in table \ref{tab:fit1}. Obviously, the lowest QNM dominates the oscillation in the current.
\begin{table}[H]\centering
	\begin{tabular}{*{6}{c}}
		\toprule
		$\omega$ &	$\delta$ & $A$ & $B$ & error($A$) & error($B$)\\
		\midrule
		8.37068 &-0.0000905 & -10.8562& -4.15408& $10^{-13}$ & $10^{-14}$\\
			11.23659 &-0.132746 & -1.86093& 0.524401& $10^{-11}$ & $10^{-11}$\\
		\bottomrule\end{tabular}\caption{Fit data for fitting eq. \eqref{eq:fit2} to the data presented in figure \ref{fig:fitqnm}. \label{tab:fit1}}
\end{table}
 Furthermore, we demonstrate, that the damping, presented in figure \ref{fig:latetime}, corresponds exactly to the imaginary part of the lowest QNM.
 Therefore, we consider the following fit model
 \begin{equation}
 g(v)=A_1\,\e^{A_2\,v},\label{eq:fit3}
 \end{equation}
 which we fit to the envelope of the oscillatory current. The results are depicted in table \ref{tab:fit2} and the r.h.s. of figure \ref{fig:latetime}, respectively. The damping parameter matches perfectly the imaginary part of the QNM. Therefore, we conclude, that the damping of the oscillation is determined by the lowest QNM.
 \begin{table}[H]\centering
 	\begin{tabular}{*{4}{c}}
 		\toprule
 		$\kappa\tilde B$ &	Im$[\omega]$ & Im$[\omega_\text{fit}]$ & error($\omega_\text{fit}$)\\
 		\midrule
 		0.5 &-0.43758 & -0.43749& $10^{-5}$ \\
 		0.75 &-0.13078 &-0.13120& $10^{-5}$ \\	
 		1 &-0.021148 & -0.021147& $10^{-6}$ \\
 		1.5 &-9.1 $10^{-5}$ & -9.1 $10^{-5}$& $10^{-6}$ \\
 		\bottomrule\end{tabular}\caption{Fit data for fitting eq. \eqref{eq:fit3} to the data presented in figure \ref{fig:fitqnm}. \label{tab:fit2}}
 \end{table}
\begin{figure}[H] 
	\centering
	\includegraphics[width=7.05cm]{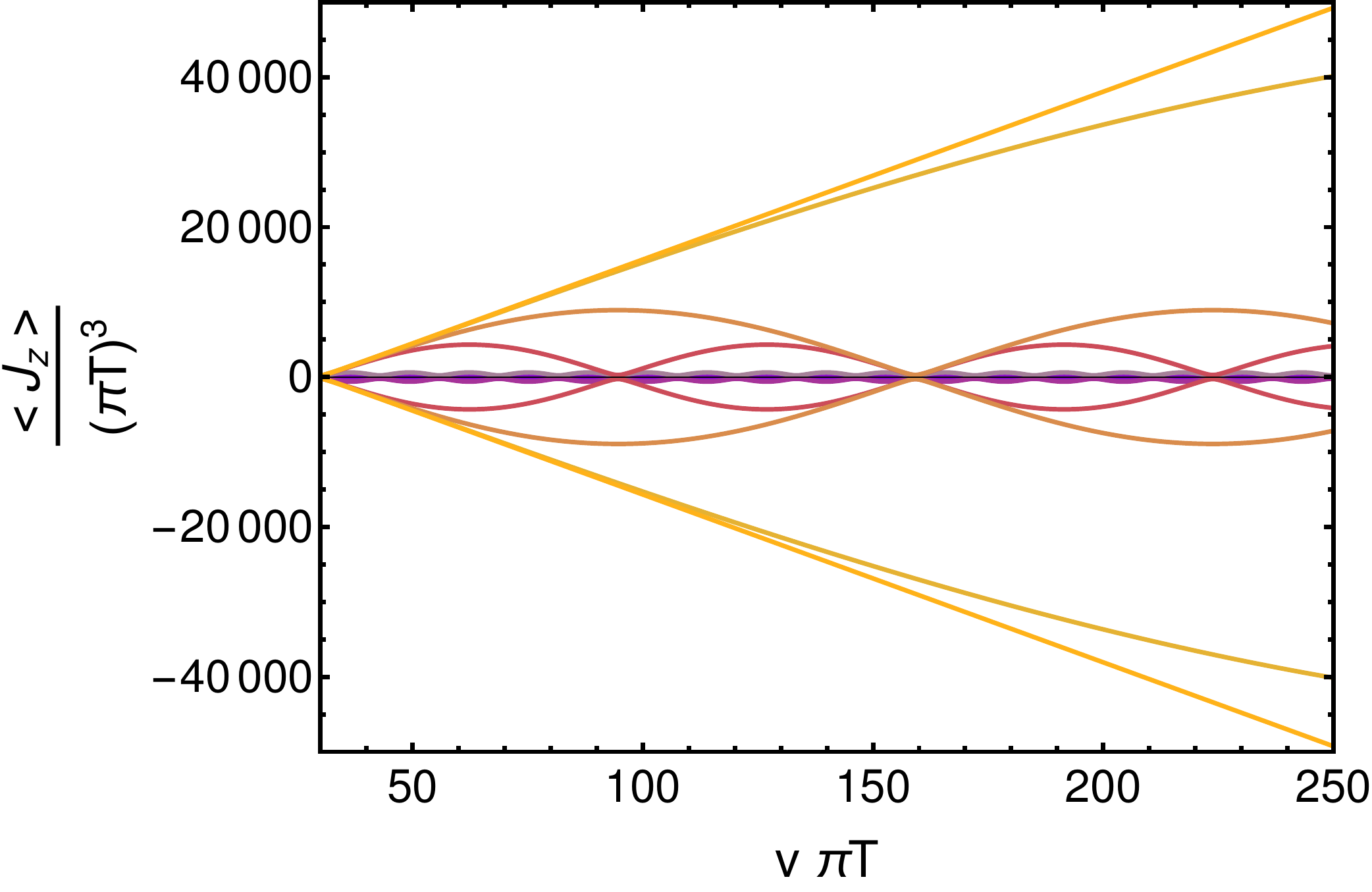}
	\hspace{0.8cm}
	\includegraphics[width=6.55cm]{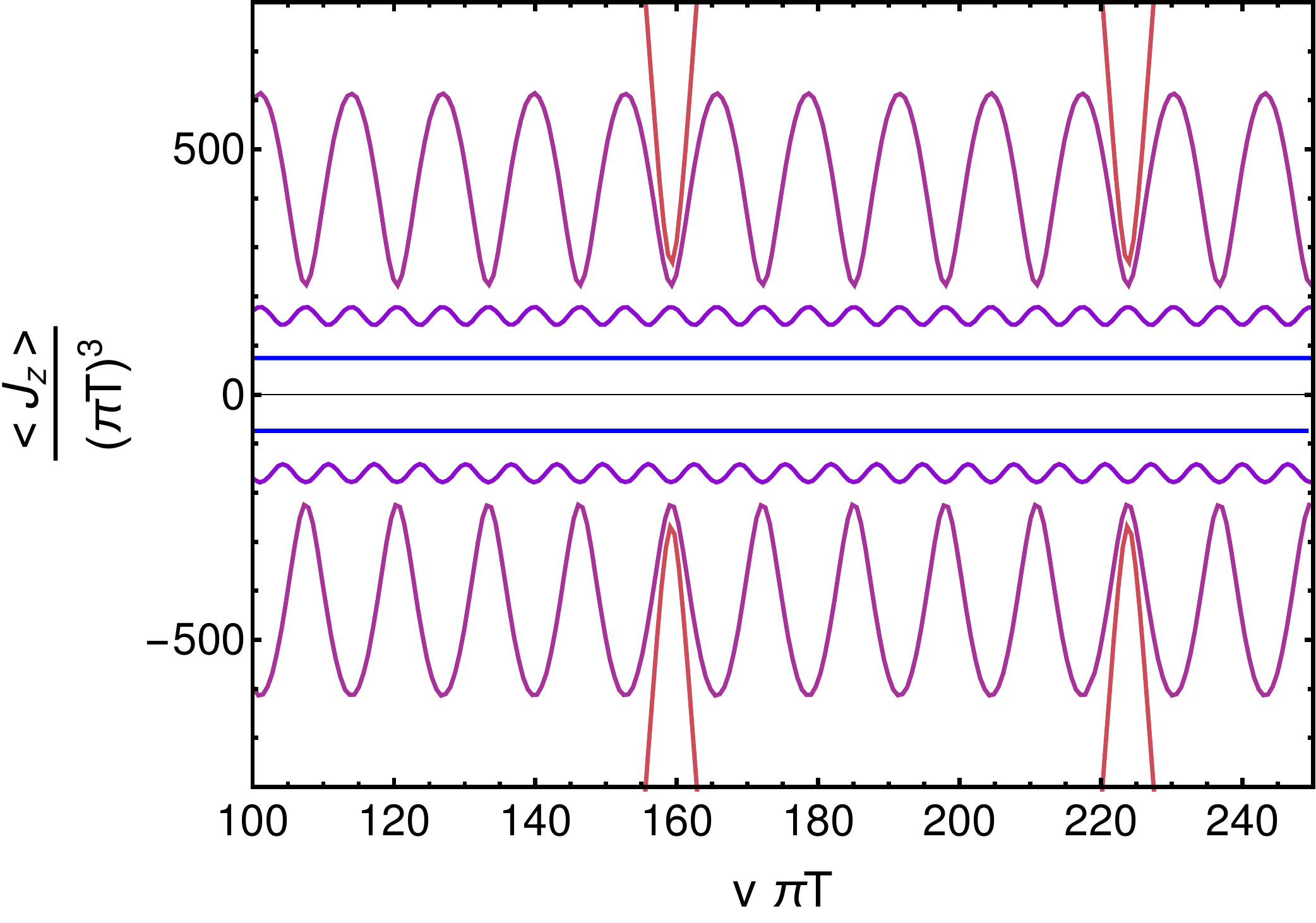}
	\caption{\label{fig:resonance}Left: Envelope of the current versus time for an oscillatory source in \eqref{eq:tanhsin} for $\omega\in\{0.5\,\omega_{\rm c},\, 0.9\,\omega_{\rm c},\,0.95\,\omega_{\rm c},\,0.99\,\omega_{\rm c},\,0.995\,\omega_{\rm c},\,0.999\omega_{\rm c},\,\omega_{\rm c}\}$  (blue to yellow) for $\kappa\tilde B=2$. Right: Zoom in the picture, shown on the l.h.s.}
\end{figure}
\begin{figure}[H] 
	\centering
\includegraphics[width=7.1cm]{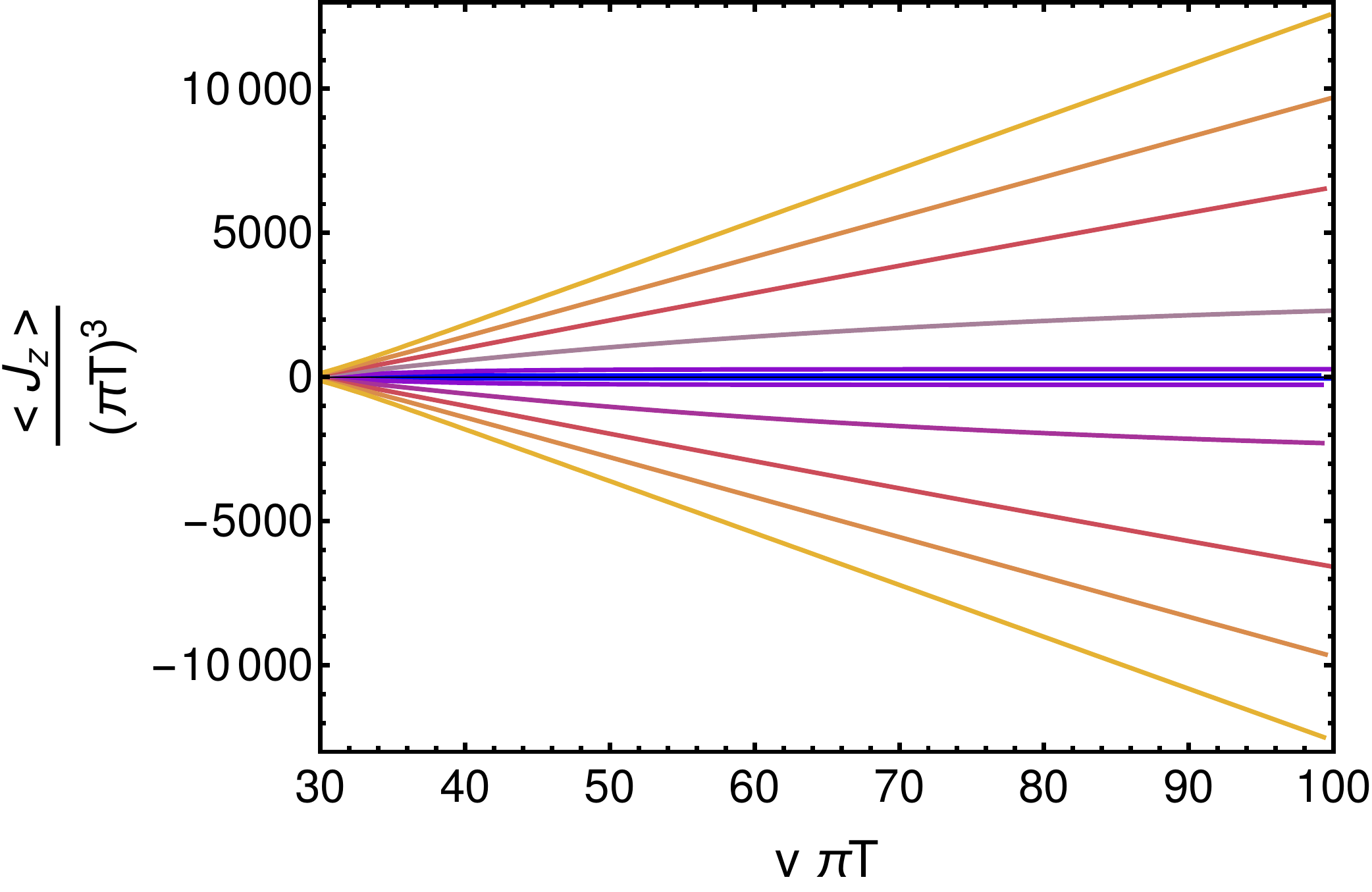}
	\caption{\label{fig:resonance2}Envelope of the current vs time for an oscillatory source \eqref{eq:tanhsin} for \mbox{$\omega=\omega_{\rm c}(\kappa \tilde B)$}  for \mbox{$\kappa\tilde B\in\{0.5,0.75,1,1.25,1.5,1.75\}$} (blue to yellow).}
\end{figure}
\newpage\noindent As a first step, we try to excite the resonances of the system directly. To do so we consider an oscillatory source, given by
\begin{equation}
V_0(v)=(1+\tanh(v-v_i))\,\sin(\omega v),\label{eq:tanhsin}
\end{equation}
where the factors differing from the sin term are chosen for numerical convenience. We investigate the response of the system for frequencies $\omega$ approaching the resonance frequency $\omega_c$ of the system. In order to determine the resonance frequency of the system, we have to make again use of the QNMs. \newline\newline
In figure \ref{fig:resonance} we tuned the frequency of the source from $\omega=\omega_\text{c}/2$ to $\omega=\omega_\text{c}$, in order to determine, whether the system shows a resonant behaviour or not. To obtain a resonant behaviour the frequency has to be very close to the resonance frequency, namely up to 99.9\%. This is obvious, considering the r.h.s. of figure \ref{fig:resonance}. All frequencies of the source, which are too far away from the resonance frequency induce a periodically oscillating current.\newline\newline
On the right hand side of figure \ref{fig:resonance2}, we investigated the dependence of the current for different $\kappa\tilde  B$ and a resonant source. The current grows in the case $\kappa \tilde B\gtrsim 1$  unbounded while for smaller $\kappa \tilde B$ the current oscillates with a constant amplitude. The reason why we can not observe resonances for $\kappa \tilde B<1$ will be clear in section \ref{section:Landau}.
Finally, we considered higher QNMs as frequencies for the source with results being qualitatively the same as for the lowest QNM.
%%%%%%%%%%%%%%%%%%%%%%%%%%%%%%%%%%%%%%%%%%%%%%
\subsection{Quasi-normal modes with backreaction}\label{sec:qnmwb}
\begin{figure}[t!] 
	\centering
	\includegraphics[width=6.7cm]{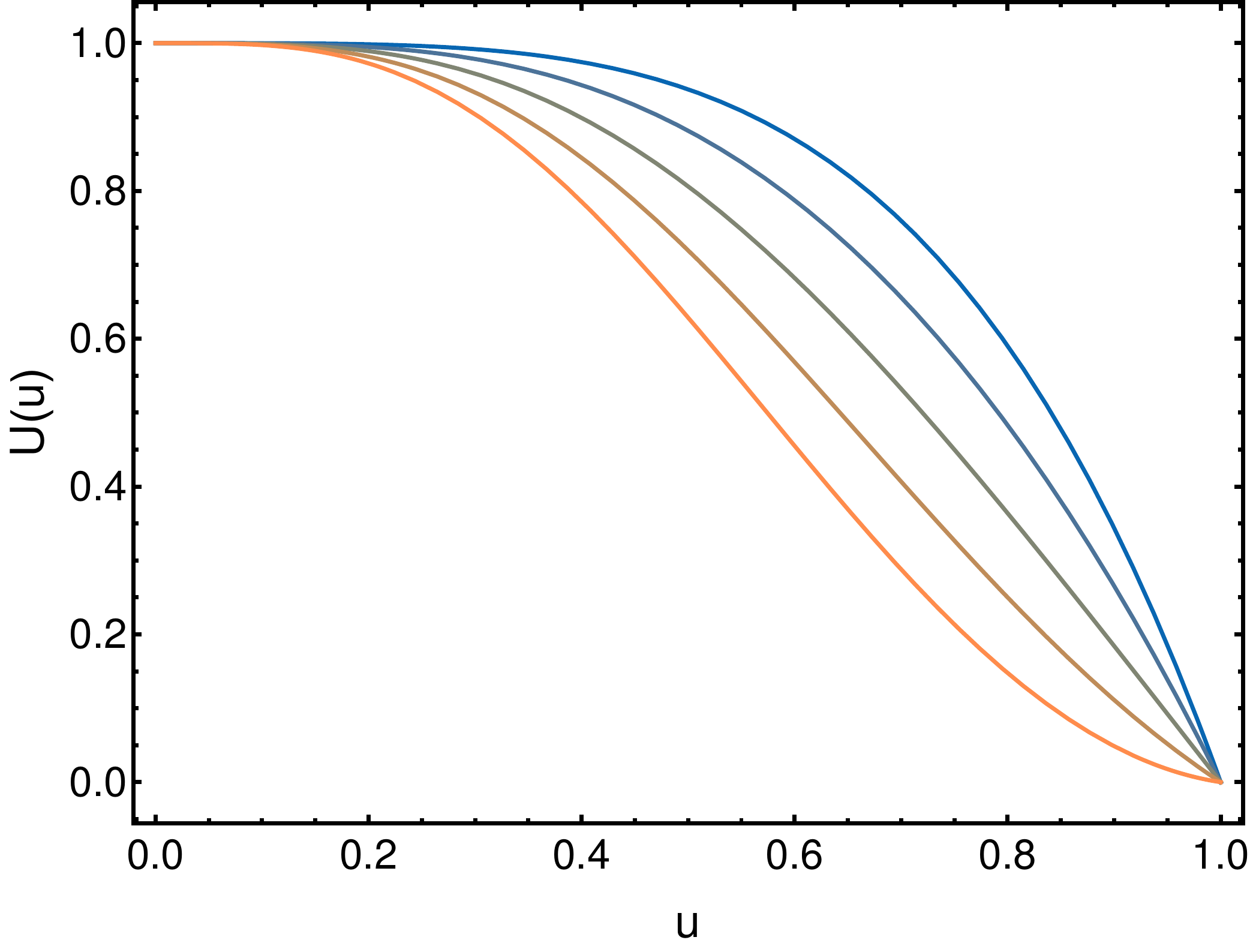}
	\hspace{1cm}
	\includegraphics[width=6.7cm]{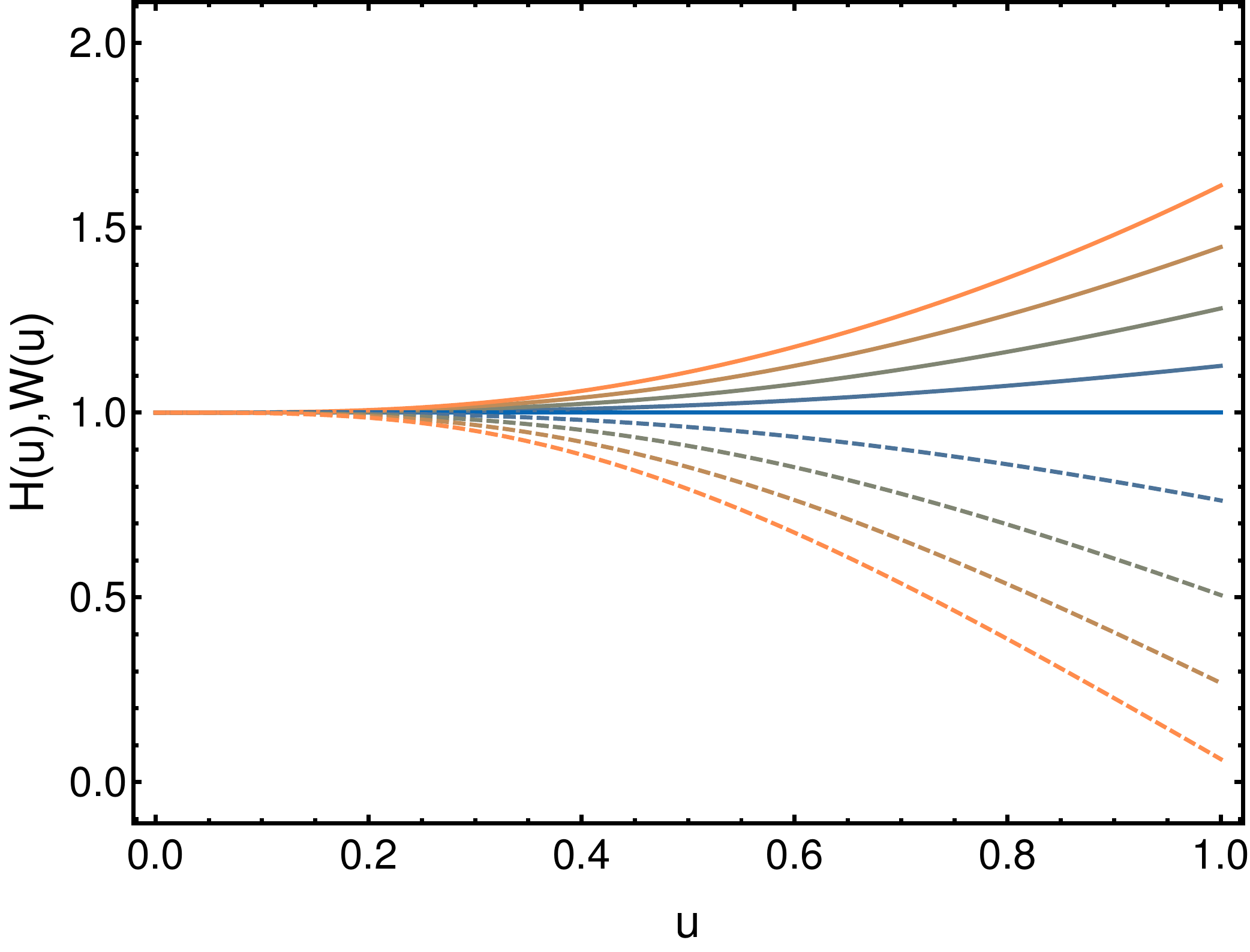}
	\caption{\label{fig:background}Left: Metric function $U(u)$ for  $\tilde{B}\in\{0,3,5,7,9\}$ (blue-orange). Right: Metric function $W(u)$ and $H(u)$ (dashed) for $\tilde{B}\in\{0,3,5,7,9\}$ (blue-orange).}
\end{figure}
%%%%%%%%%%%%%%%%%%%%%%%%%%%%%%%%%%%%%%%%%%%%%%
We now will investigate whether the resonances, observed in the previous section, are really a footprint of the anomaly. Therefore, we have to vary $\kappa$ and $B$ independent from each other. Due to this demand and since we are interested in strong magnetic fields, we have to go beyond the probe approximation and take backreaction into account. The magnetic field, pointing in $z$-direction, induces a spatial anisotropy and therefore breaks the spatial rotational invariance. Therefore, the metric function multiplied by $\dd z^2$ has to be distinct from the ones multiplying $\dd x^2$ and $\dd y^2$. 
This is manifest in the metric ansatz \cite{D'Hoker:2009mm}
\begin{equation}
 \dd s^2=\frac{1}{u^2}\left(-U(u)\, \dd v^2 - 2\,\dd v\,\dd u+W(u)^2\,(\dd x^2+\dd y^2)+H(u)^2\,\dd z^2\right).
\end{equation}
The action for the model under consideration reads
\begin{align}
S&= \frac{1}{2\tau^2}\int\dd^{5}X\,\sqrt{-g}\,\left(R+12\right)\\& +\int\dd^{5}X\,\sqrt{-g}\,\Big(-\frac{1}{4}F^{\mu\nu}F_{\mu\nu}-\frac{1}{4}H^{\mu\nu}H_{\mu\nu}+ \frac{\kappa}{2}\,\varepsilon^{\,\mu\alpha\beta\rho\lambda}A_\mu\left(F_{\alpha\beta}F_{\rho\lambda}+
3H_{\alpha\beta}H_{\rho\lambda}\right)\Big)\,.\nonumber
\end{align}
The Chern-Simons term does not contribute to the energy-momentum tensor since the energy-momentum tensor is given by the variation of the matter action with respect to the metric. Hence, the energy-momentum tensor is the usual one obtained in electrodynamics
\begin{equation}
T_{\mu\nu}=-\frac 14\,g_{\mu\nu}\,\left(F^{\rho\sigma}F_{\rho\sigma}+H^{\rho\sigma}H_{\rho\sigma}\right)+g^{\rho\sigma}\left(F_{\mu\rho}F_{\nu\sigma}+H_{\mu\rho}H_{\nu\sigma}\right),
\end{equation}
with the trace 
\begin{equation}
T^\mu_\mu = g^{\mu\nu}T_{\mu\nu}=-\frac 14\,\left(F^{\rho\sigma}F_{\rho\sigma}+H^{\rho\sigma}H_{\rho\sigma}\right).
\end{equation}
The trace of Einstein equations reads
\begin{equation}
0=g^{\mu\nu}\,(R_{\mu\nu}-\frac 12 g_{\mu\nu}R+6 g_{\mu\nu}-\tau^2T_{\mu\nu})=-\frac32 R+30+ \frac {\tau^2}{4}\,\left(F^{\rho\sigma}F_{\rho\sigma}+H^{\rho\sigma}H_{\rho\sigma}\right).
\end{equation}
Solving this equation for $R$ and plug it into the Einstein equations yields to the equations of motion for the background, in the so called trace reduced form,
\begin{equation}
 R_{\mu\nu}=-4\,g_{\mu\nu}+\tau^2\,\left(-\frac 16\, g_{\mu\nu}\,\left( F_{\alpha\beta}F^{\alpha\beta}+H_{\alpha\beta}H^{\alpha\beta}\right)+g^{\alpha\beta}\,\left(F_{\mu\alpha}F_{\nu\beta}+H_{\mu\alpha}H_{\nu\beta}\right)\right).\label{eqs::efg}
\end{equation}
The equations of motion of the gauge fields (which will be trivially fulfilled for the chosen background) are given by eq. \eqref{eq:eom11} and eq. \eqref{eq:eom22}.
Due to the diffeomorphism freedom, we can choose our coordinate system in a way that the black-hole horizon is located at $u=1$. The metric function $U(u)$ can be viewed as a blackening factor and has to vanish for this reason at the horizon, i.e. $U(1)=0$. For numerical convenience we will rescale the metric functions by use of the asymptotic expansions. The general ansatz is restricted to asymptotically AdS, requiring
\begin{equation}
U(0)=1,\ W(0)=1,\ H(0)=1.
\end{equation} 
Furthermore, we want to fix the remaining diffeomorphisms, namely the shift freedom, by imposing $U'(0)=0$. With this restrictions the asymptotic expansions are given by
\begin{align}
 U(u) & =1+u^4 \,\left[\bm{u}_4+\mathcal O(u^2)\right]+u^4\,\log(u)\,\left[\frac{B^2 \tau^2}{3}+\mathcal O(u^2)\right],\label{uexp}\\
  W(u) & =1+u^4 \,\left[-\frac{\bm{h}_4}{2}+\mathcal O(u^2)\right]+u^4\,\log(u)\,\left[-\frac{B^2 \tau^2}{12}+\mathcal O(u^2)\right],\label{wexp}\\
   H(u) & =1+u^4 \,\left[\bm{h}_4+\mathcal O(u^2)\right]+u^4\,\log(u)\,\left[-\frac{B^2 \tau^2}{6}+\mathcal O(u^2)\right].\label{hexp}
\end{align}
We can now introduce the rescaled functions $\tilde U(u),\tilde W(u), \tilde H(u)$ which will be used for all numerical calculations. They are given by
 \begin{align}
 U(u) & =1+u^4 \,[-1+(1-u)\tilde{U}(u)]+\frac{B^2 \tau^2}{3}\,u^4\,\log(u),\\
 W(u) & =1+u^4 \,\tilde W(u)-\frac{B^2 \tau^2}{12}u^4\,\log(u),\\
 H(u) & =1+u^4 \,\tilde{H}(u)-\frac{B^2 \tau^2}{6}\,u^4\,\log(u).
 \end{align}
The magnetic field $B$ and $\tau$ appear always as a product in the equations; hence, we can fix $ 2\tau^2=1$ without loss of generality. We want to investigate fluctuation on top of a background containing a magnetic field. Therefore, we consider the ansatz $V_\mu=V_y(x)=Bx$ and $A_\mu=0$ for the gauge fields. With this ansatz at hand, we can solve the e.o.m. for given magnetic field using a spectral method. \newline\newline In figure \ref{fig:background} we show the background functions for different values of the magnetic field. The homogenous solution ($B=0$) is given by the Schwarzschild solution \newline\mbox{$(U(u)=1-u^4,W(u)\equiv 1\equiv H(u))$}.\newline\newline
After having solved the background, we can look at pertubations on top of it. The Chern-Simons term contains, as already mentioned in the derivation of the energy-momentum tensor, no term proportional to the metric or its determinant and is fully contracted. Therefore, the metric fluctuations will decouple from the fluctuations of the gauge fields. Since the gauge fields are five dimensional there are 5 possible sectors, whereby only the $(v)$- and $(z)$-sectors of the gauge fields are coupled. In the $k=0$ regime only $(A_v)-(V_z)$ and $(V_v)-(A_z)$ are coupled. For the former the constraint equation reads
\begin{equation}
\dot a_0'(v,u)+\frac{\lambda \,u}{H(u)\, W(u)^2}\, \dot v_3(v,u)=0.
\end{equation}
This relation can be, analogously to the non-backreacted case, easily integrated in time leading to the relation
\begin{equation}
a_0'(v,u)=-\frac{\lambda\, u}{H(u)\, W(u)^2}\, v_3(v,u)+C_1.\label{eq:integrateout}
\end{equation}
We set the constant $C_1$ to zero, as discussed in the non-backreacted case (see section \ref{sec:statsol}). 
Relation \eqref{eq:integrateout} allows us to replace $a_0$ in the $v_3$-equation which is, after a Laplace transformation, given by
\begin{equation}
\bm{\alpha}[v_{3,n}]+s_n\,\bm{\beta}[v_{3,n}]=0, \label{eqS::gevp}
\end{equation}
with 
\begin{align}
\bm{\alpha}&=-\lambda^2\, u^3\,H(u)+u\, H(u) U(u) W(u)^4 \,\frac{\dd^2}{\dd u^2}\\&+\left[H(u) W(u)^3 \left(2 u \,U(u) W'(u)\!-\!W(u) (U(u)-u\,U'(u)\right)-u\, U(u) W(u)^4 H'(u)\right]\,\frac{\dd}{\dd u}\nonumber
\intertext{and}
\bm{\beta}&= \left[H(u) W(u)^3 \left(W(u)-2 u\, W'(u)\right)\right] +2 H(u) W(u)^4 \,u\,\frac{\dd}{\dd u}\,.\end{align}
Eq. \eqref{eqS::gevp} will be solved in context of a GEVP. We observed that in strong magnetic fields the QNMs, we are interested in, converge slowly or do not converge at all. To ensure the convergence the spectral solution is improved via a mapping of the radial coordinate $u\to\tilde u^2$. This mapping provides more gridpoints near the boundary located at $\tilde u=0$ and moves the logarithms, present in the expansions \eqref{uexp}-\eqref{hexp} to higher orders
\begin{equation}
u^n\,\log(u)\mapsto 2\, u^{2n}\,\log(u).
\end{equation} 
\subsection{Landau resonances}\label{section:Landau}
\begin{figure}[H] 
	\centering
	\includegraphics[width=6.7cm]{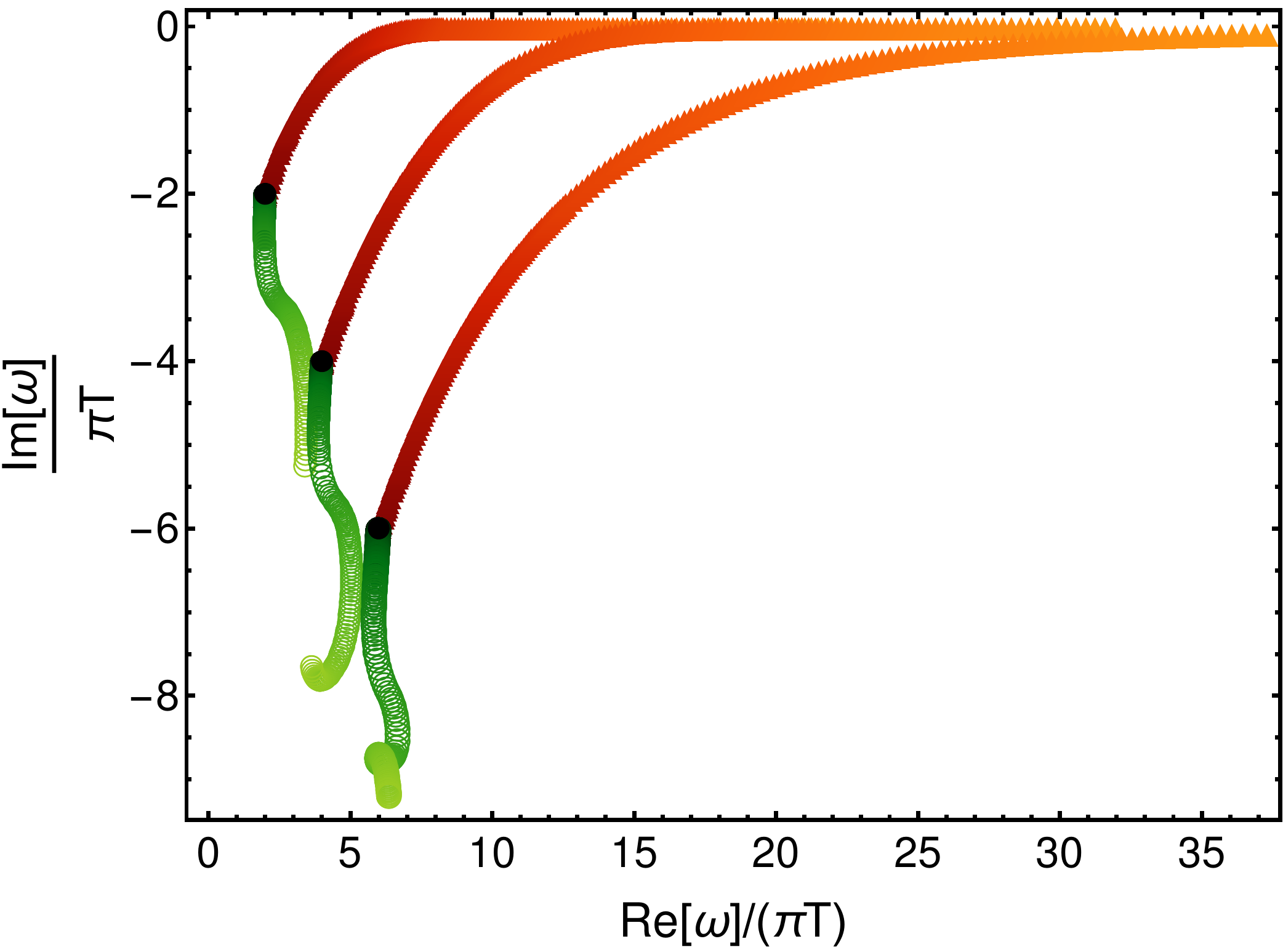}
	\hspace{1cm}
	\includegraphics[width=6.7cm]{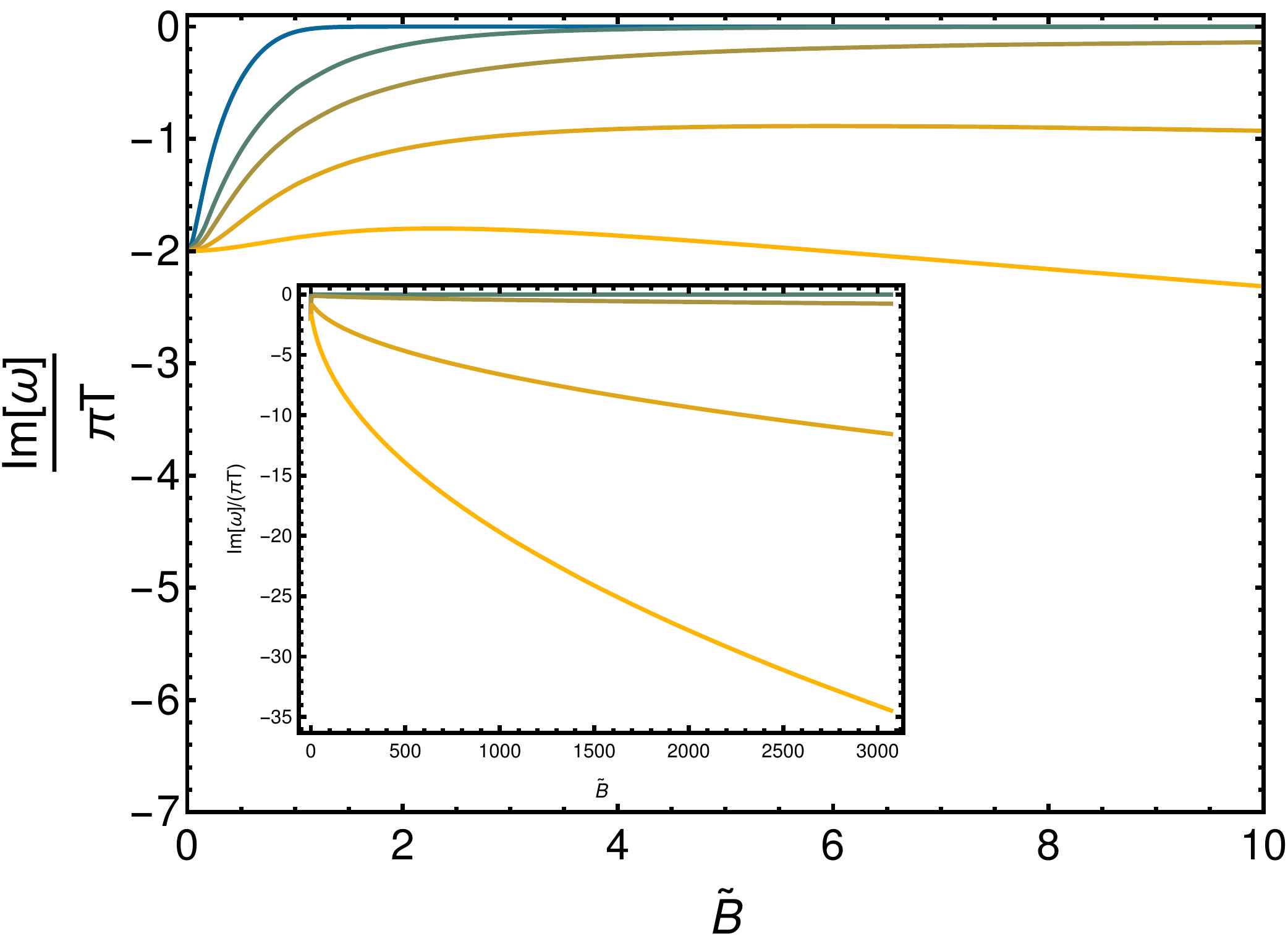}
	\caption{\label{fig:imqnm}Left: Imaginary versus real part of the three lowest QNMs at zero frequency for  $\kappa=0$, $\tilde{B}=0-6.77$ (green-yellow) and $\kappa=1$, $\tilde{B}=0-6.77$  (red-yellow). Right: QNMs' imaginary part versus magnetic field for $\kappa\in\{ 0.1, 0.2, 0.34, 0.5, 1\}$ (yellow-blue).}
\end{figure}

\begin{figure}[H] 
	\centering
	\includegraphics[width=6.7cm]{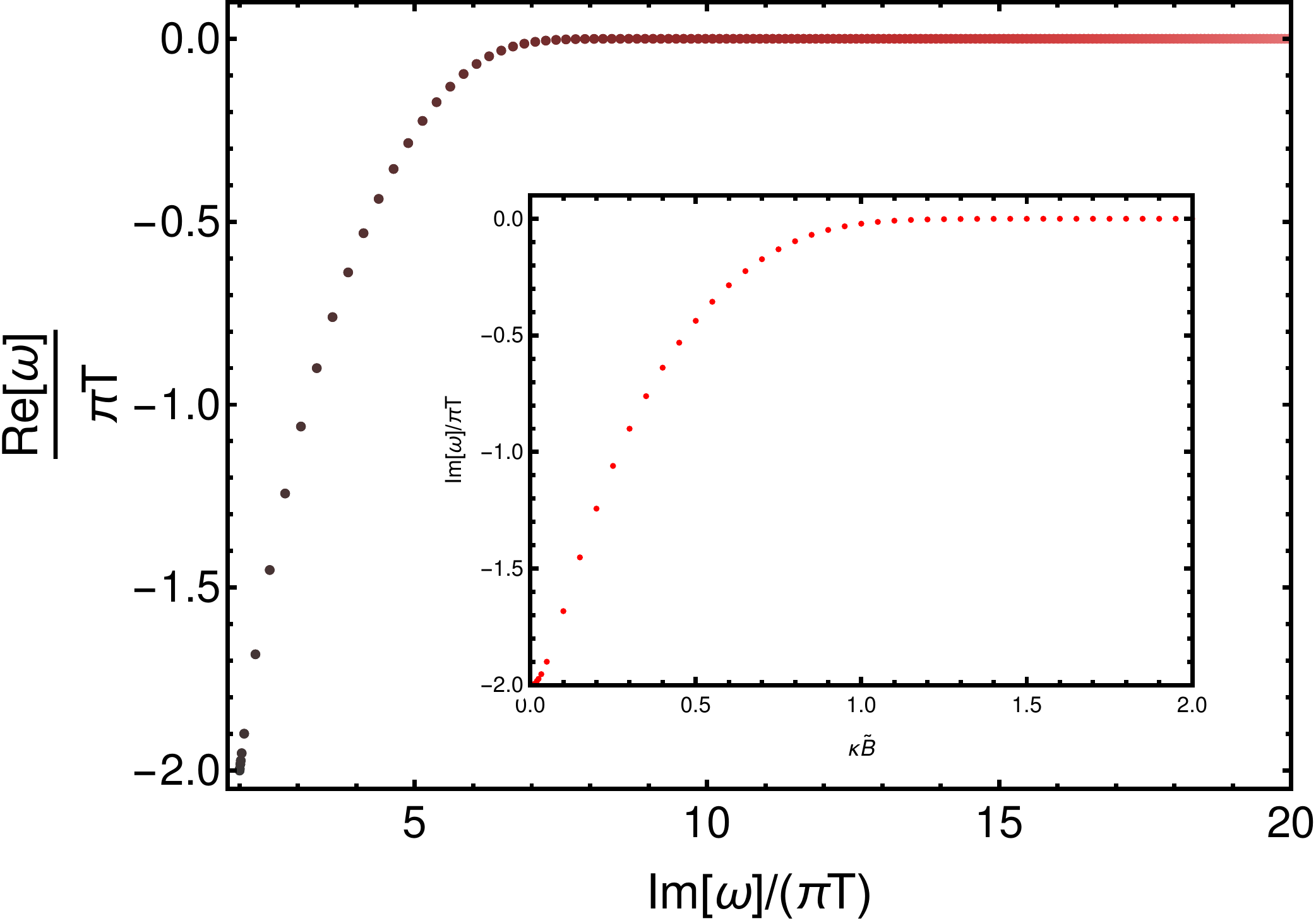}
	\hspace{1cm}
	\includegraphics[width=6.7cm]{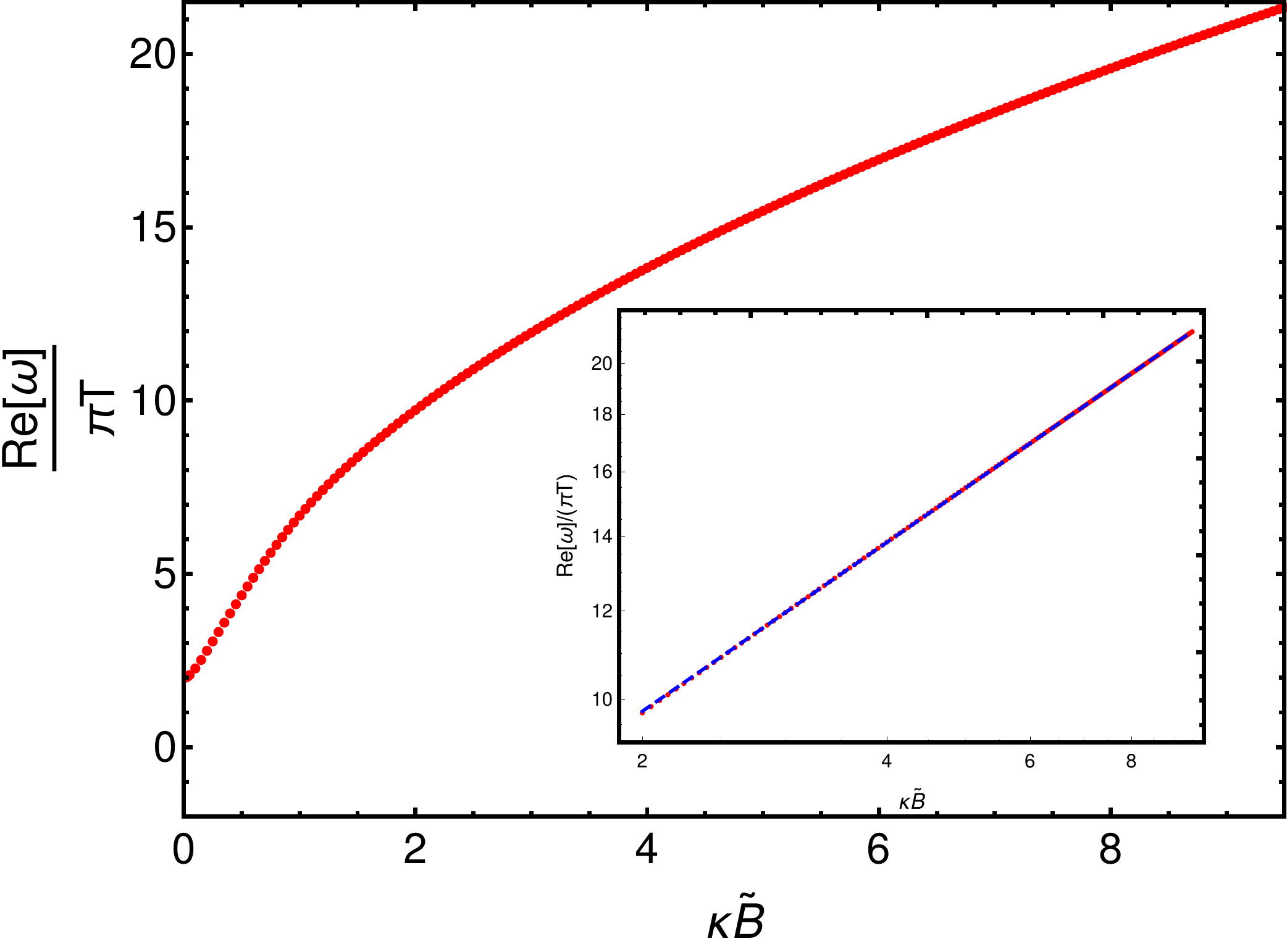}
	\caption{\label{fig:probe}Left: Imaginary versus real part of the lowest QNMs at zero frequency in the probe approximation for $\kappa\tilde B\in\{0-9\}$. Right: Real part of the QNM versus $\kappa\tilde B\in\{0-9\}$. Inset: Fit of a square root to the real part of the QNM.}
\end{figure}

Since we are now able to vary the Chern-Simons coupling $\kappa$ and the magnetic field independent from each other, we can check whether the resonances are an anomaly driven phenomenon or purely an effect of the magnetic field. Furthermore, the probe approximation restricted us to small values of the magnetic field, whereas we are interested in the fate of the QNMs at high values of the magnetic field.\newline\newline 
To check whether there is a qualitative difference in presence of the anomaly, we look at the behaviour of the three lowest QNMs while increasing $B$ with anomaly parameter switched on ($\kappa=1$) and without anomaly ($\kappa=0$). The result is depicted in figure \ref{fig:imqnm} and the difference in the behaviour is obvious. For $\kappa=1$ (and in general for $\kappa\gtrsim 0.5$) the QNMs approach the real axis, whilst for $\kappa=0$ they tend away from it. We observe an intermediate region $0<\kappa\lesssim 0.5$, as shown in the r.h.s. of figure \ref{fig:imqnm} where the absolute value of the imaginary part increases again once the magnetic field is large enough.  We notice that the bigger the value of $\kappa$, the faster the QNMs approach the real axis. The results for higher QNMs are qualitatively similar to the results for the lowest QNM. We conclude that in the case of vanishing $\kappa$ there are no resonances in the current, independent of the value of $B$ since the QNMs do not approach the real axis. In the region $\kappa\gtrsim0.5$, we observe the QNMs approaching monotonically the real axis up to very small values (Im$[\omega]/(\pi T)\sim-10^{-3}$ for $\kappa=0.5$). QNMs with vanishing real part are not damped and therefore we should observe resonances in the current. For values of $\kappa\gtrsim 0.5$  the QNMs do not go down again in the regime of $\tilde B$ allowed by our numerics. Furthermore we did not observe a mode crossing the real axis to the upper half plane, which would have been an indicator for instabilities.
%%%%%%%%%%%%%%%%%%%%%%%%%%%%%%%%%%%%%%%%%%%%%%
%%%%%%%%%%%%%%%%%%%%%%%%%%%%%%%%%%%%%%%%%%%%%%
\begin{figure}[H] 
	\centering
\includegraphics[width=6.7cm]{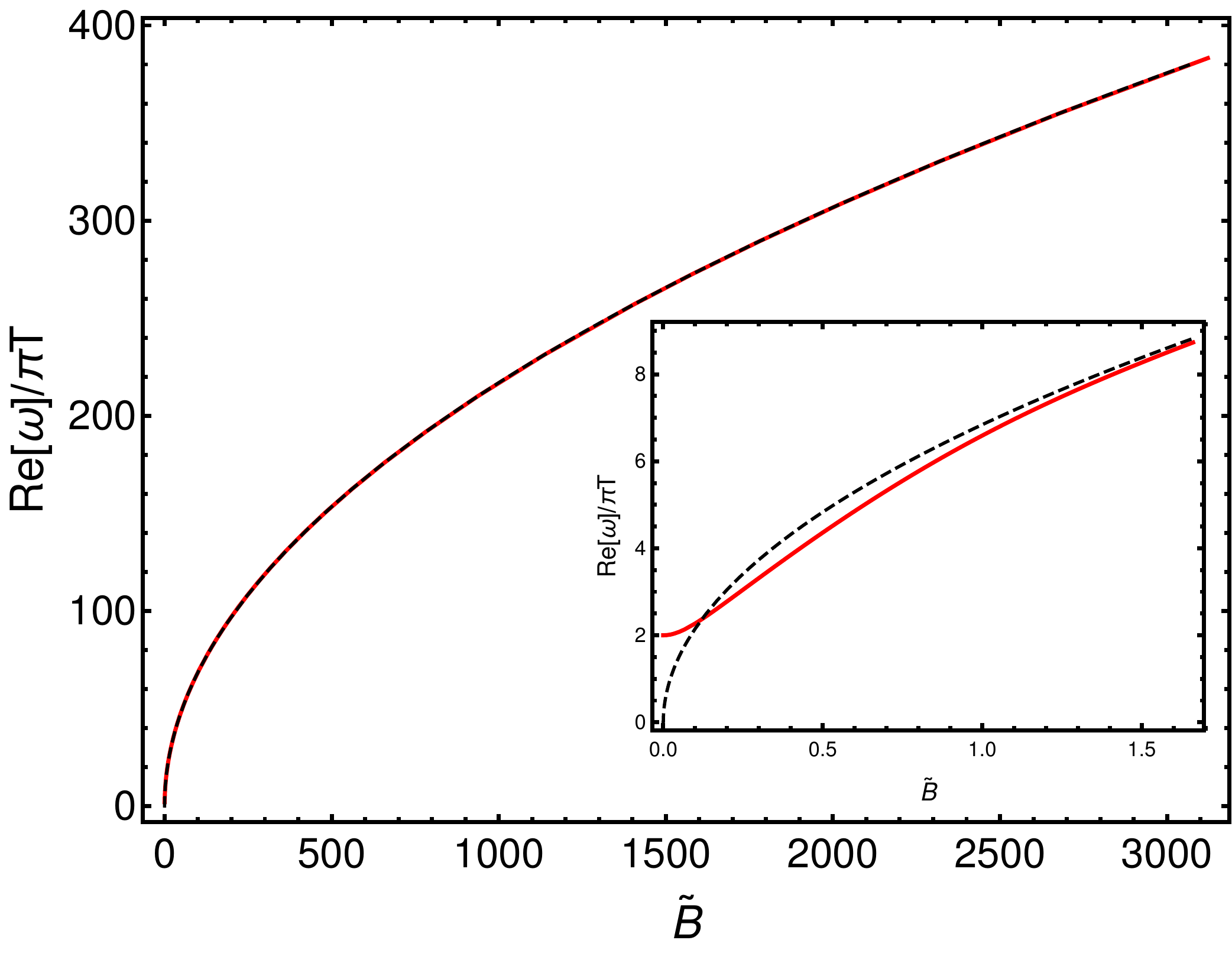}
	\hspace{1cm}
	\includegraphics[width=6.7cm]{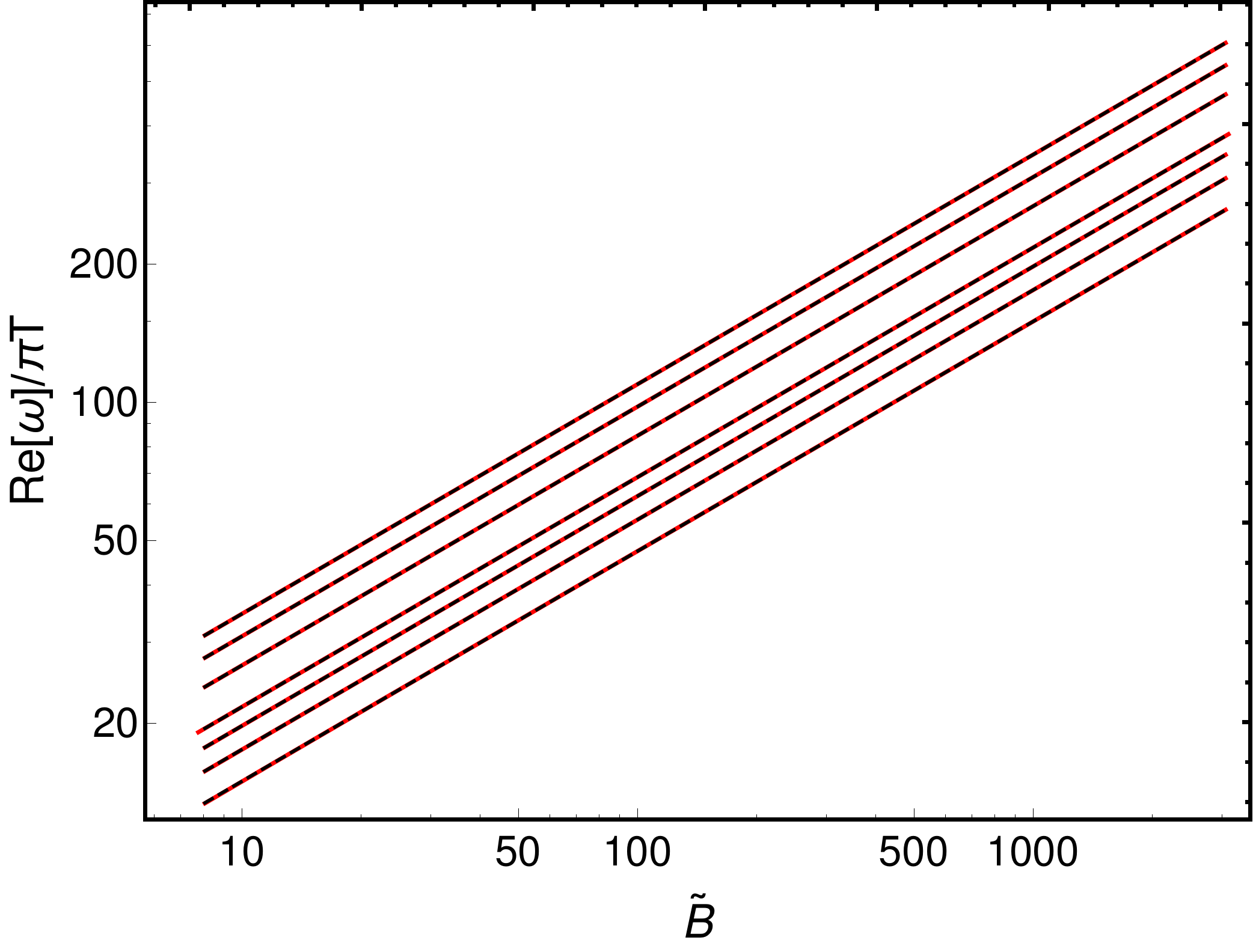}
	\caption{\label{fig:reqnm}Left: Real part of the lowest QNM against magnetic field for $\kappa=1$ (red). For $\tilde{B}> 2 $ data fits $\sqrt{\tilde{B}}$ (dashed, black). Right: Double logarithmic plot of $\text{Re}[\omega]$ of the lowest QNM against $\tilde{B}$ for several values of $\kappa\in\{0.5, 0.67, 0.83, 1, 1.5, 2, 2.5, 3.5 \}$  in the large magnetic field regime. Data (red) fits to $\sqrt{\tilde{B}}$ (dashed, black) in all cases.}
\end{figure}
%%%%%%%%%%%%%%%%%%%%%%%%%%%%%%%%%%%%%%%%%%%%%%
%
In order to characterise the behaviour of the QNMs at strong magnetic fields, we found it useful to look closer at the real part of the lowest QNM in dependence of the magnetic field. Therefore, we fitted the real part of the QNM for $\kappa\in[0.1,3]$ to a function of the form
\begin{equation}
\text{Re}[\omega]/(\pi T)=a (\tilde B-B_0)^c.\label{eq:fit1}
\end{equation}
\newpage\noindent We find that in all cases the exponent in eq. \eqref{eq:fit1} is given by $c\approx 0.5$. The biggest deviation is $c=0.499992$ (with an fitting error of $10^{-6}$) for the lowest value of $\kappa$. The bigger the value of $\kappa$, the better the data fits to a square root. In the l.h.s of figure \ref{fig:reqnm}, we show the real part of the lowest QNM in the case $\kappa=1$. We observe two regimes, namely the low $\tilde B$ regime where the behaviour deviates from a square root and the regime where it fits perfectly to a square root. In the r.h.s., we show the square root behaviour of the real part for several values of $\kappa$. The real part of QNMs is proportional to the energy; furthermore, the split-up of the energy levels of a relativistic particle in a magnetic field, the so called Landau levels, is given by $E_\text{n}\sim \text{sign(n)}\,\sqrt{\tilde B\,|n|}$. Hence, the square root behaviour indicates that the resonances in the current are the consequence of the presence of Landau levels in our set-up. \newline\newline It is obvious in figure \ref{fig:imqnm} that the real part in the case $\kappa=0$ has no square root behaviour for any value of $\tilde B$. Furthermore, we did not observe any resonances in the current in the case $\kappa=0$.
 In addition, taking the behaviour of the imaginary part into account, we can conclude that the resonances are indeed a footprint of the anomaly.
Since we considered the behaviour of the QNMs only up to a certain value of the magnetic field, it could be possible that the modes tend again away from the axis. In this case, the resonances would be restricted to a certain magnetic field regime.\newline\newline
We investigated the dependence of the real part of the QNMs on the magnetic field and found a square root behaviour. But is there also an easy connection to the value of the anomaly parameter $\kappa$?
\begin{figure}[H] 
	\centering
	\includegraphics[width=6.7cm]{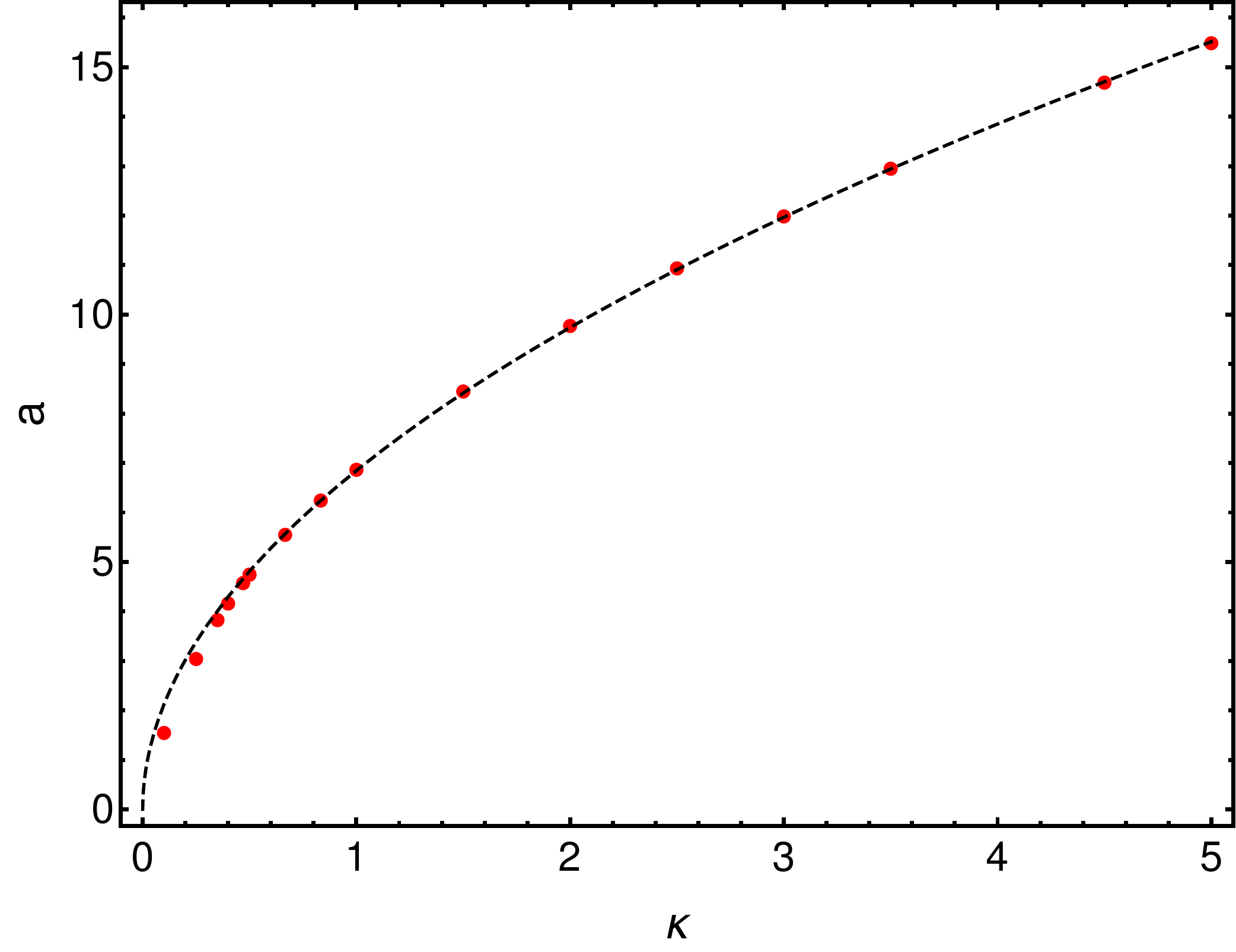}
	\caption{\label{fig:fitkB}Fit of the data, obtained by fitting \eqref{eq:fit1} to the real part for several $\kappa$.}
\end{figure}
In order to answer this question, we determined the slope of the square root $a$, defined in eq. \eqref{eq:fit1}, for different values of $\kappa$. The result is depicted in figure \ref{fig:fitkB}. For $\kappa>0.5$ we fitted the data to $a(\kappa)=s\,\kappa^x$ with the result is $a\sim \kappa^{0.510}$, where the fitting error of the exponent is of magnitude $10^{-3}$. We therefore conclude that the behaviour of the real part of the QNMs for strong enough magnetic fields follows Re$[\omega]/(\pi T)\sim\sqrt{\kappa \tilde B}$.\newline\newline This is an interesting result which can already be obtained in the probe limit. In figure \ref{fig:reqnm} we investigated the dependence of the lowest QNM on $\kappa \tilde B$. As a first observation, we notice that the absolute value of the imaginary part decreases when increasing $\kappa \tilde B$ and the QNM is approaching the real axis. Therefore, we should be able to see resonances for $\kappa \tilde B\gtrsim1$, as indicated in the inset in the r.h.s. of figure \ref{fig:probe}. This is precisely what we found in figure \ref{fig:resonance2}, where we observed resonances for $\kappa\tilde B\gtrsim1$. The real part indeed behaves, as shown in the r.h.s. of figure \ref{fig:probe}, as Re$[\omega]/(\pi T)\sim \sqrt{\kappa\tilde B}$.

%%%%%%%%%%%%%%%%%%%%%%%%%%%%%%%%%%%%%%%%%%%%%%

\newpage
\section{Conclusions}\label{sec:conclu}
In this thesis, we examined non-equilibrium dynamics in strongly coupled field theories by means of holography.
Concretely, we considered an $U(1)\times U(1)$-model, consisting of two gauge fields in presence of an anomaly, induced by a Chern-Simons term. The two external gauge fields provide parallel electromagnetic fields, namely a static magnetic field and a time-dependent electric field. In presence of a chirality imbalance, the system generates a current, parallel to the electromagnetic fields, which is known as the CME current. Throughout the thesis, we computed the current in order to characterise the response of the system to quenches of the electric field.\newline\newline As a first step, we considered the system in the so called probe approximation, neglecting the backreaction of the fields on the metric. This limits us to small values of the energy-momentum tensor compared to the temperature of the black hole, implying that we are restricted to sufficiently ``small'' values of the magnetic field and sufficiently ``slow'' quenches. Nevertheless, we made progress in including backreaction to study the full non-linear response. Within the probe approximation, we were able to reduce the system to a single linear hyperbolic PDE. We integrated this PDE, by means of a fully spectral code~\cite{Macedo:2014bfa}, in time direction. \newline\newline
Motivated by the work of \cite{Buchel:2013lla}, we focused on the early time response of the system in the first part of the thesis. Concretely, we examined the dependence of the CME current on different abruptnesses of the quench and how the anomaly affects the results. We found that our system shows an universal scaling behaviour in the early-time response. 
Throughout the investigation, we noticed that in case of bigger values of $\kappa \tilde B$ faster quenches are needed in order to get to the universal regime. Furthermore, we did a similar study, as the authors of \cite{Buchel:2013lla}. Therefore, we defined an ``excitation'' time, as the first time, where the response deviates 1\% from the adiabatic response. Similar to the authors of \cite{Buchel:2013lla} we found an universal scaling behaviour. In addition, the response is independent of $\kappa \tilde B$ for sufficient fast quenches. Since we excite in some cases long living oscillations in the current, a similar discussion for a relaxation time, defined as the last time, where the response deviates 1\% from the adiabatic response, is not useful. Nevertheless, our results for the early response indicate, that the relaxation time gets smaller, when increasing $\kappa \tilde B$. This coincides with $\kappa \tilde B$ being the coupling of the current to the electric field.
\newline\newline
We noticed that the quench excites, for sufficiently large $\kappa \tilde B$, long-living oscillations in the current \footnote{Note that the resonances were also observed in recent lattice calculations \cite{Buividovich:2016ulp}.}. We traced the frequencies of the oscillations back to the real part of the corresponding QNMs and the damping to the imaginary part. Therefore, we conclude that an almost undamped oscillation corresponds to a QNM with an almost vanishing imaginary part. Furthermore, we were able to excite the resonances directly by a source, oscillating with the real part of the corresponding QNM.\newline\newline To point out whether these ``resonances'' are an effect of the magnetic field or the anomaly, we computed the fate of the QNMs for strong magnetic fields, with and without anomaly. The magnetic field $B$ and the anomaly parameter $\kappa$ always appear as a product in the probe limit. Furthermore, the probe approximation is in the case of strong magnetic fields not valid.\newline\newline
In the last part of the thesis, we included backreaction into the system and computed the corresponding QNMs. The real part of the QNMs matches for $\kappa>0$ perfectly to Re$[\omega]/(\pi T)\sim\sqrt{\tilde B}$. Therefore, we traced back the resonances as presence from Landau levels in our set-up. The absence of the resonances in the case $\kappa=0$ implies, that there are no dual fermionic degrees of freedom charged under the symmetries we took into account.
\newline\newline
The next step is to characterise the behaviour of the imaginary part of the QNMs. We observe that for $\kappa\gtrsim 0.5$ the QNMs approach the imaginary axis up to very small values. The stronger the Chern-Simons coupling, the faster and closer the QNM approach the real axis. In conclusion, for $\kappa=0.5$ the lowest QNM, which determines the oscillation behaviour of the current, has for all values of $B$ a non vanishing imaginary part and therefore we are not able to observe resonances in the current. One could speculate whether the value $\kappa=0.5$ is special in regards to the quantum critical point, present at $\kappa=1/2$ \cite{D'Hoker:2010rz, D'Hoker:2010ij, Ammon:2016szz}. Furthermore, we examined the dependence of the real part on the anomaly parameter $\kappa$. Concretely, we found that Re$[\omega]/(\pi T)\sim\sqrt{\kappa\tilde B}$. This can be  already seen from the probe limit since $\kappa \tilde B$ appears again as a product.\newline\newline
Although our discussion is motivated by Dirac/Weyl semimetals, we chose for our work a simpler model. Nevertheless, in our opinion the simpler model captures the core properties, we wanted to explore. However, it would be interesting to perform an analogous study in the holographic model, established in \cite{Landsteiner:2015lsa}.\newline\newline
Furthermore, it would be interesting to include backreaction in the discussion of the time-dependent phenomena. In the probe limit, we observed resonances in the current for $\kappa \tilde B\gtrsim1$. The QNM analysis, including backreaction, reveals that resonances should be visible for $\kappa\gtrsim0.5$. Nevertheless, for the $\kappa=1$ curve, the mode becomes at $\tilde B\approx1$ massless which matches the results obtained in the probe limit. Furthermore, our system is linear, due to the probe approximation. Therefore, it would be interesting to know, whether the full non-linear behaviour shows an universal behaviour too. We found that the response is independent of $\kappa \tilde B$ for fast enough quenches. In the full system it is possible to distinguish between magnetic and anomaly effects, and to examine the response for fixed magnetic field and different values of the Chern-Simons coupling.\newline\newline
But not only the generalisation of out set-up including backreaction is interesting. As a first step in order to study an (possible) universal behaviour in presence of an anomaly, it would be interesting to include a Chern-Simons term to the system, discussed in \cite{Fuini:2015hba}. In this paper the authors consider a Gaussian spatial anisotropy, propagating in radial direction. For this Gaussian one could perform a similar study regarding the initial response, as we did. An further possible extension would be an additional boundary source, for example a time dependent electric field.\newline\newline
In our discussion of the QNMs with backreaction, we traced back the resonances to the presence of Landau levels. But are the Landau levels populated? Are the fundamental excitations populating the Landau levels, of fermionic or bosonic nature? Is it possible with this knowledge to construct microscopic Fermi surfaces in context of AdS/CFT? To answer questions in this direction, we have to switch on a finite density.
Since it is not clear, whether Dirac/Weyl semimetals are strongly coupled, it would be interesting to look at our studies from a weak coupling perspective. Nevertheless, it is worth mentioning that Landau level resonances were found experimentally in Dirac semimetals \cite{2016arXiv160102316Y, PhysRevLett.115.176404}. 
\newpage
\begin{appendix}
	\section{Equations of motions of the gauge fields}\label{app::gaugeeq}
The Lagrangian \eqref{eq::lagr} of our model reads
\begin{align}
\mathcal L&=-\frac 14 F^{\mu\nu}F_{\mu\nu}-\frac 14 H^{\mu\nu}H_{\mu\nu}+\frac{\kappa}{2}\varepsilon^{\mu\alpha\beta\rho\lambda}\,A_\mu\,(F_{\alpha\beta}F_{\rho\lambda}+3H_{\alpha\beta}H_{\rho\lambda})\\
&=(\partial_\alpha A_\beta\,\partial_\rho A_\lambda+\partial_\beta A_\alpha\,\partial_\lambda A_\rho-\partial_\beta A_\alpha\,\partial_\rho A_\lambda-\partial_\alpha A_\beta\,\partial_\rho A_\lambda)\left(\frac{\kappa}{2}\varepsilon^{\mu\alpha\beta\rho\lambda}A_\mu-\frac 14 g^{\alpha\rho}g^{\beta\lambda}\right)\nonumber\\&
+\! (\partial_\alpha V_\beta\,\partial_\rho V_\lambda\!+\!\partial_\beta V_\alpha\,\partial_\lambda V_\rho\!-\!\partial_\beta V_\alpha\,\partial_\rho V_\lambda\!-\!\partial_\alpha V_\beta\,\partial_\rho V_\lambda)\left(\frac{3\kappa}{2}\varepsilon^{\mu\alpha\beta\rho\lambda}A_\mu-\frac{g^{\alpha\rho}g^{\beta\lambda}}{4} \right).
\end{align}
Therefore, we obtain with $F_{\mu\nu}=\partial_\mu A_\nu-\partial_\nu A_\mu=\nabla_\mu A_\nu-\nabla_\mu A_\mu$
\begin{align}
\nabla_\eta\frac{\partial\mathcal L}{\partial(\nabla_\eta A_\gamma)}&=-\nabla_\eta F^{\eta\gamma}-\varepsilon^{\gamma\sigma\eta\rho\lambda}F_{\sigma\eta}F_{\rho\lambda}\label{eq:a1}\\
\frac{\partial\mathcal L}{\partial A_\gamma}& = \frac \kappa 2 \varepsilon^{\gamma\alpha\beta\rho\lambda}(F_{\alpha\beta}F_{\rho\lambda}+3H_{\alpha\beta}H_{\rho\lambda})\label{eq:a2}\\
\nabla_\eta\frac{\partial\mathcal L}{\partial(\nabla_\eta V_\gamma)}&=-\nabla_\eta H^{\eta\gamma}-3\varepsilon^{\gamma\sigma\eta\rho\lambda}F_{\sigma\eta}H_{\rho\lambda}\label{eq:a3}\\
\frac{\partial\mathcal L}{\partial V_\gamma}& =0,\label{eq:a4}
\end{align}
leading by setting \eqref{eq:a1}=\eqref{eq:a2} and \eqref{eq:a3}=\eqref{eq:a4} directly to the equation of motion.
\section{Holographic renormalisation}\label{app::ren1}
In order to compute the one-point-functions, we have to renormalise the model considered in eq. \eqref{eq::lagr}. Therefore, we follow the procedure explained in section \ref{sec:holo}. In our case the non-normalisable modes of the induced metric and the gauge fields behave asymptotically as
\begin{equation}
\gamma_{ij}\sim \e^{2r}g_{(0)ij}(x), \ \ A_i\sim A_{(0)i}(x), \ \ V_i\sim V_{(0)i}(x),
\end{equation}
and their radial derivatives as
\begin{equation}
\dot{\gamma}_{ij}\sim 2\gamma_{ij}, \ \ E_i=\dot A_i
=\mathcal O(\e^{-r}), \ \  \Sigma_i=\dot V_i
=\mathcal O(\e^{-r}),\end{equation}
implying for the radial derivative $\partial_r$ and the dilatation operator $\delta_\text{D}$ defined in eq. \eqref{eq:rad}
\begin{equation}
\partial_r=\int\dd^4x\ 2\gamma_{ij}\frac{\delta}{\delta\gamma_{ij}}+\mathcal O(\e^{-r})\equiv\delta_\text{D}+\mathcal O(\e^{-r}).
\end{equation}
At first, we derive the equation of motion for the gauge fields in Gaussian coordinates
\begin{equation}
\dd s^2=\dd r^2+\e^{2r}\dd \bm{x}^2, \ \sqrt{-g}=\e^{4r}.
\end{equation}
Rewriting the first term
\begin{align}
\nabla_\mu F^{\mu j}&=\partial_\mu F^{\mu j}+\Gamma^\mu_{\mu\lambda}F^{\lambda j}+\Gamma_{\mu\nu}^jF^{\mu\nu}=\partial_r(g^{rr}g^{ij}F_{ri})+g^{jk}\partial^iF_{ik}+4g^{rr}g^{jk}F_{rk}\\
&=g^{ij}\,\left(\partial_r F_{ri}-2F_{ri}+4F_{ri}+\partial^kF_{ki}\right)=g^{ij}\left(\dot E_i+2E_i+\partial^k F_{ki}\right),
\end{align}
where we set $A_r=0$, since we are working in the axial gauge. The same holds for $\nabla_\mu H^{\mu j}$. For the Chern-Simons part we get
\begin{equation}
\varepsilon^{j\alpha\beta\rho\lambda}F_{\alpha\beta}F_{\rho\lambda}=4\varepsilon^{jrikl}\,F_{ri}F_{kl}=4\varepsilon^{jrikl}\,E_{i}F_{kl}
\end{equation}
and analogous for the Chern-Simons part in the e.o.m for $H$
\begin{equation}
\varepsilon^{j\alpha\beta\rho\lambda}F_{\alpha\beta}H_{\rho\lambda}=2\varepsilon^{jrikl}\,(F_{ri}H_{kl}+F_{kl}H_{ri})=2\varepsilon^{jrikl}\,(E_{i}H_{kl}+\Sigma_{i}F_{kl}).
\end{equation}
Summarising the e.o.m. read
\begin{align}
&\dot E_j+2E_j+\partial^kF_{kj}+\frac{6\kappa\,g_{jm}}{\sqrt{-g}}\tilde\varepsilon^{mrikl}\,(E_iF_{kl}+\Sigma_iH_{kl})=0,\label{eq:reneq1}\\
& \dot{\Sigma}_j+2\Sigma_j+\partial^kH_{kj}+\frac{6\kappa\,g_{jm}}{\sqrt{-g}}\tilde\varepsilon^{mrikl}\,(E_iH_{kl}+\Sigma_iF_{kl})=0.
\end{align}
All terms without radial derivative have a well defined unique weight, whereas the radial derivative has to be expanded. A not well defined weight will be denoted by $\alpha$.
Solving eq. \eqref{eq:reneq1} to first order results in (since $E_{i(0)}$ and $\Sigma_{i(0)}$ are zero)
\begin{equation}
\dot E_i\Big|_{(1)}+2E_{i(1)}=\delta_\text{D}E_{i(1)}+2E_{i(1)}=E_{i(1)}\overset{!}{=}0,
\end{equation}
analogous we get $\Sigma_{i(1)}=0$.
The equation for the on-shell action reads
\begin{align}
\dot{\lambda}+4\lambda=\mathcal L_\text{OS}
&=-\frac 12 \gamma^{ij}\,E_iE_j\Big|_{(2+2\alpha)}-\frac 14 \gamma^{ik}\gamma^{jl}\,F_{ij}F_{kl}\Big|_{(4)}-\frac 12 \gamma^{ij}\,\Sigma_i\Sigma_j\Big|_{(2+2\alpha)}\nonumber\\&\ \ -\frac 14 \gamma^{ik}\gamma^{jl}\,H_{ij}H_{kl}\Big|_{(4)}-\frac{2\kappa\,\tilde\varepsilon^{lijk}}{\sqrt{-g}}\,A_l\,(E_i\,F_{jk}+3\Sigma_i\,H_{jk})\Big|_{(4+\alpha)},
\end{align}
where we implement the on-shellness of the Lagrangian using the asymptotic expansions. Since $E_{i(0)}=E_{i(1)}=\Sigma_{i(0)}=\Sigma_{i(1)}=0$ we conclude  $\alpha\ge2$. To first non-trivial order we have
\begin{equation}
\dot{\lambda}\Big|_{(2)}+4\lambda_{(2)}=\delta_\text{D}\lambda\Big|_{(2)}+4\lambda_{(2)}=2\lambda_{(2)}\overset{!}{=}0.
\end{equation}
The next non-vanishing order is given by
\begin{align}
\dot{\lambda}\Big|_{(4)}+4\lambda_{(2)}&=\delta_\text{D}\lambda\Big|_{(4)}+4\lambda_{(4)}+\delta_{(2)}\lambda_{(2)}-=2\tilde\lambda_{(4)}\overset{!}{=}-\frac 14\,F^{ij}F_{ij}+\frac 14 H^{ij}H_{ij}\\
\Leftrightarrow \tilde \lambda_{(4)}&=\frac 18\,F^{ij}F_{ij}+\frac 18 H^{ij}H_{ij}\ \Rightarrow\ \lambda_\text{div}=\left(\frac 18\,F^{ij}F_{ij}+\frac 18 H^{ij}H_{ij}\right)\,\log(r^{-2}).
\end{align}
Therefore, the counter-term in the $u$-patch reads
\begin{equation}
S_\text{ct}=-\frac14\int\dd^4x\,\sqrt{\gamma}\,\left(F^{ij}F_{ij}+ H^{ij}H_{ij}\right)\,\log(u).
\end{equation}
	\section{One-point-functions}\label{app::ren}
We consider the Lagrangian (where latin letters run in this section from 0-4)
\begin{equation}
\mathcal L=\sqrt{-g}\left(-\frac 14\,F^{mn}F_{mn}-\frac 14\, H_{mn}H^{mn}+\frac \kappa 2 \,\varepsilon^{abcde}\,A_a\,\left(F_{bc}F_{de}+3H_{bc}H_{de}\right)\right).
\end{equation}
We now expand the terms in the Lagrangian to first order. We obtain for the first term
\begin{align}
F^{mn}F_{mn}&=\sqrt{-g}\,g^{ma}g^{nb}\,F_{ab}F_{mn}\nonumber\\
&=\sqrt{-g}\,g^{ma}g^{nb}\,\big(\partial_a(A_b+a_b)\,\partial_m(A_n+a_n)-\partial_a(A_b+a_b)\,\partial_n(A_m+a_m)\nonumber\\&\ \ -\partial_b(A_a+a_a)\,\partial_m(A_n+a_n) +\partial_b(A_a+a_a)\,\partial_m(A_n+a_n)\big)\nonumber\\
&=\sqrt{-g}\,\left(\tilde F_{mn}\tilde F^{mn}+g^{ma}g^{nb}\,(\partial_a a_b\,\partial_m A_n+\partial_a A_b\,\partial_m a_n+\ldots)\right)\nonumber\\&
=\sqrt{-g}\,\left(\tilde F_{mn}\tilde F^{mn}+2 \,f^{mn}\tilde F_{mn}+\mathcal O(a^2)\right)\nonumber\\
&=\sqrt{-g}\,\left(\tilde F_{mn}\tilde F^{mn}+4\,\partial_m a_n\,\tilde F^{mn}+\mathcal O(a^2)\right)\nonumber\\
&=\sqrt{-g}\,\left(\tilde F_{mn}\tilde F^{mn}+4\,\nabla_m \left(a_n\,\tilde F^{mn}\right)-4\,a_n\,\nabla_m\tilde F^{mn}+\mathcal O(a^2)\right).
\end{align}
The first part of the third term leads to (using $\tilde\varepsilon=\varepsilon(\sqrt{-g})$)
\begin{align}
\frac \kappa 2 \,\tilde\varepsilon^{abcde}\tilde A_aF_{bc}F_{de}&=\frac \kappa 2 \,\tilde\varepsilon^{abcde}\tilde A_a\,(\partial_b \tilde A_c\,\partial_d\tilde A_e-\partial_b\tilde A_c\,\partial_e\tilde A_d-\partial_c\tilde A_b\,\partial_d\tilde A_e+\partial_c\tilde A_b\,\partial_e\tilde A_d)\nonumber\\&=2\kappa \,\tilde\varepsilon^{abcde}\tilde A_a\,\partial_b\tilde A_c\,\partial_d\tilde A_e\nonumber\\
&=2\kappa \,\tilde\varepsilon^{abcde}\,\left(A_a\,(\partial_bA_c \,\partial_da_e+\partial_ba_c\,\partial_dA_e)+a_a\,\partial_bA_c\,\partial_dA_e\right)+\mathcal O(a^2)\nonumber\\
&=2\kappa \,\tilde\varepsilon^{abcde}\,\left(2 A_a\,\partial_bA_c \,\partial_da_e+a_a\,\partial_bA_c\,\partial_dA_e\right)+\mathcal O(a^2)\nonumber\\
&=2\kappa \,\tilde\varepsilon^{abcde}\,\left(2\, \nabla_d(A_a\,\partial_bA_c \,a_e)-2\,a_e\,\nabla_d( A_a\,\partial_bA_c)+a_a\,\partial_bA_c\,\partial_dA_e\right)+\mathcal O(a^2)\nonumber\\
&=2\kappa \,\tilde\varepsilon^{abcde}\,\left(2\, \nabla_d(A_a\,\partial_bA_c \,a_e)-2\,a_e\,\partial_d A_a\,\partial_bA_c+a_a\,\partial_bA_c\,\partial_dA_e\right)+\mathcal O(a^2)\nonumber\\
&=6\kappa \,\tilde\varepsilon^{abcde}a_a\,\partial_bA_c\,\partial_dA_e+4\kappa \,\tilde\varepsilon^{abcde} \nabla_d(A_a\,\partial_bA_c \,a_e)+\mathcal O(a^2)\nonumber\\
&=\frac 32\kappa \,\tilde\varepsilon^{abcde}a_a\,\tilde F_{bc}\tilde F_{de}+2\kappa \,\tilde\varepsilon^{abcde} \nabla_b(A_a\,\tilde F_{cd} \,a_e)+\mathcal O(a^2).
\end{align}
The second part gives rise to
\begin{align}
\frac{3\kappa}{2}\tilde\varepsilon^{abcde}\tilde A_aH_{bc}H_{de}&=\frac{3\kappa}{2}\tilde\varepsilon^{abcde}\tilde A_a\,(\partial_b \tilde V_c\,\partial_d\tilde V_e-\partial_b\tilde V_c\,\partial_e\tilde V_d-\partial_c\tilde V_b\,\partial_d\tilde V_e+\partial_c\tilde V_b\,\partial_e\tilde V_d)\nonumber\\&=6\kappa \,\tilde\varepsilon^{abcde}\tilde A_a\,\partial_b\tilde V_c\,\partial_d\tilde V_e\nonumber\\
&=6\kappa \,\tilde\varepsilon^{abcde}\,\left(a_a\,\partial_b\tilde V_c\,\partial_d\tilde V_e+2 A_a\,\partial_bv_c\,\partial_d\tilde V_e\right)\nonumber\\
&=\frac{3\kappa}{2} \,\tilde\varepsilon^{abcde}\,a_a\,\tilde H_{bc}\tilde H_{de}+6\kappa \,\tilde\varepsilon^{abcde}\,A_a\partial_bv_c\,\tilde H_{de}\nonumber\\
&=\frac{3\kappa}{2} \,\tilde\varepsilon^{abcde}\,a_a\,\tilde H_{bc}\tilde H_{de}+6\kappa \,\tilde\varepsilon^{abcde}\,\left(\partial_b\,(A_av_c\,\tilde H_{de})-v_c\,\partial_b(\tilde A_a\tilde H_{de})\right)\nonumber\\
&=\frac{3\kappa}{2}\, \tilde\varepsilon^{abcde}(a_a\,\tilde H_{bc}\tilde H_{de}+v_c\,\tilde F_{ab}\tilde H_{de})\nonumber\\&\ \ +6\kappa \,\tilde\varepsilon^{abcde}\left(\partial_b\,(A_av_c\,\tilde H_{de})-v_c\,\tilde A_a\partial_b\tilde H_{de}\right)\nonumber\\
&=\frac{3\kappa}{2} \,\tilde\varepsilon^{abcde}\,(a_a\,\tilde H_{bc}\tilde H_{de}+v_a\,\tilde F_{bc}\tilde H_{de})+6\kappa \,\tilde\varepsilon^{abcde}\,\partial_b\,(A_av_c\,\tilde H_{de}).
\end{align}
Therefore, we can write the Lagrangian, expanded to first order as
\begin{align}
\mathcal L^{(1)}&=\sqrt{-g}\,\Big(-\frac 14\,\tilde F_{mn}\tilde F^{mn}-\frac 14\,\tilde H_{mn}\tilde H^{mn}-\,\nabla_m \left(a_n\,\tilde F^{mn}+h_n\,\tilde H^{mn}\right)+\,a_a\,\nabla_b\tilde F^{ba}
\nonumber\\&+\frac 32\,\kappa \,\varepsilon^{abcde}a_a\,(\tilde F_{bc}\tilde F_{de}+\tilde H_{bc}\tilde H_{de}) +\,v_a\,\nabla_b\tilde H^{ba}+\frac{3\kappa}{2} \,\varepsilon^{abcde}\,v_a\,\tilde F_{bc}\tilde H_{de}\nonumber\\&+6\kappa \,\varepsilon^{abcde}\,\nabla_b\,(A_av_c\,\tilde H_{de})+2\kappa \,\varepsilon^{abcde} \nabla_b(A_a\,F_{dc} \,a_e)\Big).
\end{align}
We now put the Lagrangian on-shell and obtain, using the EOMs and the fact that the Christoffel symbols are symmetric while the Levi Civita tensor is antisymmetric
\begin{align}
\mathcal L^{(1)}&=\sqrt{-g}\,\left(-\,\nabla_m \left(a_n\,\tilde F^{mn}+v_n\,\tilde H^{mn}\right)+2\kappa \,\varepsilon^{abcde}\,\nabla_b\,\left(A_a\tilde F_{cd} \,a_e+3\,A_av_c\,\tilde H_{de}\right)\right)\nonumber\\&=
-\nabla_m \left[\sqrt{-g}\,\left(a_n\,\tilde F^{mn}+v_n\,\tilde H^{mn}\right)\right]+2\kappa \,\tilde\varepsilon^{abcde}\,\nabla_b\,\left(A_a\tilde F_{cd} \,a_e+3\,A_av_c\,\tilde H_{de}\right).
\end{align}
Integrating the full divergence gives rise to boundary terms
\begin{align}
S^{(1)}&=-\int\dd^dx\,\sqrt{-\gamma}\, n_m\left(a_n\,\tilde F^{mn}+v_n\,\tilde H^{mn}\right)\nonumber\\&\ +2\kappa \,\tilde\varepsilon^{abcde}\int\dd^dx\,n_b\,\frac{\sqrt{-\gamma}}{\sqrt{-g}}\left(A_a\tilde F_{cd} \,a_e\!+\!3A_av_c\,\tilde H_{de}\right),
\end{align}
where $n$ is the outward pointing unit vector which reads
\begin{equation}
n_m \dd x^m= n_u \dd u.
\end{equation}
Due to the normalisation $n^mn_m=1$, we obtain
\begin{equation}
g^{mn}n_m n_n=g^{uu}n_u n_u=f(u)\,u^2 (n_u)^2=1 \ \Rightarrow \ n_u=-\frac{1}{u\,\sqrt{f(u)}}.
\end{equation}
The determinant of the induced metric on the $u=\varepsilon$ plane is given by $\sqrt{-\gamma}=\frac{\sqrt{f(\varepsilon)}}{\varepsilon^4}$ and $\sqrt{-g}=\frac{1}{\varepsilon^5}$ and therefore $\frac{\sqrt{-\gamma}}{\sqrt{-g}} n_u=-1$. The action now reads
\begin{align}
S^{(1)}&=\frac{1}{\varepsilon^5}\int\dd^dx\,g^{u a}g^{n b}\left(a_n\,\tilde F_{a b}+v_n\,\tilde H_{ab}\right)-2\kappa \,\tilde\varepsilon^{au nde}\int\dd^dx\,\left(A_a\tilde F_{nd} \,a_e+3A_av_n\,\tilde H_{de}\right)
\end{align}
With this at hand and the ansatz $V=Bx \,\dd y+V_z(v,u)\,\dd z$, we are able to compute one point functions 
\begin{align}
\frac{\delta S^{(1)}}{\delta v_z}&=\frac{1}{\varepsilon^5}\,\left(g^{u u}g^{zz}\,\left(\partial_u V_z-\partial_z V_u\right)+g^{u v}g^{zz}\,\left(\partial_v V_z-\partial_z V_v\right)\right)-12\kappa\,\tilde{\varepsilon}^{au z de}A_a\,\partial_d V_e\nonumber\\
&=\frac{1}{\varepsilon^5}\,\left(g^{u u}g^{zz}\,\partial_u V_z+g^{u v}g^{zz}\,\partial_v V_z\right)-12\kappa\,\tilde{\varepsilon}^{au z de}A_a\,\partial_d V_e\nonumber\\
&=\frac{1}{\varepsilon^5}\,(\varepsilon^4\, f(\varepsilon) \,\partial_u V_z-\varepsilon^4\,\partial_v V_z)-12\kappa\,\tilde{\varepsilon}^{vu z xy}A_v\,\partial_x V_y\nonumber\\
&=\frac{1}{\varepsilon}\,(\dot V_{(0)}+2\varepsilon V_{(2)}+\varepsilon \log(\varepsilon)\,\ddot V_{(0)}+ \frac12\,\varepsilon\,\ddot V_{(0)}-\dot V_{(0)}-\varepsilon \ddot V_{(0)})-12B\kappa\, A_{v,(0)}\nonumber\\
&=2 V_{(2)}+ \log(\varepsilon)\,\ddot V_{(0)}-\frac12 \ddot V_{(0)}-12B\kappa\, A_{v,(0)}.\label{eq:logcontr}
\end{align}
After adding the counter-terms, the one-point function reads
\begin{equation}
\langle J^z\rangle=\frac{\delta S^{(1)}_\text{ren}}{\delta v_z}=2 V_{(2)}-12B\kappa\, A_{v,(0)}.
\end{equation}
We now derive the Ward identities. The vector Ward identity reads
\begin{align}
\partial_i\frac{\delta S^{(1)}}{\delta v_i}&= \partial_i\left(\sqrt{-g}\,\tilde H^{u i}\right)-3\kappa \,\tilde\varepsilon^{au ide}\tilde F_{ia}\,\tilde H_{de}\nonumber\\
&=-\sqrt{-g}\left(\frac{1}{\sqrt{-g}}\, \partial_i\left(\sqrt{-g}\,\tilde H^{iu}\right)+3\kappa \,\varepsilon^{u a ide}\tilde F_{ai}\,\tilde H_{de}\right)\nonumber\\
&=0,
\end{align}
since the equation corresponds to the $u$ component of the EOMs. Furthermore we obtain for the axial Ward identity 
\begin{align}
\partial_i\frac{\delta S^{(1)}}{\delta a_i}&=\partial_i\left(\sqrt{-g}\,\tilde F^{u i}\right)-\kappa \,\tilde\varepsilon^{au ndi}\tilde F_{ia}\tilde F_{nd} \nonumber\\
&=-\sqrt{-g}\left(\frac{1}{\sqrt{-g}}\,\partial_i\left(\sqrt{-g}\,\tilde F^{iu}\right)+\kappa \,\varepsilon^{u ia nd}\tilde F_{ia}\tilde F_{nd}\right) \nonumber\\
&=-\sqrt{-g}\left(\nabla_i\tilde F^{iu}+\kappa \,\varepsilon^{u ia nd}\tilde F_{ia}\tilde F_{nd}\right) \nonumber\\
&=-\frac 12\,\sqrt{-g}\left(-3\kappa \,\varepsilon^{u ia nd}(\tilde F_{ia}\tilde F_{nd}+\tilde H_{ia}\tilde H_{dn})+2\kappa \,\varepsilon^{u ia nd}\tilde F_{ia}\tilde F_{nd}\right) \nonumber\\
&=\frac 12\,\kappa \,\tilde\varepsilon^{u ia nd}(\tilde F_{ia}\tilde F_{nd}+3\tilde H_{ia}\tilde H_{dn}).
\end{align}
\section{Numerical methods}\label{num}
Within this thesis, we solve the arising equations by means of pseudo-spectral methods. Let $u(x)$ be the solution to the differential equation \cite{Boyd00,lrr-2009-1,canuto2006erratum}
\begin{equation}
\mathcal L_x u(x)=f(x),
\end{equation}
where $\mathcal L_x$ is a differential operator, $f(x)$ a known function and the coordinate $x$ is bounded by $x\in[x_a,x_b]$. Let ${\phi_j(x)}$ a set of globally defined, orthogonal basis functions. We can expand the unknown solution function $u$ in this basis $u(x)=\sum_{j=0}^{\infty}c_j\phi_j(x)$. Theoretically, we can plug this ansatz in the differential equation and solve for the coefficients. In praxis, we have to make an approximation in order to obtain a problem, which we can tackle numerically. The spectral approximation is to truncate the series at a certain finite value $N$. Furthermore, we discretise the coordinate $[x_a,x_b]$ interval in $N+1$ non-equidistant gridpoints. 
Since the derivatives of the basis functions are known, we can rearrange the sum so that a derivative of the function acts only on the expansion coefficients $u'(x)=\sum_{j=0}^{\infty}\tilde c_j\phi_j(x)$. 
Plugging the expansion in the differential equation and evaluate the corresponding equation at the gridpoints yields to an algebraic equation system, with unknown coefficients. In the context of pseudo-spectral methods, we do not solve for the coefficients, but convert the expansion into a weighted sum over grid points and solve directly for the field values. Throughout this thesis, we use Chebyshev polynomials as basis functions and a Gauss-Lobatto grid in the spatial direction. Since a fully spectral code is not the usual procedure to integrate in time, we give a brief recipe of the method.\newline\newline
\textbf{Fully Pseudo-spectral time evolution}\newline
To evolve a 1+1 dimensional system from an initial time $v_\text{inital}$ to a final time $v_\text{final}$ with a fully spectral code we have to introduce spectral coordinates in time and space direction. To do so we change coordinates to map the physical domain onto a unit square and expand the function in terms of a truncated series of Chebyshev polynomials. To solve our concrete problem we use the following recipe~\cite{Macedo:2014bfa,Hennig:2008af,Hennig:2012zx}.
\begin{enumerate}
	\item Solve the static equation to determine the initial data $V_z(v_0,u)=V_\text{z,in}(u)$
	\item Decompose the time interval in appropriate time domains\newline \mbox{$(v_\text{inital},v_\text{final}]=(v_\text{inital},v_1]\cup(v_1,v_2]\cup\ldots\cup(v_\text{k},v_\text{final}]$}
	\item Given initial data $V_\text{z,in}(u)$ at $v=v_\text{in}$, introduce an auxiliary field $W(v,u)$ 
	\begin{equation}
	V_z(v,u)=V_\text{z,in}(u)+(v-v_\text{in})\,W(v,u)
	\end{equation}
	\item introduce a Chebyshev-Lobatto grid in $u$-direction ($u\in[0,1]$) and a right-hand side Chebyshev-Radau grid (at some generic time interval $v\in[v_\text{a},v_\text{b}]$)
	\begin{equation}
	u_j=\frac12 \left(1+\cos\frac{\pi j}{N_z}\right), \  v_k=\frac 12 (v_\text{a}-v_\text{b})+(v_\text{a}-v_\text{b})\,\cos\frac{2\pi i}{2N_v+1},
	\end{equation}
	where $j=0,\ldots,N_u,$ and  $i=0,\ldots,N_v$.
	\item Replace the derivatives by their discrete version: $\partial_z\rightarrow D_z^\text{Lobatto}$ and $\partial_v\rightarrow D_v^\text{Radau}$ and impose the discrete version of the equation of motion on every grid point of the unit square
	\item Solve the resulting linear system
	\item Use the numerical solution $V_z(v_1,v)$ as new initial value for the next time domain to proceed iteratively in time until $v_\text{final}$
\end{enumerate}
Note that we can adjust the size of each time sub domain individually according to the quench profile.

\section{Convergence and numerical accuracy}\label{conv}
\subsection{Time evolution}
\begin{figure}[H] 
	\centering
	\includegraphics[width=11cm]{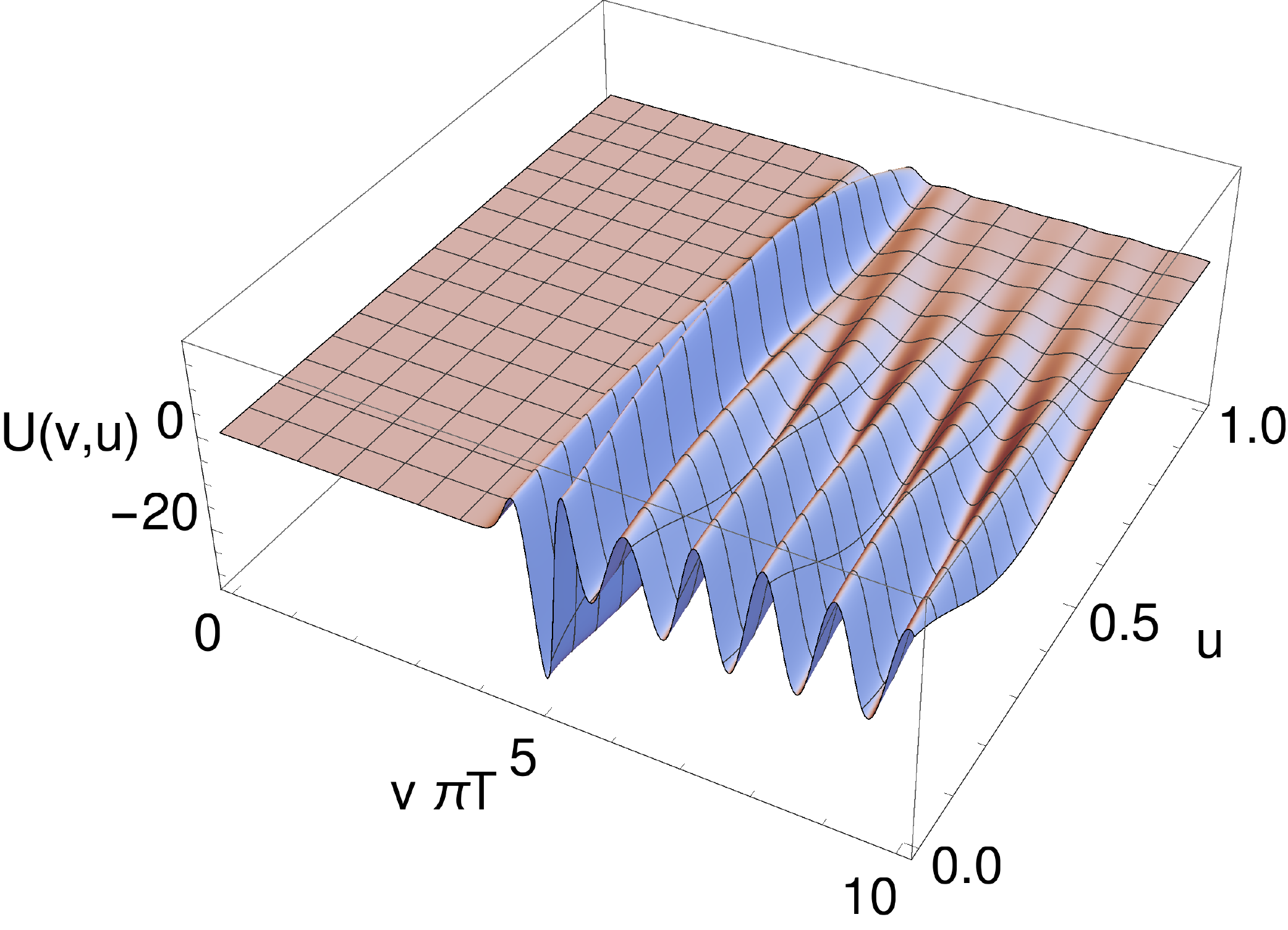}
	\caption{\label{fig:3dfield}Gaussian quench with center at $v_0$=$5\pi T$.}
\end{figure}

\begin{figure}[H] 
	\centering
	\includegraphics[width=11cm]{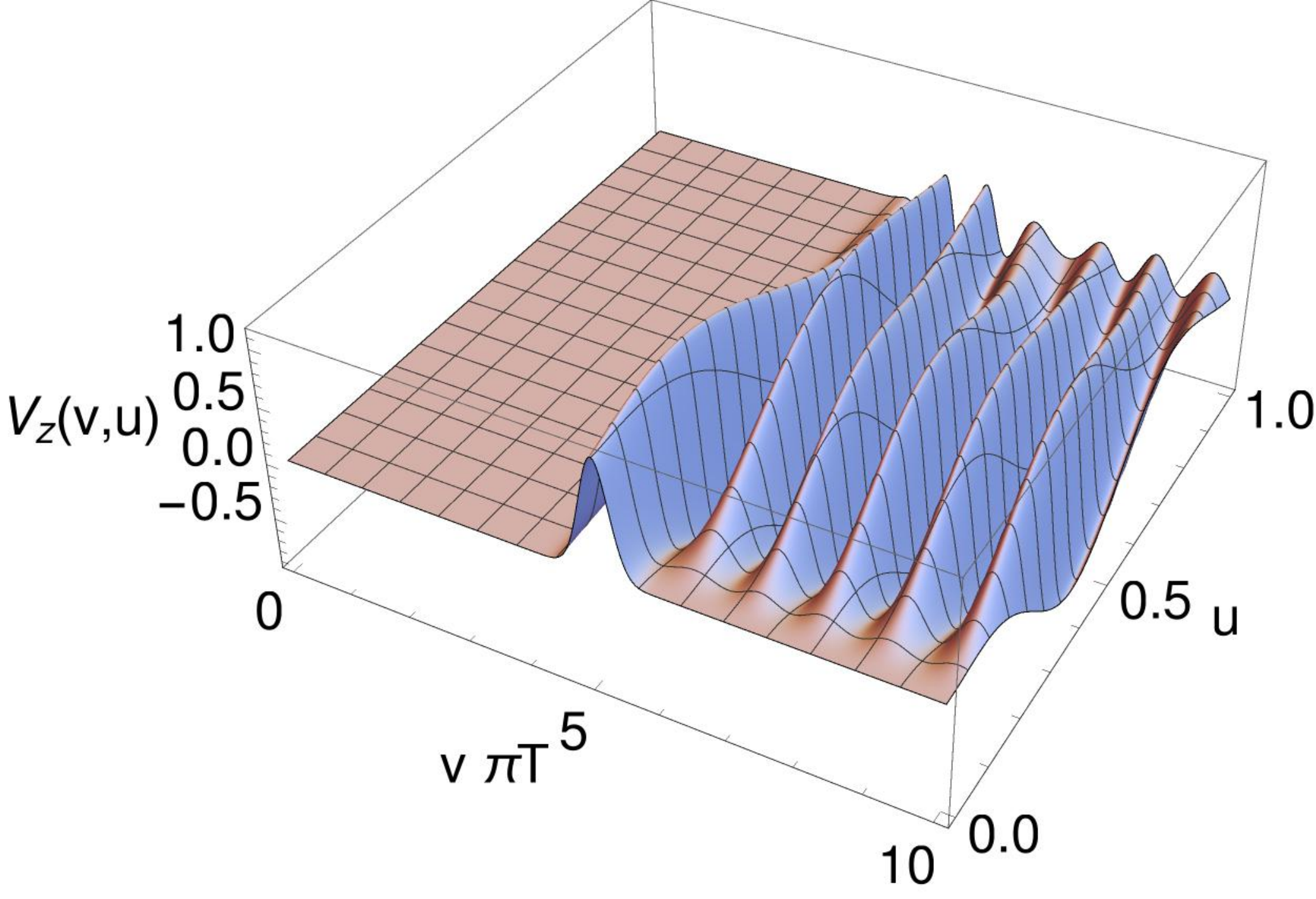}
	\caption{\label{fig:3dfield2}Gaussian quench with center at $v_0$=$5\pi T$.}
\end{figure}

\begin{figure}[H] 
	\centering
	\includegraphics[width=7cm]{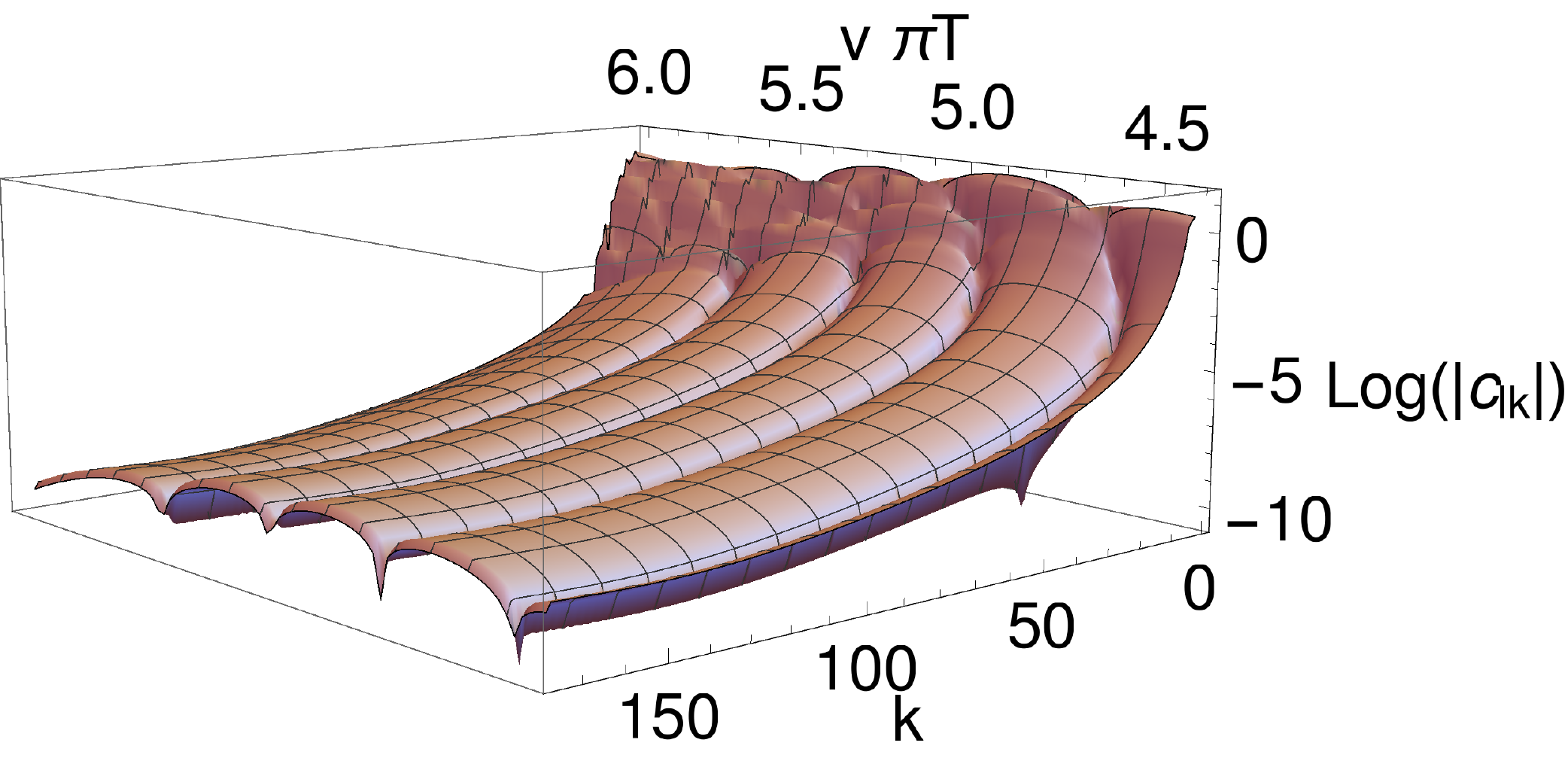}
	\hspace{0.4cm}
		\includegraphics[width=7cm]{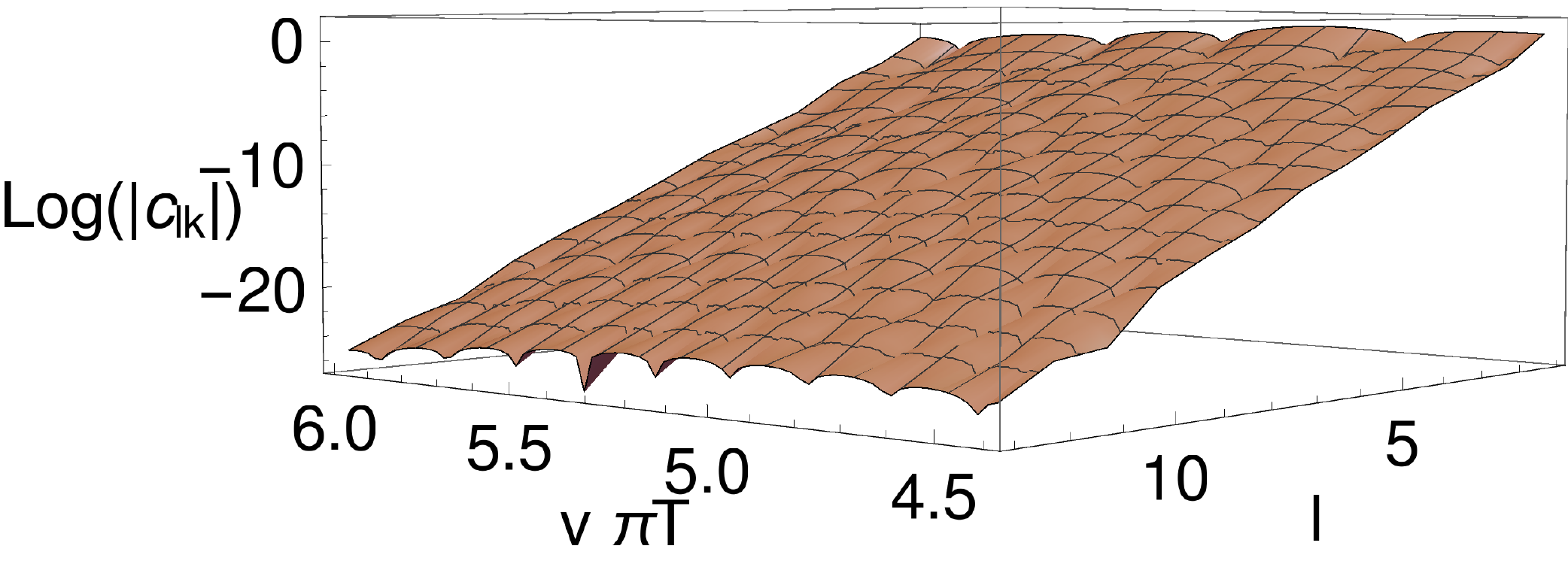}\newline
			\centering
			\includegraphics[width=6.7cm]{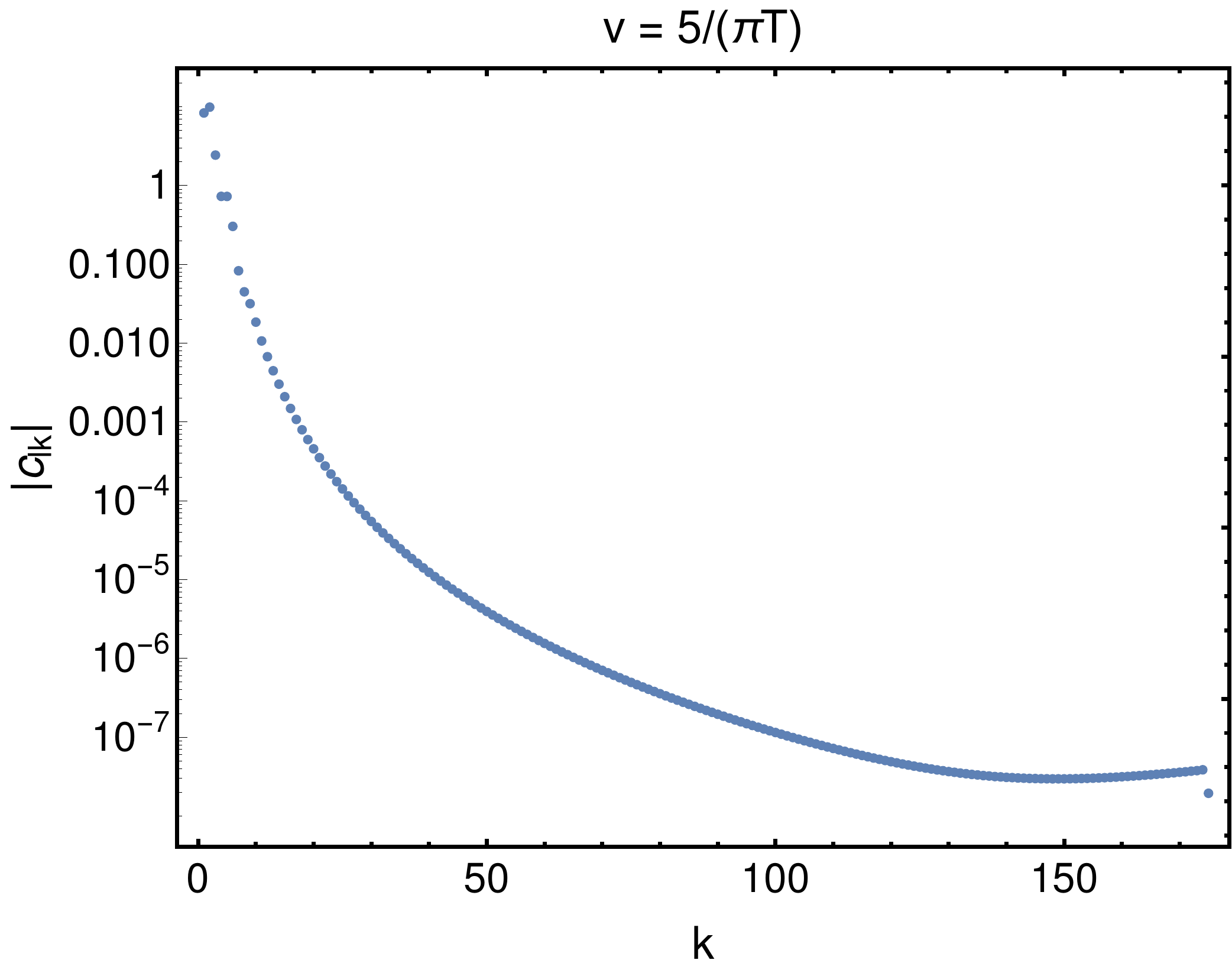}
			\hspace{1cm}
			\includegraphics[width=6.7cm]{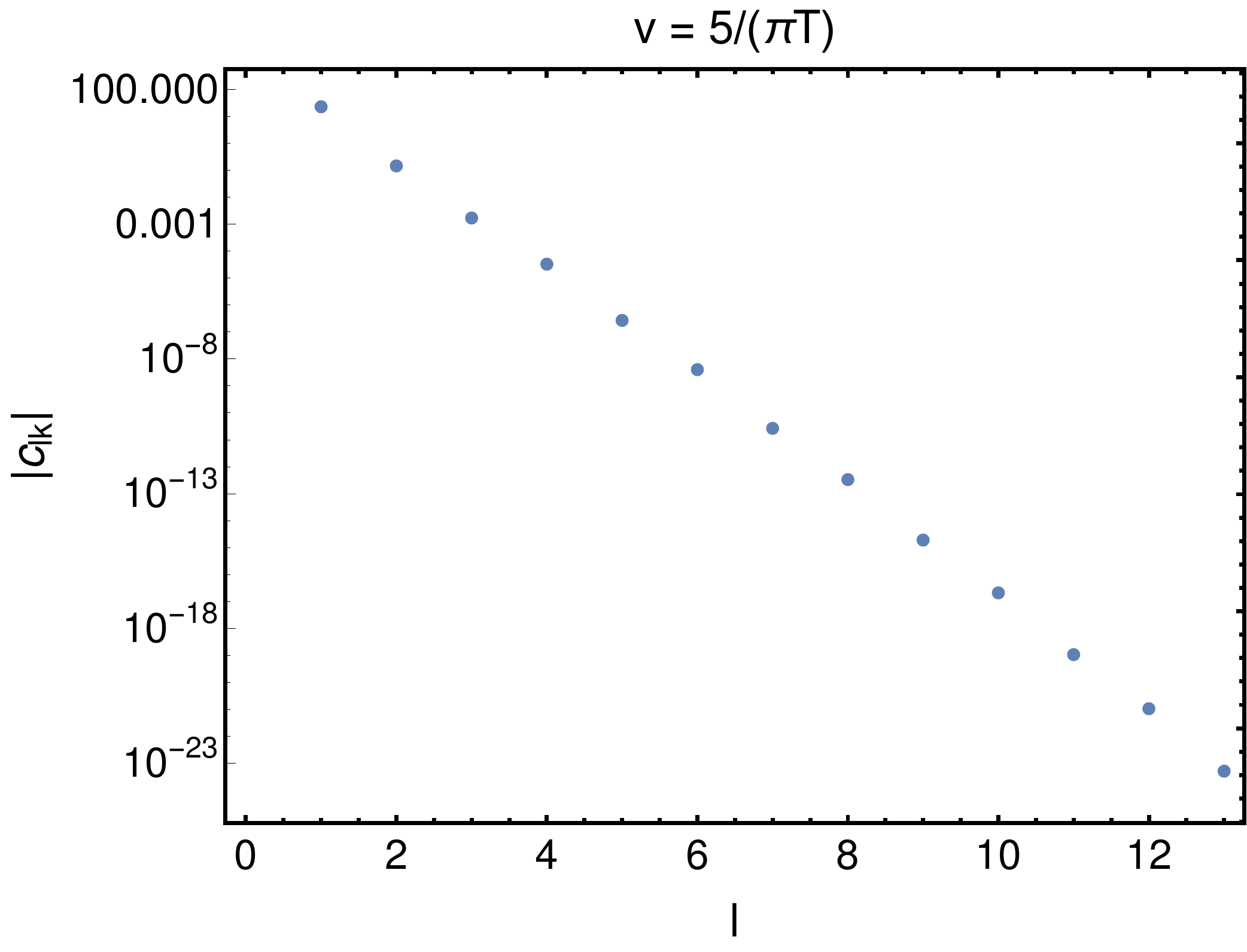}
	\caption{\label{fig:convq12}Left: Chebyshev coefficients in radial direction. Right: Chebyshev coefficients in the time direction.}
\end{figure}

In figure \ref{fig:3dfield} we depicted the auxiliary field during the time evolution for a Gaussian quench. We chose the quench parameter so that we excite with the quench an oscillation. The impact of quench propagates from $u=0$ to $u=1$. But unlike near the boundary the field near the horizon is damped to zero. This matches exactly our expectation, since we chose ingoing boundary condition and the field content is absorbed by the black hole.
In figure \ref{fig:3dfield2} we depicted the field $V_z$. The field contains the source, which determines is asymptotic behaviour. We notice again, that the field is damped in direction of the horizon.
\newpage
Since a fully spectral code is not the usual procedure to integrate in time direction, we show in figure the composition of the current with respect to the time intervals. Thereby each color represents one timestep. The dots of the same color within on time step represent the grid in time direction. In figure \ref{fig:convq12} we present the Chebyshev coefficients throughout the time evolution. It is apparent that near the quench the convergence is the worst. At this point we present the coefficients in time and the spatial direction. Furthermore, we checked our numerics using implicit (Crank-Nicolson) and explicit (4/5th order Runge-Kutta, 5th Adams–Bashforth) time propagators.
\begin{figure}[H] 
	\centering
	\includegraphics[width=6cm]{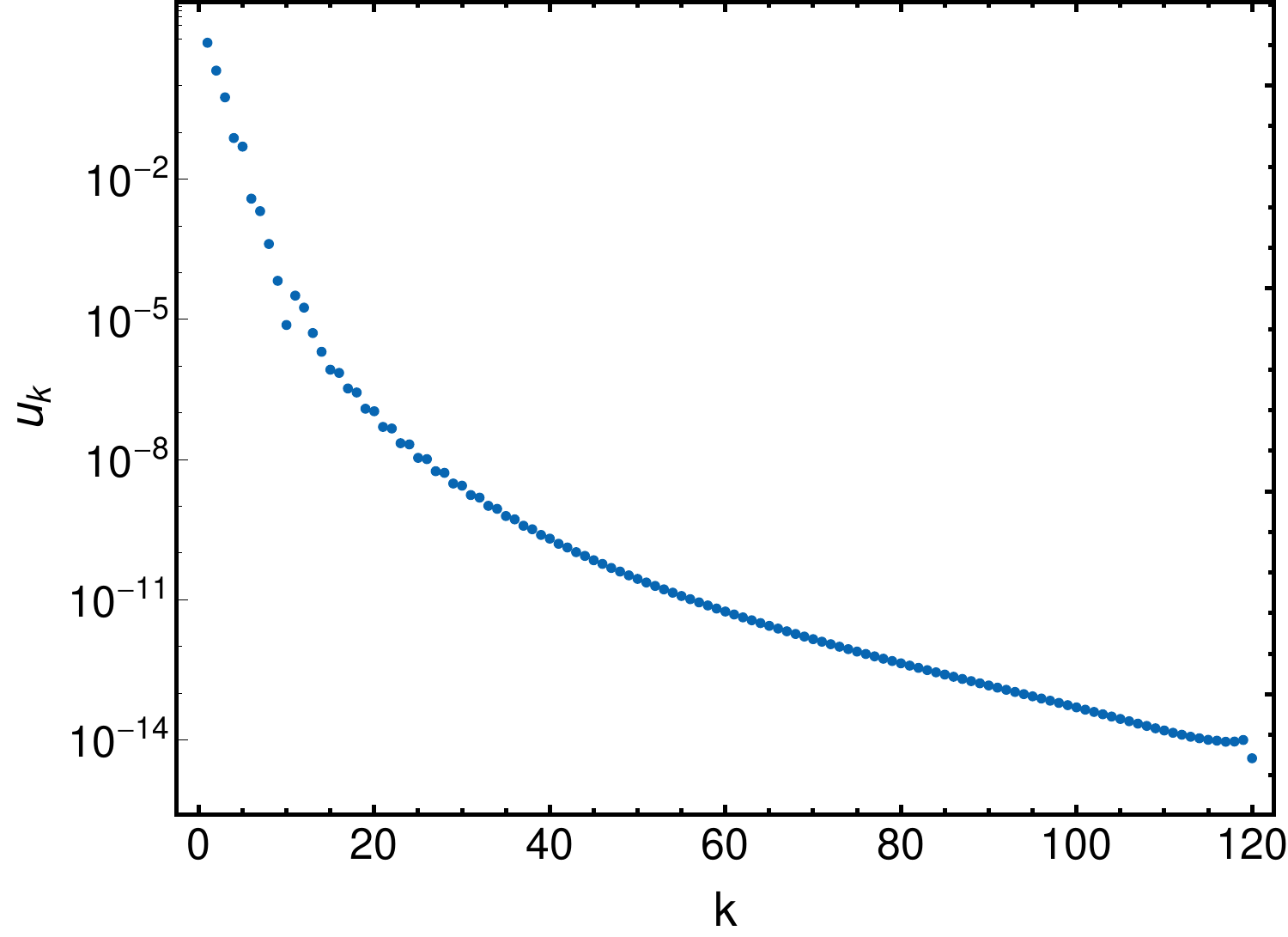}
	\hspace{0.4cm}
		\includegraphics[width=6cm]{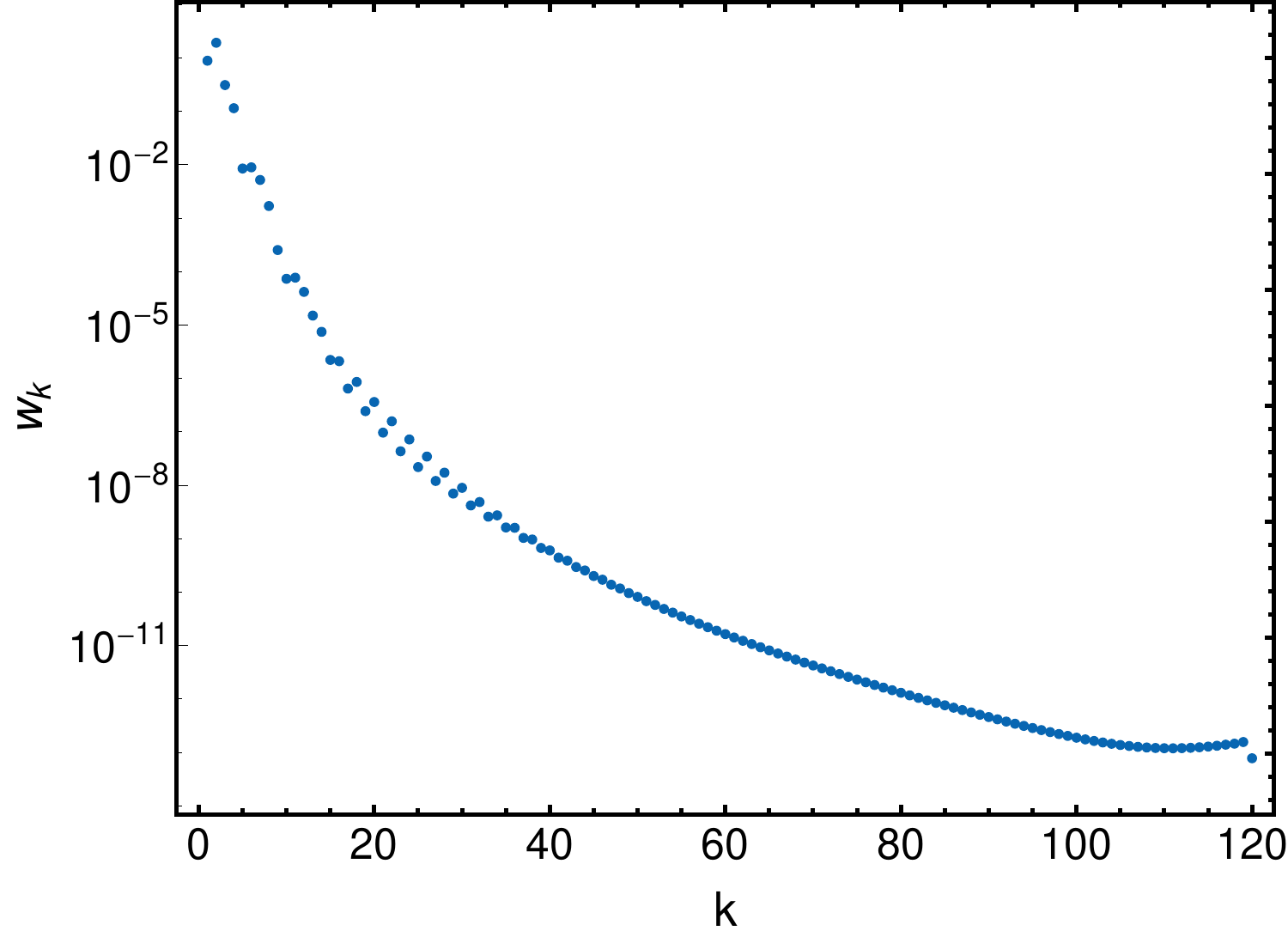}\newline\centering
			\includegraphics[width=6cm]{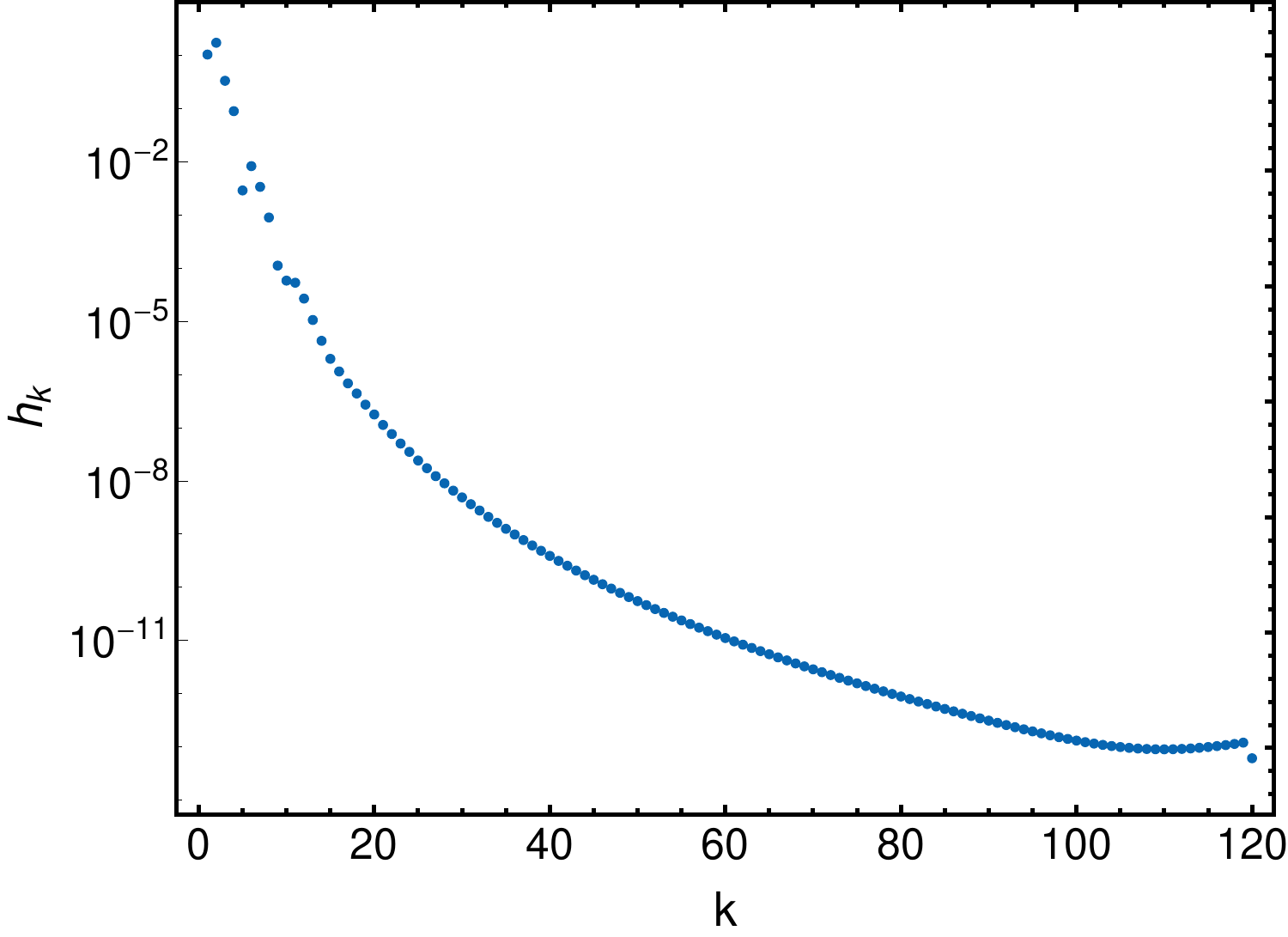}
	\caption{Chebyshev coefficients of the background solutions for $B=9$.\label{bac2}}
\end{figure}\begin{figure}[H] 
\centering
\includegraphics[width=6cm]{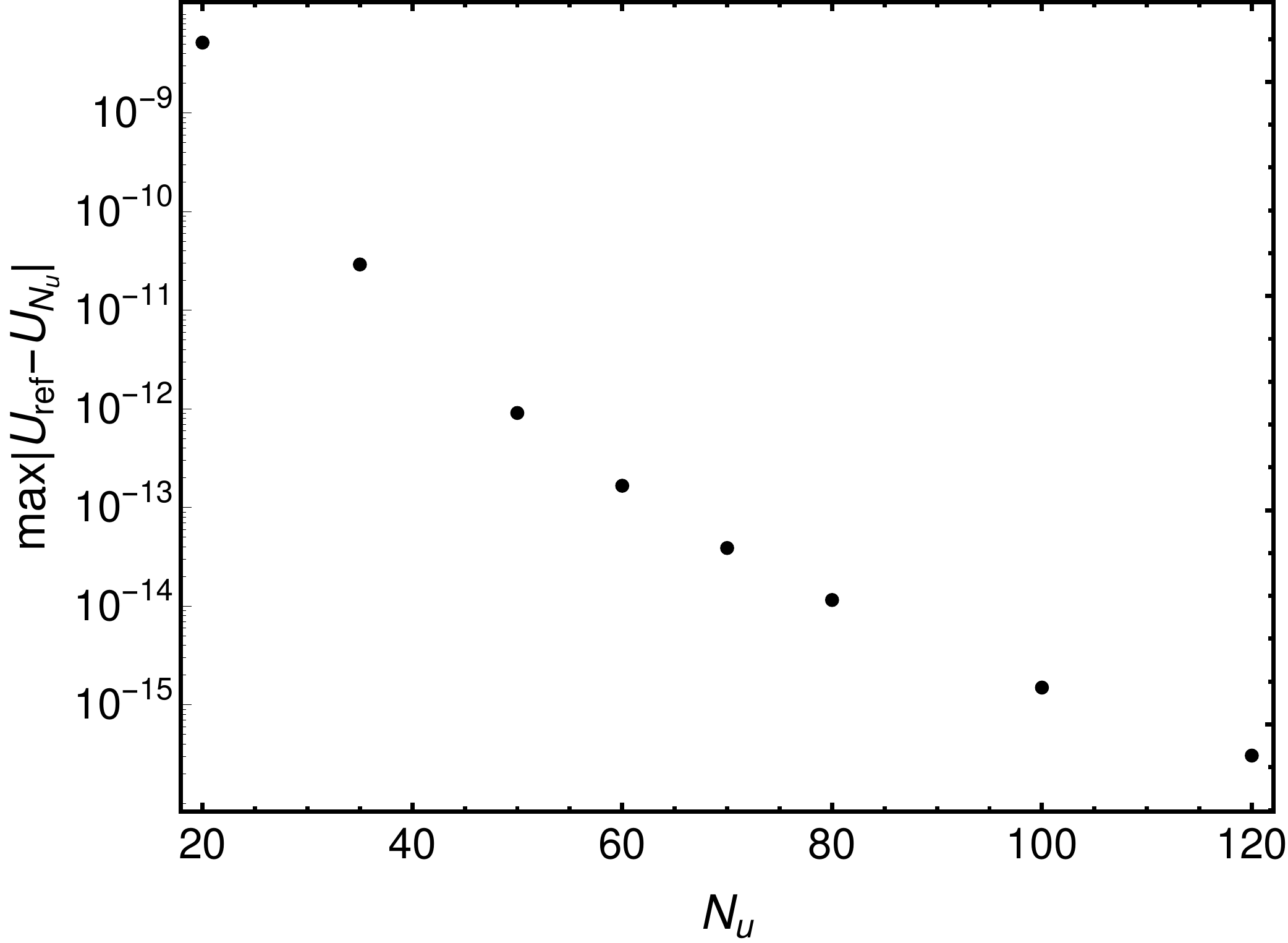}
\hspace{0.4cm}
\includegraphics[width=6cm]{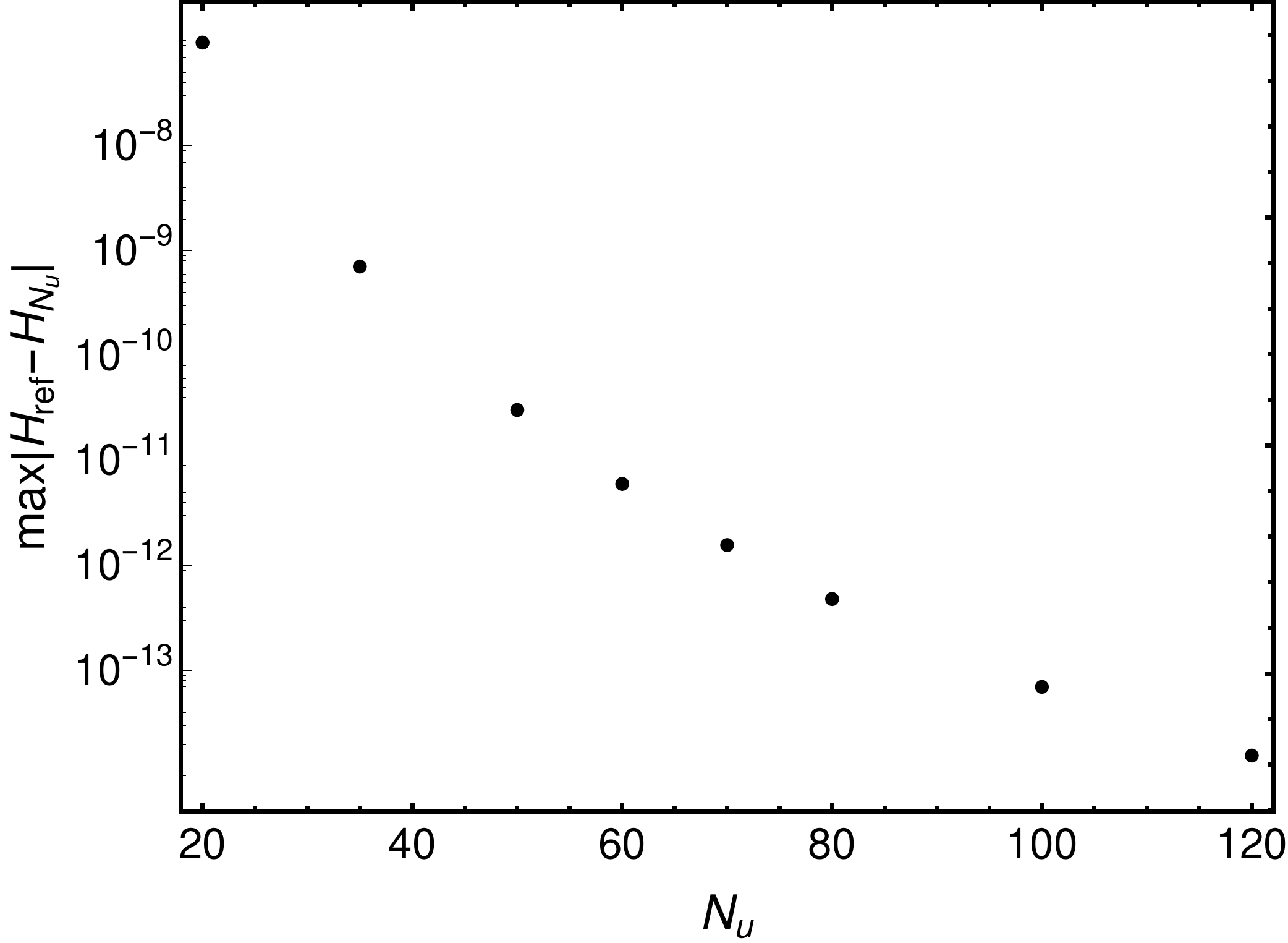}\newline\centering
\includegraphics[width=6cm]{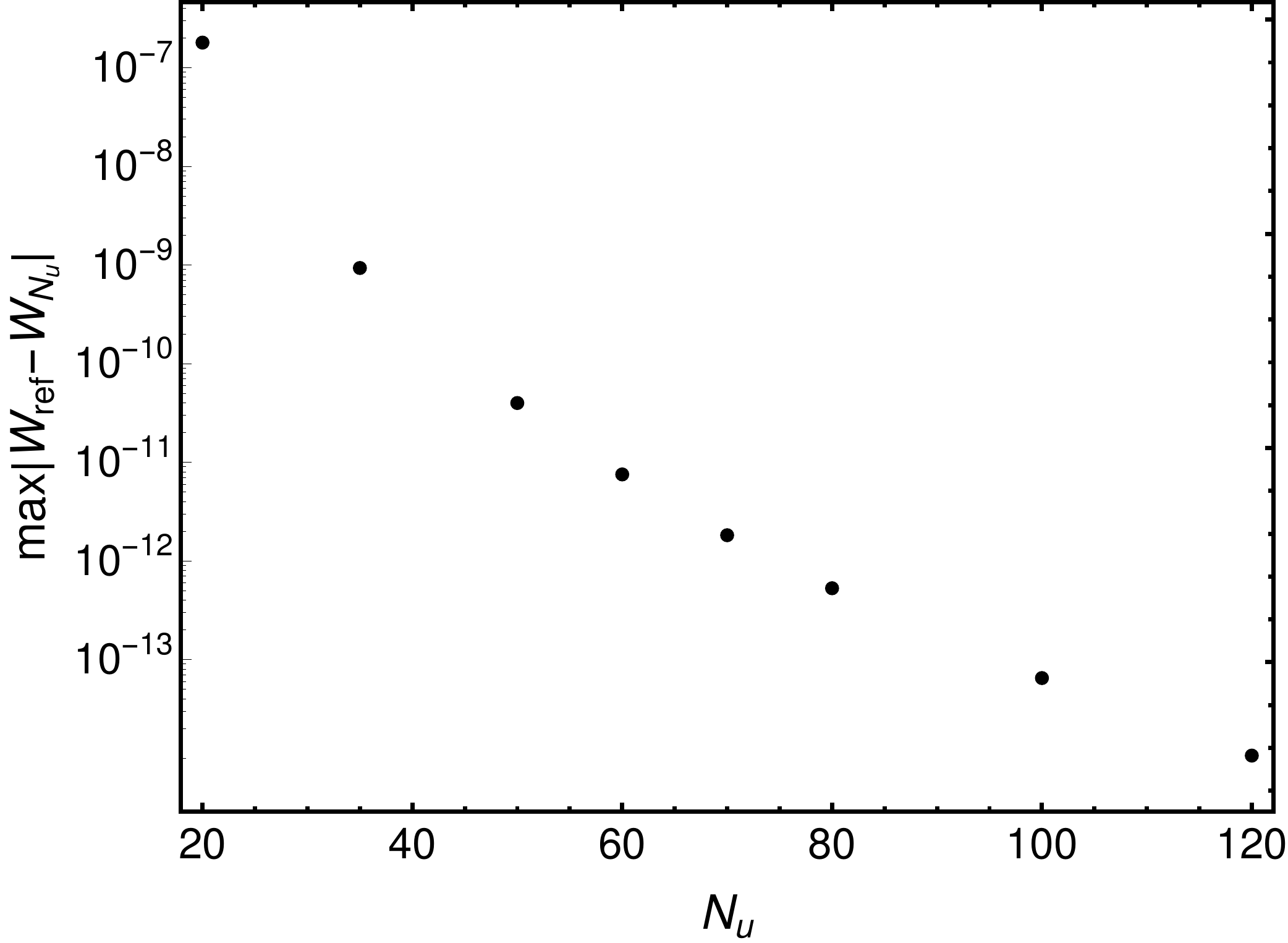}
\caption{Convergency test of the background solutions for $B=9$.\label{bac}}
\end{figure}
\subsection{Quasi-normal modes with backreaction}

\begin{figure}[H] 
	\centering
	\includegraphics[width=6.5cm]{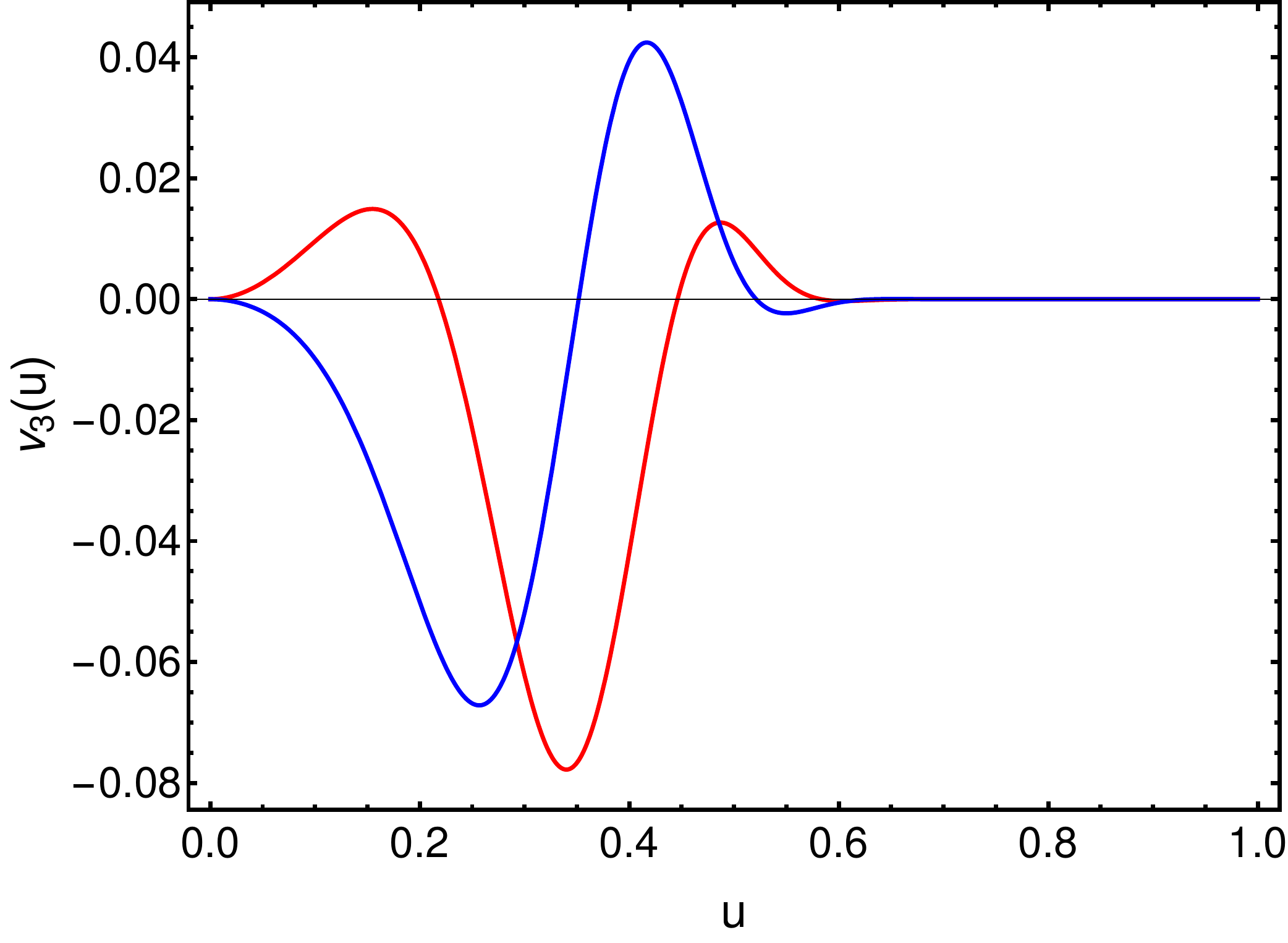}
	\hspace{1cm}
		\includegraphics[width=6.5cm]{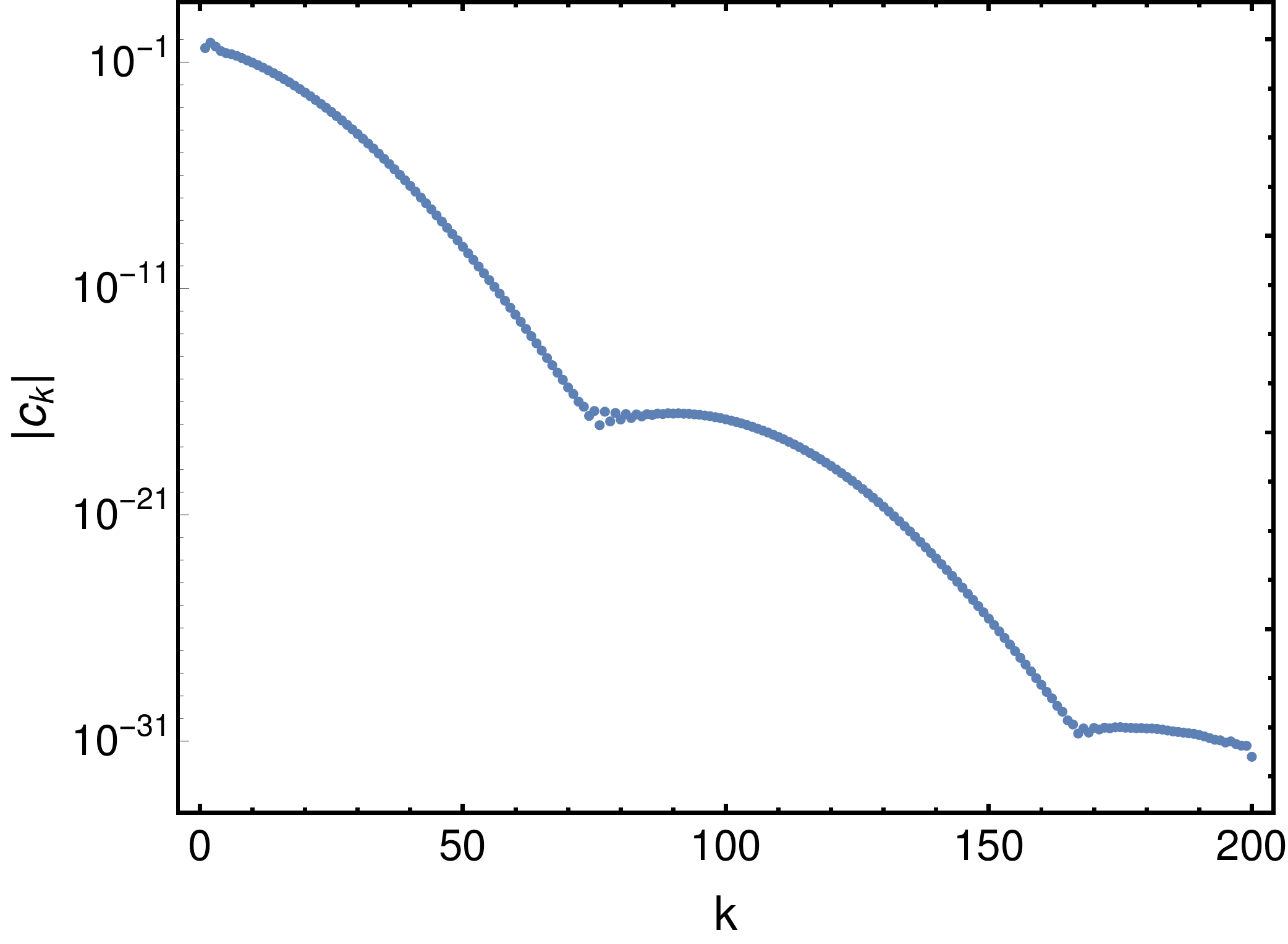}\newline
			\centering
			\includegraphics[width=6.5cm]{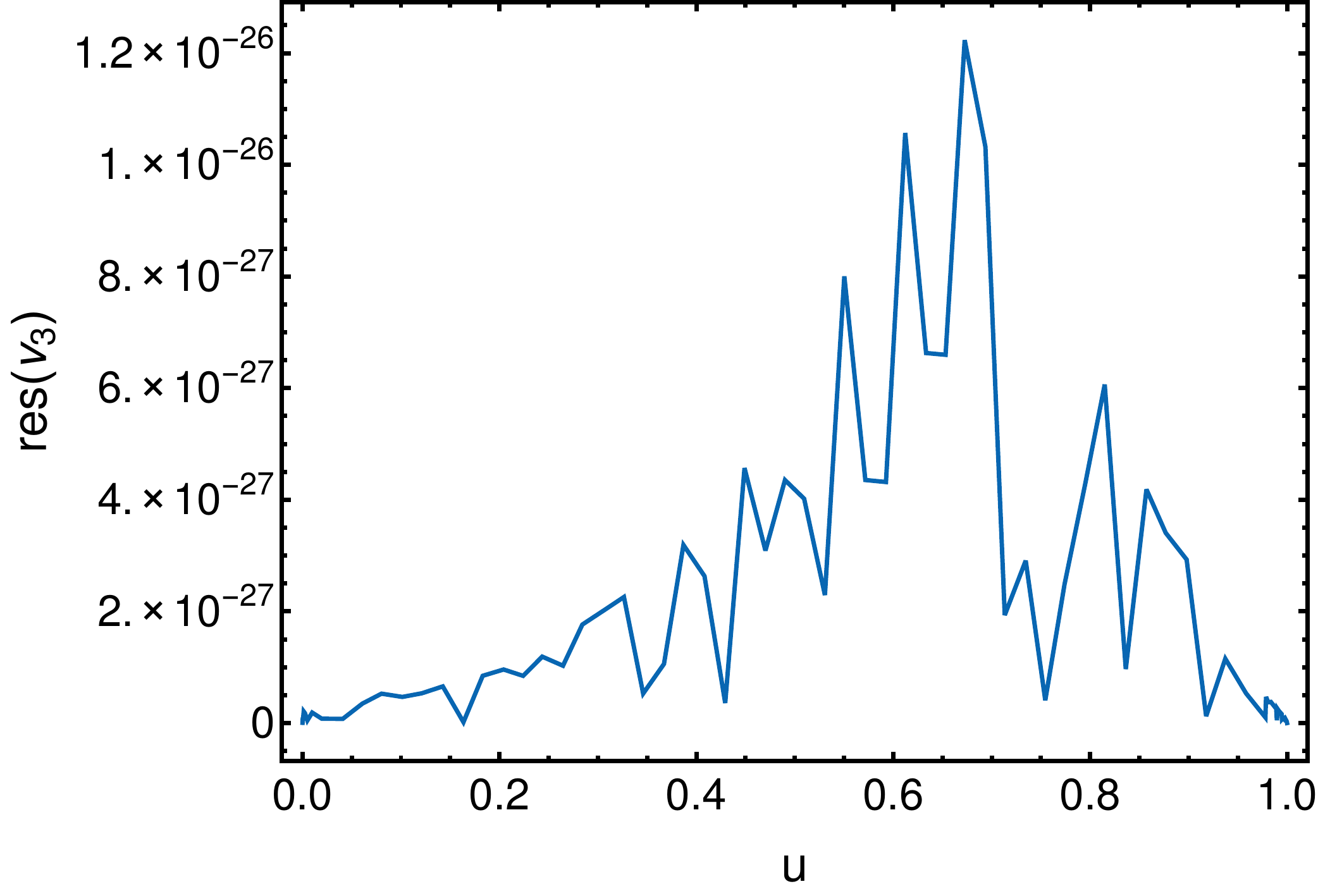}
			\hspace{1cm}
			\includegraphics[width=6.5cm]{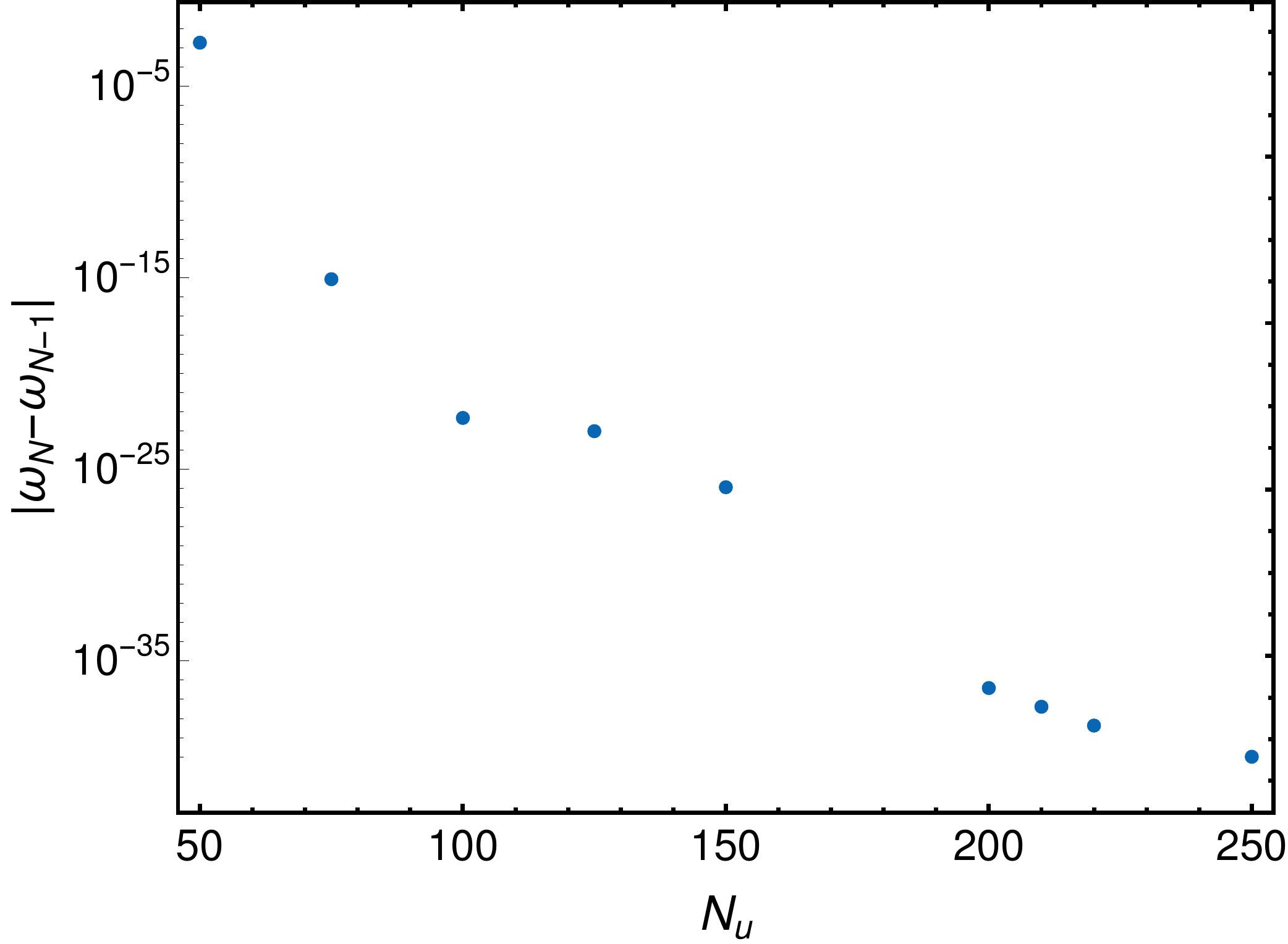}
	\caption{\label{fig:convq1}Eigenfunction of the lowest QNM in the case $B=9$, and their coefficients and convergence.}
\end{figure}

We now present the convergence plots for the QNMs in the backreacted case. In figure \ref{bac2} we show the convergence in the numerical worst case background functions. Concretely, we considered a magnetic field of magnitude $B=9$. The coefficients of the background solutions drop down to $10^{-14}$ in the worst case. The supremums-norm, which is the maximal deviation from the considered solution to a reference solution ($N_u=150$) is depicted in figure \ref{bac} and reveals that for $N_u=150$ the maximal deviation is $10^{-14}$. Furthermore, we investigated the convergence for the pertubations. In figure \ref{fig:convq1}, we show the convergence in the numerical worst case for the lowest QNM. The QNMs vary for $N_u\ge$ 80 less than machine precision. The differential equation for the pertubations is in the case $N_u=200$ gridpoints fulfilled up to $10^{-26}$. 
\end{appendix}\newpage
\addcontentsline{toc}{section}{References}
\bibliography{./Thesis_Arxiv.bbl}

\begin{thebibliography}{10}

\bibitem{Ammon:2016fru}
M.~Ammon, S.~Grieninger, A.~Jimenez-Alba, R.~P. Macedo, and L.~Melgar,
  ``{Holographic quenches and anomalous transport},'' 2016.

\bibitem{Maldacena:1997re}
J.~M. Maldacena, ``{The Large N limit of superconformal field theories and
  supergravity},'' {\em Int. J. Theor. Phys.}, vol.~38, pp.~1113--1133, 1999.
\newblock [Adv. Theor. Math. Phys.2,231(1998)].

\bibitem{Fukushima:2008xe}
K.~Fukushima, D.~E. Kharzeev, and H.~J. Warringa, ``{The Chiral Magnetic
  Effect},'' {\em Phys. Rev.}, vol.~D78, p.~074033, 2008.

\bibitem{Son:2009tf}
D.~T. Son and P.~Surowka, ``{Hydrodynamics with Triangle Anomalies},'' {\em
  Phys. Rev. Lett.}, vol.~103, p.~191601, 2009.

\bibitem{Kharzeev:2013ffa}
D.~E. Kharzeev, ``{The Chiral Magnetic Effect and Anomaly-Induced Transport},''
  {\em Prog. Part. Nucl. Phys.}, vol.~75, pp.~133--151, 2014.

\bibitem{Kharzeev:2010gd}
D.~E. Kharzeev and H.-U. Yee, ``{Chiral Magnetic Wave},'' {\em Phys.Rev.},
  vol.~D83, p.~085007, 2011.

\bibitem{Nielsen:1983rb}
H.~Nielsen and M.~Ninomiya, ``The adler-bell-jackiw anomaly and weyl fermions
  in a crystal,'' {\em Physics Letters B}, vol.~130, no.~6, pp.~389 -- 396,
  1983.

\bibitem{Heinz:2008tv}
U.~W. Heinz, ``{The Strongly coupled quark-gluon plasma created at RHIC},''
  {\em J. Phys.}, vol.~A42, p.~214003, 2009.

\bibitem{Shuryak:2008eq}
E.~Shuryak, ``{Physics of Strongly coupled Quark-Gluon Plasma},'' {\em Prog.
  Part. Nucl. Phys.}, vol.~62, pp.~48--101, 2009.

\bibitem{Li:2014bha}
Q.~Li, D.~E. Kharzeev, C.~Zhang, Y.~Huang, I.~Pletikosic, A.~V. Fedorov, R.~D.
  Zhong, J.~A. Schneeloch, G.~D. Gu, and T.~Valla, ``{Observation of the chiral
  magnetic effect in ZrTe5},'' {\em Nature Phys.}, vol.~12, pp.~550--554, 2016.

\bibitem{Li2015}
C.-Z. {Li}, L.-X. {Wang}, H.~{Liu}, J.~{Wang}, Z.-M. {Liao}, and D.-P. {Yu},
  ``{Giant negative magnetoresistance induced by the chiral anomaly in
  individual Cd$_{3}$As$_{2}$ nanowires},'' {\em Nature Communications},
  vol.~6, p.~10137, Dec. 2015.

\bibitem{2015PhRvX...5c1023H}
X.~{Huang}, L.~{Zhao}, Y.~{Long}, P.~{Wang}, D.~{Chen}, Z.~{Yang}, H.~{Liang},
  M.~{Xue}, H.~{Weng}, Z.~{Fang}, X.~{Dai}, and G.~{Chen}, ``{Observation of
  the Chiral-Anomaly-Induced Negative Magnetoresistance in 3D Weyl Semimetal
  TaAs},'' {\em Physical Review X}, vol.~5, p.~031023, July 2015.

\bibitem{2016NatCo...710301L}
H.~{Li}, H.~{He}, H.-Z. {Lu}, H.~{Zhang}, H.~{Liu}, R.~{Ma}, Z.~{Fan}, S.-Q.
  {Shen}, and J.~{Wang}, ``{Negative magnetoresistance in Dirac semimetal
  Cd$_{3}$As$_{2}$},'' {\em Nature Communications}, vol.~7, p.~10301, Jan.
  2016.

\bibitem{Yee:2009vw}
H.-U. Yee, ``{Holographic Chiral Magnetic Conductivity},'' {\em JHEP}, vol.~11,
  p.~085, 2009.

\bibitem{Landsteiner:2013aba}
K.~Landsteiner, E.~Megias, and F.~Pena-Benitez, ``{Frequency dependence of the
  Chiral Vortical Effect},'' {\em Phys. Rev.}, vol.~D90, no.~6, p.~065026,
  2014.

\bibitem{Fukushima:2015tza}
K.~Fukushima, ``{Simulating net particle production and chiral magnetic current
  in a $CP$-odd domain},'' {\em Phys. Rev.}, vol.~D92, no.~5, p.~054009, 2015.

\bibitem{Iwazaki:2009bg}
A.~Iwazaki, ``{Pair Creation in Electric Flux Tube and Chiral Anomaly},'' {\em
  Phys. Rev.}, vol.~C80, p.~052202, 2009.

\bibitem{Lin:2013sga}
S.~Lin and H.-U. Yee, ``{Out-of-Equilibrium Chiral Magnetic Effect at Strong
  Coupling},'' {\em Phys. Rev.}, vol.~D88, no.~2, p.~025030, 2013.

\bibitem{Basar:2013iaa}
G.~Basar, D.~E. Kharzeev, and H.-U. Yee, ``{Triangle anomaly in Weyl
  semimetals},'' {\em Phys. Rev.}, vol.~B89, no.~3, p.~035142, 2014.

\bibitem{Buchel:2013lla}
A.~Buchel, L.~Lehner, R.~C. Myers, and A.~van Niekerk, ``{Quantum quenches of
  holographic plasmas},'' {\em JHEP}, vol.~05, p.~067, 2013.

\bibitem{Bertlmann:1996xk}
R.~Bertlmann, {\em Anomalies in Quantum Field Theory}.
\newblock International Series of Monographs on Physics, Clarendon Press, 2000.

\bibitem{SUTHERLAND1967433}
D.~Sutherland, ``Current algebra and some non-strong mesonic decays,'' {\em
  Nuclear Physics B}, vol.~2, no.~4, pp.~433 -- 440, 1967.

\bibitem{10.2307/2415932}
M.~Veltman, ``I. theoretical aspects of high energy neutrino interactions,''
  {\em Proceedings of the Royal Society of London. Series A, Mathematical and
  Physical Sciences}, vol.~301, no.~1465, pp.~107--112, 1967.

\bibitem{PhysRev.177.2426}
S.~L. Adler, ``Axial-vector vertex in spinor electrodynamics,'' {\em Phys.
  Rev.}, vol.~177, pp.~2426--2438, Jan 1969.

\bibitem{Bell:1969ts}
J.~S. Bell and R.~Jackiw, ``{A PCAC puzzle: pi0 --> gamma gamma in the sigma
  model},'' {\em Nuovo Cim.}, vol.~A60, pp.~47--61, 1969.

\bibitem{PhysRev.182.1517}
S.~L. Adler and W.~A. Bardeen, ``Absence of higher-order corrections in the
  anomalous axial-vector divergence equation,'' {\em Phys. Rev.}, vol.~182,
  pp.~1517--1536, Jun 1969.

\bibitem{BARDEEN1984421}
W.~A. Bardeen and B.~Zumino, ``Consistent and covariant anomalies in gauge and
  gravitational theories,'' {\em Nuclear Physics B}, vol.~244, no.~2, pp.~421
  -- 453, 1984.

\bibitem{amadeo}
A.~Jimenez, ``Broken symmetries and transport in holography,'' 2015.

\bibitem{Fukushima:2012vr}
K.~Fukushima, ``{Views of the Chiral Magnetic Effect},'' {\em Lect. Notes
  Phys.}, vol.~871, pp.~241--259, 2013.

\bibitem{Susskind:1994vu}
L.~Susskind, ``{The World as a hologram},'' {\em J. Math. Phys.}, vol.~36,
  pp.~6377--6396, 1995.

\bibitem{Stephens:1993an}
C.~R. Stephens, G.~'t~Hooft, and B.~F. Whiting, ``{Black hole evaporation
  without information loss},'' {\em Class. Quant. Grav.}, vol.~11,
  pp.~621--648, 1994.

\bibitem{1975CMaPh..43..199H}
S.~W. {Hawking}, ``{Particle creation by black holes},'' {\em Communications in
  Mathematical Physics}, vol.~43, pp.~199--220, Aug. 1975.

\bibitem{PhysRevD.7.2333}
J.~D. Bekenstein, ``Black holes and entropy,'' {\em Phys. Rev. D}, vol.~7,
  pp.~2333--2346, Apr 1973.

\bibitem{Gubser:1998bc}
S.~S. Gubser, I.~R. Klebanov, and A.~M. Polyakov, ``{Gauge theory correlators
  from noncritical string theory},'' {\em Phys. Lett.}, vol.~B428,
  pp.~105--114, 1998.

\bibitem{Witten:1998qj}
E.~Witten, ``{Anti-de Sitter space and holography},'' {\em Adv. Theor. Math.
  Phys.}, vol.~2, pp.~253--291, 1998.

\bibitem{Ammon:2015wua}
M.~Ammon and J.~Erdmenger, {\em {Gauge/gravity duality}}.
\newblock Cambridge, UK: Cambridge Univ. Pr., 2015.

\bibitem{Nastase}
H.~Nastase, {\em {Introduction to the AdS/CFT Correspondence}}.
\newblock Cambridge, UK: Cambridge Univ. Pr., 2015.

\bibitem{Schalm}
J.~Zaanen, Y.~Liu, Y.-W. Sun, and K.~Schalm, {\em {Holographic Duality in
  Condensed Matter Physics}}.
\newblock Cambridge, UK: Cambridge Univ. Pr., 2016.

\bibitem{Ramallo:2013bua}
A.~V. Ramallo, ``{Introduction to the AdS/CFT correspondence},'' 2013.

\bibitem{Hartnoll:2009sz}
S.~A. Hartnoll, ``{Lectures on holographic methods for condensed matter
  physics},'' {\em Class.Quant.Grav.}, vol.~26, p.~224002, 2009.

\bibitem{D'Hoker:2002aw}
E.~D'Hoker and D.~Z. Freedman, ``{Supersymmetric gauge theories and the AdS /
  CFT correspondence},'' pp.~3--158, 2002.

\bibitem{Freedman:1998tz}
D.~Z. Freedman, S.~D. Mathur, A.~Matusis, and L.~Rastelli, ``{Correlation
  functions in the CFT(d) / AdS(d+1) correspondence},'' {\em Nucl.Phys.},
  vol.~B546, pp.~96--118, 1999.

\bibitem{Papadimitriou:2004ap}
I.~Papadimitriou and K.~Skenderis, ``{AdS / CFT correspondence and geometry},''
  {\em IRMA Lect. Math. Theor. Phys.}, vol.~8, pp.~73--101, 2005.

\bibitem{papadimitriou2005aspects}
I.~Papadimitriou, ``Aspects of the gauge/gravity duality and holography,''
  2005.

\bibitem{Henningson:1998ey}
M.~Henningson and K.~Skenderis, ``{Holography and the Weyl anomaly},'' {\em
  Fortsch. Phys.}, vol.~48, pp.~125--128, 2000.

\bibitem{deHaro:2000vlm}
S.~de~Haro, S.~N. Solodukhin, and K.~Skenderis, ``{Holographic reconstruction
  of space-time and renormalization in the AdS / CFT correspondence},'' {\em
  Commun. Math. Phys.}, vol.~217, pp.~595--622, 2001.

\bibitem{Bianchi:2001kw}
M.~Bianchi, D.~Z. Freedman, and K.~Skenderis, ``{Holographic
  renormalization},'' {\em Nucl. Phys.}, vol.~B631, pp.~159--194, 2002.

\bibitem{Skenderis:2002wp}
K.~Skenderis, ``{Lecture notes on holographic renormalization},'' {\em Class.
  Quant. Grav.}, vol.~19, pp.~5849--5876, 2002.

\bibitem{PhysRev.116.1322}
R.~Arnowitt, S.~Deser, and C.~W. Misner, ``Dynamical structure and definition
  of energy in general relativity,'' {\em Phys. Rev.}, vol.~116,
  pp.~1322--1330, Dec 1959.

\bibitem{lrr-1999-2}
K.~D. Kokkotas and B.~G. Schmidt, ``Quasi-normal modes of stars and black
  holes,'' {\em Living Reviews in Relativity}, vol.~2, no.~2, 1999.

\bibitem{Kovtun:2005ev}
P.~K. Kovtun and A.~O. Starinets, ``{Quasinormal modes and holography},'' {\em
  Phys. Rev.}, vol.~D72, p.~086009, 2005.

\bibitem{Sachs:2003zj}
I.~Sachs, ``{Quasinormal modes},'' {\em Fortsch. Phys.}, vol.~52, pp.~667--671,
  2004.

\bibitem{Horowitz:1999jd}
G.~T. Horowitz and V.~E. Hubeny, ``{Quasinormal modes of AdS black holes and
  the approach to thermal equilibrium},'' {\em Phys. Rev.}, vol.~D62,
  p.~024027, 2000.

\bibitem{Berti:2009kk}
E.~Berti, V.~Cardoso, and A.~O. Starinets, ``{Quasinormal modes of black holes
  and black branes},'' {\em Class. Quant. Grav.}, vol.~26, p.~163001, 2009.

\bibitem{Gynther:2010ed}
A.~Gynther, K.~Landsteiner, F.~Pena-Benitez, and A.~Rebhan, ``{Holographic
  Anomalous Conductivities and the Chiral Magnetic Effect},'' {\em JHEP},
  vol.~02, p.~110, 2011.

\bibitem{Jimenez-Alba:2014iia}
A.~Jimenez-Alba, K.~Landsteiner, and L.~Melgar, ``{Anomalous magnetoresponse
  and the Stueckelberg axion in holography},'' {\em Phys. Rev.}, vol.~D90,
  p.~126004, 2014.

\bibitem{Jimenez-Alba:2014pea}
A.~Jimenez-Alba and L.~Melgar, ``{Anomalous Transport in Holographic Chiral
  Superfluids via Kubo Formulae},'' {\em JHEP}, vol.~10, p.~120, 2014.

\bibitem{kharzeev2014strongly}
D.~Kharzeev, K.~Landsteiner, A.~Schmitt, and H.~Yee, {\em Strongly Interacting
  Matter in Magnetic Fields}.
\newblock Lecture Notes in Physics, Springer Berlin Heidelberg, 2014.

\bibitem{Buchel:2012gw}
A.~Buchel, L.~Lehner, and R.~C. Myers, ``{Thermal quenches in N=2* plasmas},''
  {\em JHEP}, vol.~08, p.~049, 2012.

\bibitem{Das:2014hqa}
S.~R. Das, D.~A. Galante, and R.~C. Myers, ``{Universality in fast quantum
  quenches},'' {\em JHEP}, vol.~02, p.~167, 2015.

\bibitem{Kokkotas:1999bd}
K.~D. Kokkotas and B.~G. Schmidt, ``{Quasinormal modes of stars and black
  holes},'' {\em Living Rev. Rel.}, vol.~2, p.~2, 1999.

\bibitem{PhysRevD.45.2617}
H.-P. Nollert and B.~G. Schmidt, ``Quasinormal modes of schwarzschild black
  holes: Defined and calculated via laplace transformation,'' {\em Phys. Rev.
  D}, vol.~45, pp.~2617--2627, Apr 1992.

\bibitem{D'Hoker:2009mm}
E.~D'Hoker and P.~Kraus, ``{Magnetic Brane Solutions in AdS},'' {\em JHEP},
  vol.~10, p.~088, 2009.

\bibitem{Macedo:2014bfa}
R.~Panosso~Macedo and M.~Ansorg, ``{Axisymmetric fully spectral code for
  hyperbolic equations},'' {\em J. Comput. Phys.}, vol.~276, pp.~357--379,
  2014.

\bibitem{Buividovich:2016ulp}
P.~V. Buividovich and S.~N. Valgushev, ``{First experience with
  classical-statistical real-time simulations of anomalous transport with
  overlap fermions},'' 2016.
\newblock [PoSLATTICE2016,253(2016)].

\bibitem{D'Hoker:2010rz}
E.~D'Hoker and P.~Kraus, ``{Holographic Metamagnetism, Quantum Criticality, and
  Crossover Behavior},'' {\em JHEP}, vol.~05, p.~083, 2010.

\bibitem{D'Hoker:2010ij}
E.~D'Hoker and P.~Kraus, ``{Magnetic Field Induced Quantum Criticality via new
  Asymptotically AdS5 Solutions},'' {\em Class. Quant. Grav.}, vol.~27,
  p.~215022, 2010.

\bibitem{Ammon:2016szz}
M.~Ammon, J.~Leiber, and R.~P. Macedo, ``{Phase diagram of 4D field theories
  with chiral anomaly from holography},'' {\em JHEP}, vol.~03, p.~164, 2016.

\bibitem{Landsteiner:2015lsa}
K.~Landsteiner and Y.~Liu, ``{The holographic Weyl semi-metal},'' {\em Phys.
  Lett.}, vol.~B753, pp.~453--457, 2016.

\bibitem{Fuini:2015hba}
J.~F. Fuini and L.~G. Yaffe, ``{Far-from-equilibrium dynamics of a strongly
  coupled non-Abelian plasma with non-zero charge density or external magnetic
  field},'' {\em JHEP}, vol.~07, p.~116, 2015.

\bibitem{2016arXiv160102316Y}
X.~{Yuan}, P.~{Cheng}, L.~{Zhang}, C.~{Zhang}, J.~{Wang}, Y.~{Liu}, Q.~{Sun},
  P.~{Zhou}, D.~W. {Zhang}, Z.~{Hu}, X.~{Wan}, H.~{Yan}, Z.~{Li}, and F.~{Xiu},
  ``{Direct observation of Landau level resonance and mass generation in Dirac
  semimetal Cd3As2 thin films},'' {\em ArXiv e-prints}, Jan. 2016.

\bibitem{PhysRevLett.115.176404}
R.~Y. Chen, Z.~G. Chen, X.-Y. Song, J.~A. Schneeloch, G.~D. Gu, F.~Wang, and
  N.~L. Wang, ``Magnetoinfrared spectroscopy of landau levels and zeeman
  splitting of three-dimensional massless dirac fermions in
  ${\mathrm{zrte}}_{5}$,'' {\em Phys. Rev. Lett.}, vol.~115, p.~176404, Oct
  2015.

\bibitem{Boyd00}
J.~P. Boyd, {\em Chebyshev and Fourier Spectral Methods (Second Edition,
  Revised)}.
\newblock New York: Dover Publications, 2001.

\bibitem{lrr-2009-1}
P.~Grandclément and J.~Novak, ``Spectral methods for numerical relativity,''
  {\em Living Reviews in Relativity}, vol.~12, no.~1, 2009.

\bibitem{canuto2006erratum}
C.~Canuto, M.~Y. Hussaini, A.~Quarteroni, and T.~A. Zang, ``Spectral methods:
  Fundamentals in single domains,'' {\em Spectral Methods: Fundamentals in
  Single Domains}, pp.~e1--e4, 2006.

\bibitem{Hennig:2008af}
J.~Hennig and M.~Ansorg, ``{A Fully Pseudospectral Scheme for Solving Singular
  Hyperbolic Equations on Conformally Compactified Space-Times},'' {\em J.
  Hyperbol. Diff. Equat.}, vol.~6, p.~161, 2009.

\bibitem{Hennig:2012zx}
J.~Hennig, ``{Fully pseudospectral time evolution and its application to 1+1
  dimensional physical problems},'' {\em J. Comput. Phys.}, vol.~235,
  pp.~322--333, 2013.

\end{thebibliography}
\bibliographystyle{ieeetr}

\end{document}